%% file: tft.tex
\documentclass[10pt]{elsart}
\usepackage{axodraw}
\input{pix.sty}
\input{psfig.sty}

\newcommand{\comment}[1]{}
\newcommand{\cj}[1]{{\bf [j]}}
\newcommand{\beq}{\begin{equation}}
\newcommand{\eeq}{\end{equation}}
\newcommand{\bqa}{\begin{eqnarray}}
\newcommand{\eqa}{\end{eqnarray}}
\renewcommand{\thefootnote}{\fnsymbol{footnote}}
\protect

\def\sumint{\hbox{$\sum$}\!\!\!\!\!\!\,{\int}}

\begin{document}

\begin{frontmatter}

\title{Resummation in Hot Field Theories}

\author[jens]{Jens O. Andersen}
\ead{jensoa@nordita.dk}
\address[jens]{Nordita, Blegdamsvej 17, 2100-Copenhagen, Denmark}

\author[mike]{Michael Strickland}
\ead{mike@hep.itp.tuwien.ac.at}
\address[mike]{Institut f\"ur Theoretische Physik, Technische Universit\"at Wien,
	Wiedner Hauptstrasse 8-10, A-1040 Vienna, Austria}

\begin{abstract}
There has been significant progress in our understanding of finite-temperature
field theory over the past decade.
In this paper, we review the progress in perturbative thermal field
theory focusing on thermodynamic quantities. 
We first discuss the breakdown of naive perturbation theory at finite
temperature and the need for an effective expansion that resums an infinite
class of diagrams in the perturbative expansion. This effective expansion
which is due to Braaten and Pisarski, can be used to systematically calculate
various static and dynamical quantities as a weak-coupling expansion
in powers of $g$. However, it turns that the weak-coupling expansion for
thermodynamic quantities are useless unless the coupling constant is very small.
We critically 
discuss various ways of reorganizing the perturbative series for
thermal field theories in order to improve its convergence. 
These include screened perturbation theory (SPT),
hard-thermal-loop perturbation theory (HTLPT),
the $\Phi$-derivable approach, dimensionally reduced (DR) SPT, and 
the DR $\Phi$-derivable approach.
\end{abstract}


\end{frontmatter}

\section{Introduction}	

A detailed understanding of the properties of the deconfined phase of QCD
is important in several areas of physics. Examples are the evolution of the
early universe \cite{dominik}, heavy-ion collisions at RHIC (Brookhaven) and LHC 
(CERN)~\cite{heavyion1,heavyion2,heavyion3,heavyion4,heavyion5}, and 
astrophysical objects such as quark stars and supernovae ~\cite{mhj,fraga,quarkjm}.
In both cases, since the QCD running coupling decreases with increasing 
momentum transfer, the naive expectation is that weak-coupling techniques will
be sufficient to describe the relevant physics.  In the case of the RHIC
and LHC heavy-ion experiments the expectation is that during the collisions
that temperatures on the order of 0.2-1 GeV will
be generated corresponding to a strong coupling constant $\alpha_s \sim 0.2$ 
or $g_s \sim 2$.  
For cold astrophysical objects the relevant scale is 
set by the density which, in the case of quark stars, is related to the 
quark chemical potential.  At densities expected at the center of neutron 
stars the quark chemical potential is again on the order of 0.2-1 GeV and
the corresponding strong coupling constant is again $\alpha_s \sim 0.2$.
Naively one might expect that since $\alpha_s \sim 0.2$ that weak coupling
expansions could provide a systematic method for calculating observables;
however, in practice we find that weak-coupling 
approximations to even thermodynamic observables have a small (perhaps vanishing) 
radius of convergence.  This has motivated consideration of methods for
systematically reorganizing perturbative expansions in order to obtain
more convergent successive approximations to observables.  
In this paper we will
review theoretical progress in the perturbative thermodynamics of scalar
and gauge theories and provide a critical comparison of various methods for
improving the convergence of the resulting approximants via non-perturbative
variational and effective field theory methods.

In the past 15 years we have improved our understanding of thermal field
theories considerably. A major problem that spurred progress in the
late 1980s was the apparent
gauge dependence of the gluon damping rate $\gamma$.
The gluon damping rate had been calculated in ordinary perturbation theory
by a number of authors in various 
gauges \cite{damping1,damping2,damping3,damping4,damping5,damping6,damping7,damping8,damping9,damping10,damping11,damping12,damping13,damping14,damping15,damping16}.
It turned out that the result obtained for the  
damping rate depended on the gauge-fixing condition.
Even the sign of the damping rate seemed to depend on the 
gauge which would be worrisome since, in those gauges where 
the sign is negative, this indicated a plasma instability. 
However, since the damping rate is a physical quantity,
being the inverse of the lifetime of a collective excitation
in the plasma, it cannot be gauge dependent.
The problem was solved by Pisarski \cite{dampingpis},
who pointed out that
the one-loop calculations which had been performed to date 
were incomplete and that in order to obtain the correct leading-order
result in the coupling $g$, one must sum an infinite subset of diagrams. 
A detailed account of the problem has been given in
Ref.~\cite{robbie}.
The general program of resummation
was later developed by Braaten and Pisarski in Ref.~\cite{er}, where they
showed that effective propagators and vertices must be used whenever
the momentum characterizing the external lines are of order $gT$.

The next significant step was made a few years later, when perturbative
calculations of the free energy for high-temperature field theories
were pushed to higher order in the coupling. 
The free energy to order
$g^4$ has been calculated in massless 
$\phi^4$-theory in Ref.~\cite{fst}, 
for QED in Ref.~\cite{cor}, and for
QCD in Ref.~\cite{arnold1}. 
In particular the calculations by Arnold and Zhai~\cite{arnold1}
were important since they were completely analytic.
Since $g^4$ is the first order at which renormalization 
effects enter, these corrections are important beyond determining
another term in the perturbative expansion.
Calculations were later pushed to
order $g^5$ in Ref.~\cite{singh,ea1} 
for $\Phi^4$-theory, in Refs.~\cite{parwani,jensqed}
for QED, and in Refs.~\cite{zhai,ea2} for nonabelian gauge theories.

If one is interested in static quantities such as the pressure or screening
lengths, there exists an efficient alternative to explicit resummation.
It is based on the idea of dimensional reduction~\cite{gins,gpy,appel,nad}
and effective field theory methods \cite{lepage}. The basic idea is that
equilibrium properties of a 3+1 dimensional field theory at high 
temperature can be calculated using an effective field theory in three spatial
dimensions. This follows from the fact that in the imaginary time formalism,
the Matsubara frequencies act like masses in the propagators and 
the nonstatic modes decouple according to the Appelquist-Carrazone
theorem \cite{cara}. This approach was developed into tool for quantitative
calculations by Braaten and Nieto \cite{ea1,ea2},
and by Farakos, Kajantie, Rummukainen, and
Shaposhnikov \cite{farakos}. 

However, these results showed that the weak-coupling expansion for the
pressure of a thermal field theory only converges if the coupling
constant is very small. 
As successive terms in the weak-coupling expansion
are added, the predictions change wildly and the sensitivity to the
renormalization scale $\mu$ grows. In the case of QCD, the order-$g^3$
contribution is smaller than the order-$g^2$ contribution only if
$\alpha_s\leq1/20$, which corresponds to a temperature of 
$10^5$ GeV. This is orders of magnitude higher than 
the temperature expected to be generated
at RHIC and LHC.
It is clear that a reorganization of the perturbation series
is essential if perturbative calculations are to be of any quantitative
use at temperatures accessible in heavy-ion collisions.

There are some proposals for reorganizing perturbation theory in hot field 
theories that are basically mathematical manipulations of the weak coupling 
expansion. The methods include {\it Pad\'e approximates} \cite{Pade}, {\it Borel 
resummation} \cite{Parwani}, and {\it self-similar approximates} \cite{Yukalov}. 
Additionally we note that in Ref.~\cite{cvetic} Pade and Borel resummation 
techniques have been applied separately to the soft and hard contributions to 
the QCD free energy. The methods listed above can be used to construct more stable 
sequences of successive approximations which agree with the weak-coupling 
expansion when expanded in powers of $g$.  However, there are two major 
limitations of this approach:  the first is that these methods can only be 
applied to quantities for which several orders in the weak-coupling expansion 
are known, so they are limited in practice to the thermodynamic functions; the 
second is that selecting which Pade approximant is used is a non-trivial task 
with different Pade approximants sometimes giving dramatically different results 
with the difference between resulting approximants being on the order of the 
variations of the successive perturbative approximations.

The free energy can also be calculated nonperturbatively using
lattice gauge theory \cite{Karsch}.
The thermodynamic functions for pure-glue QCD have been calculated with
high precision by Boyd et al \cite{lattice-0}. There have also been
calculations with $N_f=2$ and 4 flavors of dynamical quarks \cite{lattice-Nf}.
These lattice studies find that the 
free energy is very close to zero near $T_c$ and, 
as the temperature increases, the free energy increases rapidly from
$T_c$ to $2\, T_c$
and then slowly approaches that of an ideal gas of massless quarks and gluons.
Continued advances in lattice gauge theories will provide even more precise
information about the QCD phase transition.
In practice, however, lattice gauge theories have 
some limitations. For instance, the method 
can only be applied to static quantities and many of the more promising 
signatures for 
a quark-gluon plasma involve non-equilibrium dynamical quantities. 
One would therefore like a coherent framework for calculating 
both static and dynamical quantities that would reproduce lattice data.
A second major limitation of lattice methods is that the Monte Carlo 
approach used fails
at nonzero baryon number density due to the sign problem of the action.
It is therefore important to have perturbative and non-perturbative
analytical techniques which can be used in this region of the QCD phase
diagram.  Note, however, that recent advances have made it possible to
use lattice methods to calculate the pressure at small baryon 
densities \cite{Fodor:2001au,Fodor:2001pe,deForcrand:2002ci,Allton:2002zi,Fodor:2002km,D'Elia:2002gd,Gavai:2003mf,Allton:2003vx,Fodor:2004nz}.

One analytical approach to the thermodynamics of QCD is the application
of so-called quasi-particle
models \cite{quasi0,quasi1,quasi2}. In these models, the quark-gluon plasma
is simply an ideal gas of quasi-particles with temperature-dependent masses.
The zero-point energy of the quasi-particles is replaced by a bag constant
which is necessary to enforce thermodynamic consistency.
By fitting the parameters to finite-temperature 
lattice data one can obtain a quite
reasonable description of the thermodynamics of QCD.  In
addition, using thermodynamic self-consistency results obtained at $T=0$
can be extended to nonzero chemical potential \cite{quasi3} (see also 
Ref.~\cite{szabo}).  
While these approaches seem
to do a good job at fitting the existing lattice data, it would be
preferable to have a framework which would incorporate the same
physics, namely quasi-particle degrees of freedom, in a more systematic
way.

One way to accomplish this goal is to reorganize the weak-coupling 
expansion based on a variational approach.
The free energy ${\mathcal F}$ is then expressed as the variational
minimum of a thermodynamic potential $\Omega(T,\alpha_s;m^2)$
that depends on one or more variational parameters
that are collectively denoted by $m^2$.
One such approach is the $\Phi$-derivable approximation in which the
exact propagator  
is used as an infinite set of variational parameters \cite{lw,baym,CJT-74}.
The variational principle leads to a gap equation for the self-energy which
is, in general, extremely difficult to solve. Only in simple cases where
the self-energy is independent of the external momentum, is the theory
easy to solve. Approximate solutions of the two-loop
$\Phi$-derivable approximation 
in terms of HTL self-energies have been obtained for 
QCD at finite temperature~\cite{bir1,bir2,rebb,Peshier-00} and 
chemical potential \cite{paulquasi1,paulquasi2}.
Additionally, there has been progress recently
both with respect to important issues such as 
renormalization~\cite{knoll,urko}
and gauge (in-)dependence~\cite{arri,carr} 
as well as the development of calculational 
technology \cite{ep1}.  The $\Phi$-derivable 
approach has recently been reviewed in Refs.~\cite{birrev} and
\cite{rebrev}.

The intractability of the $\Phi$-derivable approach
motivates the use of simpler variational approximations.
One such strategy that involves a single variational mass parameter $m$
has been called {\it optimized perturbation theory} \cite{Stevenson-81},
{\it variational perturbation theory} \cite{varpert},
or the {\it linear $\delta$ expansion} \cite{deltaexp}.
This strategy was applied to the thermodynamics of the massless $\phi^4$
field theory by Karsch, Patk\'os and Petreczky under the name
{\it screened perturbation theory} \cite{K-P-P-97}.
The method has also been applied to spontaneously broken
field theories at finite temperature \cite{C-K-98}.
The calculations of the thermodynamics of the massless $\phi^4$
field theory using screened perturbation theory
have been extended to three loops \cite{spt}.
The calculations can be greatly simplified by using a double
expansion in powers of the coupling constant and $m/T$ \cite{AS-01}.
Screened perturbation theory has also been generalized to gauge 
theories~\cite{htl1,fermions} and is then called hard-thermal-loop (HTL)
perturbation theory. Two-loop calculations in pure-glue QCD~\cite{htl2}
and QCD with dynamical fermions have been performed \cite{aps1}.

The dimensionally reduced effective field theory can also be used
as the starting point for nonperturbative calculations of 
static properties. The coefficients in the effective Lagrangian are 
calculated using perturbation theory, but calculations in three dimensions
are performed nonperturbatively using analytic methods and/or lattice gauge theory.
Unfortunately, dimensional reduction has similar limitations to ordinary
lattice gauge theory: it can be applied only to static quantities
and only for baryon number densities which are much smaller than the temperature.
Unlike ordinary lattice gauge theory, however, light dynamical quarks do not 
require any additional computer power, because they only enter through the 
perturbatively calculated parameters of ${\mathcal L}_{\rm eff}$. This method 
has been applied to the Debye screening mass for QCD \cite{Kajantie-97} as 
well as the pressure \cite{KLRS,Vuorinen:2003fs}.

The paper is organized as follows. In Sec.~\ref{resum}, we discuss
the need for resummation
and the Braaten-Pisarski resummation program.
In Sec.~\ref{weak}, we discuss the weak-coupling expansion of
the thermodynamic quantities in detail.
Sec.~\ref{dimred} is devoted to 
dimensional reduction and 
the application of effective field theory methods to hot field theories 
In Secs.~\ref{spt} and \ref{htl}, 
we discuss screened perturbation theory and 
hard-thermal-loop perturbation theory.
The $\Phi$-derivable approach is reviewed in Sec.~\ref{phi}.
In Secs.~\ref{drspt} and \ref{drphisec} we  
cover dimensionally reduced screened perturbation theory and dimensionally
reduced $\Phi$-derivable approaches.
In Sec.~\ref{conclude}, we summarize and conclude.
There are three appendices where we list sum-integrals
and integrals needed as well as some technical details
of the calculations.

\input{resum.tex}

\input{weak.tex}
\input{dimred.tex}

\input{spt.tex}
\input{htl.tex}

\input{phi.tex}

\input{drspt.tex}

\input{pdr.tex}

\input{concl.tex}

\section*{Acknowledgments}
The authors would like to thank A.~Rebhan for 
discussions and suggestions. The authors would also like to thank 
E.~Braaten and E.~Petitgirard for a fruitful collaboration 
on which part of this paper is based.  M.S. was supported by an FWF 
Der Wissenschaftsfonds Project M689.

\input{app.tex}

\input{bibl.tex}
\end{document}

%% file: resum.tex
\section{The Need for Resummation}
\label{resum}
In this section, and in the rest of the paper, we consider thermal field 
theories
at high temperatures, which means temperatures much higher than 
all zero-temperature masses or any mass scales generated 
at zero temperature. 

It has been known for many years that naive perturbation theory, or the loop
expansion breaks, down at high temperature due to infrared divergences.
Diagrams which are nominally of higher order in the coupling constant
contribute to leading order in $g$. A consistent perturbative
expansion requires the resummation of an infinite subset of
diagrams from all orders of perturbation theory. We discuss these issues
next.
\subsection{\it Scalar field theory}
We start our discussion by considering the simplest interacting thermal
field theory,
namely that of a single massless scalar field with a $\phi^4$-interaction.
The Euclidean Lagrangian is
\bqa
{\mathcal L}&=&{1\over2}(\partial_{\mu}\phi)^2+{g^2\over24}\phi^4\;.
\label{sl}
\eqa
\comment{mj}
Perturbative calculations at zero temperature proceed by dividing the
Lagrangian into a free part and an interacting part according to
\bqa
{\mathcal L}_{\rm free}&=&{1\over2}(\partial_{\mu}\phi)^2\;,\\
{\mathcal L}_{\rm int}&=&{g^2\over24}\phi^4\;.
\eqa
\comment{mj}
Radiative corrections are then calculated in a loop expansion which is
equivalent to a power series in $g^2$. We shall see that the perturbative
expansion breaks down at finite temperature and the weak-coupling expansion
becomes an expansion in $g$ rather than $g^2$.

\begin{figure}[htb]
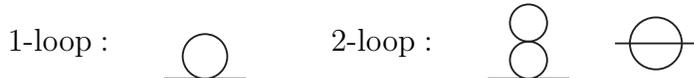

\begin{center}
$\mbox{1-loop}:\hspace{5mm} \oneloopse \hspace{1cm}\mbox{2-loop}:\hspace{5mm} \twoloopsea \;\;\; \twoloopseb$
\end{center}
\caption[a]{
One- and two-loop scalar self-energy graphs.
}
\label{thmass}
\end{figure}

We will first calculate the self-energy by evaluating the
relevant diagrams.
The Feynman diagrams that contribute to the self-energy up to two loops
are shown in Fig.~\ref{thmass}.

The one-loop diagram is independent of the 
external momentum and the resulting integral 
expression is
\bqa\nonumber
\Pi^{(1)}&=&{1\over2}g^2\sumint_P{1\over P^2} \; , \\ \nonumber
&=&{g^2\over24}T^2 \; , \\
&\equiv&m^2
\;,
\label{fm}
\eqa
\comment{mj}
where the superscript indicates the number of loops.
The sum-integral $\Sigma\!\!\!\!{\int}_{P}$ 
represents a summation over Matsubara frequencies
and integration of spatial momenta in $d=3-2\epsilon$ dimensions.
It is defined in Eq.~(\ref{sumint-def})~\footnote{
For an introduction to thermal field theory and the
imaginary time formalism see Refs.~\cite{kap} and \cite{bellac}.}.
The above sum-integral has ultraviolet power divergences that are set
to zero in dimensional regularization. We are then left with the 
finite result~(\ref{fm}), which shows that thermal fluctuations
generate a mass for the scalar field of order $gT$.
This thermal mass is analogous to the Debye mass which is 
well-known from the nonrelativistic QED plasma.

\begin{figure}[htb]
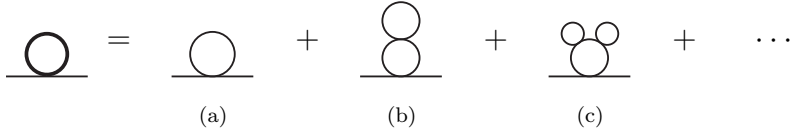

\begin{center}
\begin{tabular}{ccccccccccc}
$\oneloopseb = $ &
$\oneloopse$ & + &
$\twoloopsea$ & + &
$\threeloopsea$ & + \hspace{5mm} $\cdots$
\\
    &
{\scriptsize (a)} &&
{\scriptsize(b)} &&
{\scriptsize(c)} &&
\end{tabular}
\caption[a]{Bubble diagrams contributing to the scalar self-energy.}
\label{2lself}
\end{center}
\end{figure}

We next focus on the two-loop diagrams and first consider the double-bubble
in Fig.~\ref{2lself} (b). 

\noindent
This diagram is also independent of the external momentum
and gives the following sum-integral
\bqa
\Pi^{(2b)}&=&-{1\over4}g^4\sumint_{PQ}{1\over P^2}{1\over Q^4}\;.
\eqa
\comment{mj}
This integral is infrared divergent. The problem stems from the loop with
two propagators. 
In order to isolate the source of the divergence, we look at
the contribution from the zeroth Matsubara mode to the $Q$
integration
\bqa
-{1\over4}g^4\sumint_P{1\over P^2}T\int_q{1\over q^4}\;,
\label{g3}
\eqa 
\comment{mj}
which is quadratically infrared divergent. 
This infrared divergence indicates that
naive perturbation theory breaks down at finite 
temperature.
However,  in practice
this infrared divergence is screened by the
thermally generated mass and we must somehow take this into account.
The thermal mass can be incorporated by using an effective propagator:
\bqa
\Delta(\omega_n,p)&=&{1\over P^2+m^2}\;,
\label{impprop}
\eqa
\comment{mj}
with $m\sim gT \ll T$.

\begin{figure}[htb]
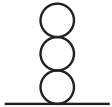

\begin{center}
 $\threeloopseb$
\end{center}
\caption[a]{ Subleading three-loop self-energy diagram.}
\label{3l}
\end{figure}

If the momenta of the propagator is of order $T$ or {\it hard}, clearly 
the thermal mass is a perturbation and can be omitted. However, if the 
momenta of the propagator is of order $gT$ or {\it soft}, the thermal 
mass is as large as the bare inverse propagator and cannot be omitted.
The mass term in the propagator~(\ref{impprop}) provides an infrared cutoff
of order $gT$. The contribution from~(\ref{g3}) would then be
\bqa
-{1\over4}g^4\sumint_P{1\over P^2}T\int_q{1\over(q^2+m^2)^2}
&=&-{1\over4}g^4\left({T^2\over 12}\right)\left({T\over8\pi m}\right)
+{\mathcal O}\left(g^4mT\right)\;.
\eqa
\comment{mj}
Since $m\sim gT$, 
this shows that the double-bubble contributes at order $g^3T^2$ to the 
self-energy and not at order $g^4T^2$ as one might have expected.
Similarly, one can show that the diagrams with any number of bubbles like
Fig.~\ref{2lself}c are all of order $g^3$.
Clearly, naive perturbation theory breaks down since the order-$g^3$
correction to the thermal mass receives contributions from all loop orders.  
On the other hand, the three-loop diagram shown in Fig.~\ref{3l}, is of order
$g^4T^2$ and thus subleading.  Therefore, we only need to resum a subset
of all possible Feynman graphs in order to obtain a consistent expansion
in $g$.

\vspace{6mm}

If we use the effective propagator to recalculate the one-loop self-energy,
we obtain
\bqa\nonumber
\Pi^{(1)}(P)&=&{1\over2}g^2\sumint_P{1\over P^2+m^2}
\\ \nonumber
&=&
{1\over2}g^2\left[
T\int_p{1\over p^2+m^2}+\sumint_P^{\prime}
{1\over P^2}
+{\mathcal O}\left(m^2\right)\right]
\\
&=&{g^2\over24}T^2\left[1-{g\sqrt{6}\over4\pi}+{\mathcal O}\left(g^2\right)
\right]\;.
\eqa
\comment{mj}
where here, and in the following, 
the prime on the sum-integral indicates that we have excluded the
$n=0$ mode from the sum over the Matsubara frequencies.
The order $g^3$ corresponds to the summation of the bubble diagrams in 
Fig.~\ref{2lself}, which can be verified by expanding the effective 
propagator~(\ref{impprop}) around $m=0$. Thus by taking the thermal mass into
account, one is resumming an infinite set of diagrams from all orders of 
perturbation theory.

The self-energy~(\ref{fm}) is the first example of a {\it hard thermal loop} (HTL).
Hard thermal loops are  loop corrections which are $g^2T^2/P^2$ times
the corresponding tree-level amplitude, where $P$ is a momentum 
that characterizes the external lines.
From this definition, we see that, whenever $P$ is hard, the loop correction
is suppressed by $g^2$ and is thus a perturbative correction.
However, for soft $P$, the hard thermal loop is ${\mathcal O}(1)$ 
and is therefore as important as the tree-level contribution to the amplitude.
These loop corrections are called ``hard'' because
the relevant integrals are dominated by momenta of order $T$.  Also note
that the hard thermal loop in the two-point function
is finite since it is exclusively due to thermal fluctuations. Quantum
fluctuations do not enter. Both properties 
are shared by all hard thermal loops.

What about higher-order $n$-point functions in scalar thermal field theory?
One can show that within scalar theory the one-loop correction
to the four-point function for high temperature behaves as~\cite{robbie}
\bqa
\Gamma^{(4)}\propto g^4\log\left(T/p\right)\;,
\eqa
\comment{jm}
where $p$ is the external momentum.
Thus the loop correction to the four-point function increases logarithmically
with temperature. It is therefore always down by one power of $g$, and it 
suffices to use a bare vertex. More generally, it can be shown that
the only hard thermal loop in scalar field theory is the tadpole 
diagram in Fig.~\ref{thmass} and resummation
is taken care of by including the thermal mass in the propagator.
In gauge theories, the situation is much more complicated as we shall see 
in the next section.

\subsection{\it Gauge theories}
\label{gres}

In the previous section, we demonstrated the need for resummation
in a hot scalar theory. 
For scalar theories, resummation simply amounts to including the
thermal mass in the propagator and since the 
running coupling depends logarithmically on the temperature,
corrections to the bare vertex are always down by powers of $g^2$.
In gauge theories, the situation is more complicated. The 
equivalent HTL self-energies
are no longer local, but depend in a nontrivial way
on the external momentum. In addition, it is also necessary to use
effective vertices that also depend on the external momentum.
It turns out that all hard thermal loops are gauge-fixing independent.
This was shown explicitly in covariant gauges, Coulomb gauges, and axial gauges.
They also satisfy tree-level like Ward identities. Furthermore, there exists
a gauge invariant 
effective Lagrangian, found independently 
by Braaten and Pisarski~\cite{gaugeer} and by Taylor and Wong \cite{wong},
that generates all of the hard thermal loop $n$-point functions. 
From a renormalization group point 
of view this is an effective Lagrangian for the soft scale $gT$ that is 
obtained by integrating out the hard scale $T$.
We return to the HTL Lagrangian in Section~\ref{htl}.

\begin{figure}[t]
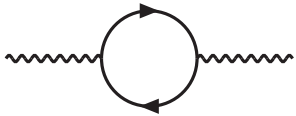

\begin{center}
 $\oneloopQEDse$
\end{center}
\caption[a]{
One-loop photon self-energy diagram.
}
\label{qed}

\end{figure}

\subsubsection{Polarization tensor}

We next discuss in some detail 
the hard thermal loop for the polarization tensor 
$\Pi_{\mu\nu}$.
For simplicity, we discuss QED. The Feynman
diagram for the one-loop self-energy is shown in Fig.~\ref{qed} and 
results in the following sum-integral
\bqa
\Pi_{\mu\nu}(\omega_n,p)&=&
e^2\,\sumint_{\{K\}}
{\rm Tr}\left[{K\!\!\!\!/\gamma_{\mu}(P\!\!\!\!/+K\!\!\!\!/)\gamma_{\nu}
\over K^2(P+K)^2}\right]\;,
\eqa
\comment{mj}
where ${\rm Tr}$ denotes the trace over Dirac indices.
After taking the trace, the self-energy becomes
\bqa\nonumber
\Pi_{\mu\nu}(\omega_n,p)&=&8e^2\sumint_{\{K\}}{K_{\mu}K_{\nu}\over K^2(P+K)^2}
-4\delta_{\mu\nu}e^2\sumint_{\{K\}}{1\over K^2}
\\ &&+
2\delta_{\mu\nu}P^2e^2\sumint_{\{K\}}{1\over K^2(P+K)^2}
+4e^2\sumint_{\{K\}}{P_{\mu}K_{\nu}+P_{\nu}K_{\mu}\over K^2(P+K)^2}
\;,
\eqa
\comment{mj}
where we have assumed, for now, that $d=3$.

We first consider the spatial components of $\Pi_{\mu\nu}(\omega_n,p)$. 
The sum over
Matsubara frequencies can be written as a contour integral in the complex
energy plane. After analytic continuation, we obtain
\bqa\nonumber
\Pi_{ij}(\omega,p)&=&
e^2\int{d\omega_0\over2\pi i}\int_k\Bigg[{4k_ik_j-(\omega^2-p^2)\delta_{ij}
+2(p_ik_j+k_ip_j)\over(k^2-\omega_0^2)
[({\bf p}+{\bf k})^2-(\omega+\omega_0)^2]}
\\&&
\hspace{3cm}
-{2\delta_{ij}\over(k^2-\omega^2_0)}\Bigg]
\tanh{\beta\omega_0\over2}
\;.
\eqa
\comment{jm}
In order to extract the hard thermal loop, we notice that terms that
contain one or more power of the external momentum can be omitted
since the external momentum $p$ is assumed to be soft. 
The self-energy then becomes
\bqa\nonumber
\Pi_{ij}(\omega,p)&=&
2e^2\int{d\omega_0\over2\pi i}\int_k\Bigg[{2k_ik_j
\over(k^2-\omega_0^2)
[({\bf p}+{\bf k})^2-(\omega+\omega_0)^2]}
\\&&
\hspace{3cm}
-{\delta_{ij}\over(k^2-\omega_0^2)}\Bigg]
\tanh{\beta\omega_0\over2}
\;.
\eqa
\comment{jm}
After integrating over the energy $\omega_0$, we obtain
\bqa\nonumber
\Pi_{ij}(\omega,p)&=&-2e^2\delta_{ij}\int_k{1\over k}\left(
1-2n_F(k)\right)
+2e^2\int_k{k_ik_j\over k|{\bf p}+{\bf k}|}
\\ &&\nonumber
\hspace{-1.3cm}\times\Bigg\{\left(1-n_F(k)-n_F(|{\bf p}+{\bf k}|)\right)
\left[{1\over k+|{\bf p}+{\bf k}|+\omega}+
{1\over k+|{\bf p}+{\bf k}|-\omega}
\right]
\\ &&
\hspace{-1.3cm}-\left[n_F(k)-n_F(|{\bf p}+{\bf k}|)\right]
\left[{1\over|{\bf p}+{\bf k}|-k+\omega}+
{1\over|{\bf p}+{\bf k}|-k-\omega}
\right]\Bigg\} \, ,
\label{omself}
\eqa
\comment{jm}
where $n_F(x)=1/(\exp(\beta x)+1)$ is the Fermi-Dirac distribution function.
The zero-temperature part of Eq.~(\ref{omself}) is logarithmically
divergent in the ultraviolet.
This term depends on the external momentum and is cancelled by standard
zero-temperature wavefunction renormalization.
We next consider the terms that depend on temperature.
In the case that the loop momentum is soft, 
the Fermi-Dirac distribution functions can be approximated by a constant.
The contribution from the integral over the magnitude of $k$ is then 
of order $g^3$ and subleading.
When the loop momentum is hard, one can expand the terms in the integrand
in powers of the external momentum. We can then make the following
approximations 
\bqa
n_F(|{\bf p}+{\bf k}|)&\approx&
n_F(k)+{dn_F(k)\over dk}{\bf p}\!\cdot\!\hat{\bf k}\;,\\
{1\over|{\bf p}+{\bf k}|+k\pm \omega}&\approx&{1\over 2k}\;,\\
{1\over|{\bf p}+{\bf k}|-k\pm\omega}&\approx&
{1\over{\bf p}\!\cdot\!\hat{\bf k}\pm\omega}
\;,
\eqa
\comment{mj}
where $\hat{\bf k}={\bf k}/k$ is a unit vector.
Thus the angular integration decouples from the 
integral over the magnitude $k$.
This implies
\bqa
\Pi_{ij}(\omega,{\bf p})&=&
{4 e^2 \over \pi^2}\int_0^{\infty}dk\;k\,n_F(k)
\int{d\Omega\over4\pi}{\omega\over{\omega-\bf p}\!\cdot\!\hat{\bf k}} \hat{k}_i\hat{k}_j \;,
\nonumber \\
&=& m_D^2 
\int{d\Omega\over4\pi}{\omega\over{\omega-\bf p}\!\cdot\!\hat{\bf k}} \hat{k}_i\hat{k}_j \;,
\eqa
\comment{mj}
where we have defined $m_D^2 = e^2 T^2/3$.

The other components of the self-energy tensor $\Pi_{\mu\nu}(\omega,{\bf p})$
are derived in the same manner.
One finds~\cite{bellac}
\bqa
\Pi_{00}(\omega,{\bf p})&=&
m_D^2 \left(
\int{d\Omega\over4\pi}{\omega\over\omega-{\bf p}\!\cdot\!\hat{\bf k}}
-1 \right)\;, 
\\ 
\Pi_{0j}(\omega,{\bf p})&=&
m_D^2 \int{d\Omega\over4\pi}{\omega \hat{k}_j\over\omega-{\bf p}\!\cdot\!\hat{\bf k}}\;.
\eqa
\comment{mj}
In $d$ dimensions, we can compactly write the self-energy tensor as
\bqa
\label{a1}
\Pi^{\mu\nu}(p)=m_D^2\left[
{\mathcal T}^{\mu\nu}(p,-p)-n^{\mu}n^{\nu}
\right]\;,
\label{scomp}
\eqa
\comment{mj}
where $n$ specifies the thermal rest frame
is canonically given by $n = (1,{\bf 0})$, we have defined 
\bqa
m_D^2&=&-4(d-1)e^2\sumint_{\{K\}}{1\over K^2}\;,
\eqa
\comment{mj}
and the tensor ${\mathcal T}^{\mu\nu}(p,q)$, which is defined only for momenta
that satisfy $p+q=0$, is
\bqa
{\mathcal T}^{\mu\nu}(p,-p)=
\left \langle y^{\mu}y^{\nu}{p\!\cdot\!n\over p\!\cdot\!y}
\right\rangle_{\bf\hat{y}} \;.
\label{T2-def}
\eqa
\comment{jm}
The angular brackets indicate averaging
over the spatial directions of the light-like vector $y=(1,\hat{\bf y})$.
The tensor ${\mathcal T}^{\mu\nu}$ is symmetric in $\mu$ and $\nu$
and the self-energy~(\ref{scomp}) satisfies the Ward identity:
\bqa
p_{\mu}\Pi_{\mu\nu}(p)&=&0\;.
\eqa
\comment{jm}
Because of this Ward identity and the rotational symmetry around the 
$\hat{p}$-axis, one can express the self-energy in terms of 
two independent functions, $\Pi_T(\omega,{\bf p})$ and 
$\Pi_L(\omega,{\bf p})$:
\bqa\nonumber
\Pi_{\mu\nu}(\omega,{\bf p})&=&
\Pi_{L}(\omega,{\bf p}){(\omega^2-p^2)g_{\mu\nu}-p_{\mu}p_{\nu}\over p^2}
\\ &&
+\left[
\Pi_T(\omega,{\bf p})-{\omega^2-p^2\over p^2}\Pi_L(\omega,{\bf p})
\right]g_{\mu i}\left(\delta_{ij}-\hat{p}_i\hat{p}_j\right)g_{j\nu}
\;,
\label{defself}
\eqa
\comment{jm}
where the functions $\Pi_T(\omega,{\bf p})$ and 
$\Pi_L(\omega,{\bf p})$ are
\bqa
\Pi_T(\omega,p)&=&{1\over2}(\delta_{ij}-\hat{p}_i\hat{p}_j)\Pi_{ij}(\omega,p)
\label{pit}
\;,\\
\Pi_L(\omega,p)&=&-\Pi_{00}(\omega,p)\;.
\label{pil}
\eqa
\comment{jm}
In three dimensions, the self-energies 
$\Pi_T(\omega,{\bf p})$ and $\Pi_L(\omega,{\bf p})$
reduce to
\bqa
\Pi_T(\omega,p)&=&{m_D^2\over2}{\omega^2\over p^2}\left[
1+{p^2-\omega^2\over2\omega p}\log{\omega+p\over\omega-p}\right]\;,
\label{redt}
\\
\Pi_L(\omega,p)&=&m_D^2\left[1-
{\omega\over2p}\log{\omega+p\over\omega-p}\right]\;.
\label{redl}
\eqa
\comment{jm}
The hard thermal loop in the photon propagator was first calculated by 
Silin more than forty years ago \cite{silin}.
The hard thermal loop in the gluon self-energy 
was first calculated by Klimov and Weldon \cite{klim,weld}.
It has the same form
as in QED, but where the Debye mass $m_D$ is replaced by
\bqa
m_D^2&=& g^2 \left[(d-1)^2 C_A\sumint_{K}{1\over K^2}
-2(d-1) N_f \sumint_{\{K\}}{1\over K^2}\right]\;,
\label{qcdmd}
\eqa
\comment{mj}
where $C_A=N_c$ is the number of colors and $N_f$ is the number of flavors.
When $d=3$ the QCD gluon Debye mass becomes
\bqa
m_D^2 = {1\over3} \left( C_A + {1\over2}N_f \right) g^2 T^2 \; .
\label{qcdmd3}
\eqa
\comment{mj}

\subsubsection{Fermionic self-energy}

The electron self-energy is given by
\bqa
\label{selfq}
\Sigma(P)=m_f^2\gamma_\mu{\mathcal T}^\mu(p)
\;,
\eqa
\comment{mj}
where
\bqa
\label{deftf}
{\mathcal T}^{\mu}(p)=
\left\langle{y^{\mu}\over p\cdot y}
\right\rangle_{\hat{\bf y}}
\;,
\eqa
\comment{mj}
and $m_f$ is the thermal electron mass
\bqa
m_f^2&=&-3e^2
\sumint_{\{K\}}
{1\over K^2}\;.
\eqa
\comment{mj}
In QCD, the quark mass is given by
\bqa
m_q^2&=&-3
C_Fg^2\sumint_{\{K\}}{1\over K^2}\;.
\eqa
\comment{mj}

\subsubsection{Higher $n$-point functions}

In gauge theories, there are also hard thermal loops involving vertices.
For instance, the one-loop correction to the three-point function in 
QED, can compactly be written as
\bqa
\Gamma^{\mu}&=&\gamma^{\mu}-m_f^2\tilde{{\mathcal T}}^{\mu}(p,q,r)\;,
\eqa
\comment{mj}
where the tensor in the HTL correction term is only defined for $p-q+r=0$:
\bqa
\tilde{{\mathcal T}}^{\mu}(p,q,r)
&=&\left\langle
y^{\mu}\left({y\!\!\!/\over q\!\cdot\!y\;\;r\!\cdot\!y}\right)
\right\rangle_{\hat{\bf y}}\;.
\label{T3-def}
\eqa
\comment{jm}
The quark-gluon vertex satisfies
the Ward identity
\bqa
p_{\mu}\Gamma^{\mu}(p,q,r)=S^{-1}(q)-S^{-1}(r)\;,
\label{qward1}
\eqa
\comment{jm}
where $S(q)$ is the resummed effective fermion propagator. 

In QED there are, in fact, infinitely many amplitudes 
with hard thermal loops. To be precise, there are hard thermal loops in
all $n$-point functions with 
two fermion lines and $n-2$ photon lines. In nonabelian gauge theories
such as QCD, 
there are in addition hard thermal loops in amplitudes with $n$ gluon 
lines \cite{er}.

%% file: weak.tex
\section{The Weak-coupling Expansion}
\label{weak}

The Braaten-Pisarski resummation program has been used to calculate the
thermodynamic functions as a weak-coupling expansion in $g$.
They have now been
calculated explicitly through order $g^5$ for massless $\phi^4$
theory \cite{singh,ea1}, for QED \cite{cor,parwani,jensqed},
and for nonabelian gauge theories \cite{arnold1,zhai,ea2}.
In this section, we review these calculations in some detail.

\subsection{\it Scalar field theory}

The simplest way of dealing with the infrared divergences in 
scalar field theory is to reorganize perturbation theory in such a way that it
incorporates
the effects of the thermally generated mass $m$ into 
the free part of the Lagrangian.
One possibility is to divide the Lagrangian~(\ref{sl}) into free 
and interacting parts according to
\bqa
{\mathcal L}_{\rm free}&=&{1\over2}(\partial_{\mu}\phi)^2+{1\over2}m^2
\phi^2\, , \\
{\mathcal L}_{\rm int}&=&{g^2\over24}\phi^4-{1\over2}m^2\phi^2\,.
\label{sint}
\eqa
\comment{jm}
Both terms in Eq.~(\ref{sint}) are treated as interaction terms of the same
order, namely $g^2$.  However, the resummation implied by the above
is rather cumbersome when it comes to calculating Green's function
with zero external energy.
Static Green's functions can always be calculated directly in imaginary
time without having to analytically continue them back to real time.
This implies that we can use Euclidean propagators with discrete 
energies when analyzing
infrared divergences which greatly simplifies the treatment.  
In particular, since only propagators
with zero Matsubara frequency have no infrared cutoff of order $T$ 
only for these modes is the thermal mass of order $gT$ relevant as an IR
cutoff.
Thus, another possibility is to add and subtract a mass term only
for the zero-frequency mode. This approach has the advantage that we
do not need to expand the sum-integrals in powers of $m^2/T^2$
in order to obtain the contribution from a given term in powers of $g^2$.
We will follow this path in the remainder of this section and write
\bqa
{\mathcal L}_{\rm free}&=&{1\over2}(\partial_{\mu}\phi)^2
+{1\over2}m^2\phi^2\delta_{p_0,0}\;,\\
{\mathcal L}_{\rm int}&=&{g^2\over24}\phi^4-{1\over2}m^2\phi^2\delta_{p_0,0}\;.
\eqa
\comment{jm} 
The free propagator then takes the form
\bqa
\Delta(\omega_n,p)&=&{1-\delta_{p_0,0}\over P^2}
+{\delta_{p_0,0}\over p^2+m^2}\;.
\eqa
\comment{jm}
This way of resumming is referred to as {\it static resummation}.
It is important to point out that this simplified resummation scheme can only
be used to calculate static quantities such as the pressure or screening
masses. Calculation of dynamical quantities requires the full Braaten-Pisarski
resummation program. The problem is that the calculation of correlation
functions with zero external frequencies cannot unambiguously be
analytically continued to real time \cite{kraemmer}.

\begin{figure}[t]
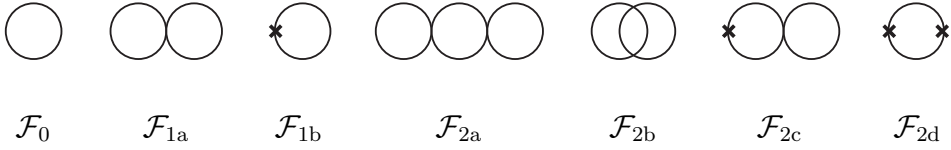


\begin{tabular}{ccccccc}
$\oneloop$ &
$\figureeight$ &
$\oneloopX$ &
$\triplebubble$ &
$\basketball$ &
$\figureeightX$ &
$\oneloopXX$
\vspace{5mm}
\\
${\mathcal F}_{\rm 0}$ &
${\mathcal F}_{\rm 1a}$ &
${\mathcal F}_{\rm 1b}$ &
${\mathcal F}_{\rm 2a}$ &
${\mathcal F}_{\rm 2b}$ &
${\mathcal F}_{\rm 2c}$ &
${\mathcal F}_{\rm 2d}$ 
\end{tabular}

\caption[a]{ 
Diagrams which contribute up to three-loop order in scalar perturbation 
theory.  
A boldfaced $\times$ indicates an insertion of $m^2$.
}  
\label{scalarpertgraphs}

\end{figure}

We next consider the calculation of the free energy through order $g^5$
in scalar field theory. This involves the evaluation of vacuum graphs
up to three-loop order.  The diagrams are those shown in Fig.~\ref{scalarpertgraphs}.

\subsubsection{One Loop}

The one-loop contribution to the free energy is
\bqa
{\mathcal F}_0&=&{1\over2}T\int_{p}\log\left(p^2+m^2\right)
+{1\over2}\sumint_P^{\prime}\log P^2\;,
\label{pf1}
\eqa
\comment{jm}
Using the integrals and sum-integrals contained in the appendices
the result for this diagram in the limit $\epsilon\rightarrow0$ is
\bqa
{\mathcal F}_0&=&-{\pi^2\over90}T^4-{Tm^3\over12\pi}\;.
\eqa
\comment{jm}

\subsubsection{Two Loops }

The two-loop contribution to the free-energy is given by 
\bqa
{\mathcal F}_1 &=& {\mathcal F}_{\rm 1a} + {\mathcal F}_{\rm 1b} \; ,
\label{pf2}
\eqa
\comment{mj}
with
\bqa
{\mathcal F}_{\rm 1a}&=&{1\over8}g^2
\left(T\int_p{1\over p^2+m^2}+\sumint_P^{\prime}{1\over P^2}\right)^2\;,
\\ 
{\mathcal F}_{\rm 1b}&=&
-{1\over2}m^2T\int_p{1\over p^2+m^2}\;.
\eqa
\comment{jm}
The result for these diagrams in the limit $\epsilon\rightarrow0$ is
\bqa\nonumber
{\mathcal F}_{\rm 1a}&=&{\pi^2T^4\over90}
\left\{{5\over4}\alpha\left[1+
\epsilon\left(4+4{\zeta^{\prime}(-1)\over\zeta(-1)}\right)\right]
\left({\mu\over4\pi T}\right)^{4\epsilon}
\right.\\&&
\hspace{-2mm}
\left.
-{5\sqrt{6}\over2}\alpha^{3/2}\left[1+\epsilon\left(4+
2{\zeta^{\prime}(-1)\over\zeta(-1)}\right)
\right]\left({\mu\over4\pi T}\right)^{2\epsilon}
\left({\mu\over2m}\right)^{2\epsilon}
+{15\over2}\alpha^2
\right\},
\\
{\mathcal F}_{\rm 1b}&=&{\pi^2T^4\over90}
{5\sqrt{6}\over2}\alpha^{3/2}\;,
\eqa
\comment{jm}
where we have kept all terms that contribute through order $\epsilon$, because
they are needed for the counterterm diagrams in the three-loop free energy.

\subsubsection{Three Loops }

The three-loop contribution is given by
\bqa
{\mathcal F}_{\rm 2}&=&
{\mathcal F}_{\rm 2a}+{\mathcal F}_{\rm 2b}+{\mathcal F}_{\rm 2c}+{\mathcal F}_{\rm 2d}+
{{\mathcal F}_{\rm 1a} \over g^2} \Delta_1 g^2\;,
\label{pf3}
\eqa
\comment{mj}
where the expressions for the diagrams are 
\bqa
{\mathcal F}_{\rm 2a}&=&-{1\over16}g^4
\left(T\int_p{1\over p^2+m^2}+\sumint_P^{\prime}{1\over P^2}\right)^2\!\!
\left(T\int_p{1\over(p^2+m^2)^2}+\sumint_P^{\prime}{1\over P^4}\right)\!,\\
\nonumber
{\mathcal F}_{\rm 2b}&=&-{1\over48}g^4\sumint_{PQR}^{\prime}
{1\over P^2}{1\over Q^2}{1\over R^2}{1\over (P+Q+R)^2}
\\ &&
\hspace{-6mm}-{1\over48}g^4T^3\int_{pqr}
{1\over p^2+m^2}{1\over q^2+m^2}{1\over r^2+m^2}
{1\over ({\bf p}+{\bf q}+{\bf r})^2+m^2}+{\mathcal O}(g^6)
\label{32b}
\;,\\
{\mathcal F}_{\rm 2c}&=&{1\over4}g^2m^2
\left(T\int_p{1\over(p^2+m^2)}+\sumint_P^{\prime}{1\over P^2}\right)
\left(T\int_p{1\over(p^2+m^2)^2}\right)\;,\\
{\mathcal F}_{\rm 2d}&=&-{1\over4}m^4T\int_p{1\over(p^2+m^2)^2}\;.
\eqa
\comment{jm}
In the appendix, we show how to rewrite the contribution from 
${\mathcal F}_{\rm 2b}$.
The result for these diagrams in the limit $\epsilon\rightarrow0$ is
\bqa\nonumber
{\mathcal F}_{\rm 2a}&=&{\pi^2T^4\over90}\left\{
-{5\sqrt{6}\over8}\alpha^{3/2}
-{5\over8}\alpha^2\left[
{1\over\epsilon}+2\gamma_E+4+4{\zeta^{\prime}(-1)\over\zeta(-1)}
\right]\left({\mu\over4\pi T}\right)^{6\epsilon}
\right.\\&&
\left.
+{5\sqrt{6}\over4}\alpha^{5/2}\left[{1\over\epsilon}
+2\gamma_E+4+{\zeta^{\prime}(-1)\over\zeta(-1)}
\right]\left({\mu\over4\pi T}\right)^{2\epsilon}
\left({\mu\over2m}\right)^{6\epsilon}
\right\}
\label{3f1}
\;,\\ \nonumber
{\mathcal F}_{\rm 2b}&=&{\pi^2T^4\over90}\left\{-{5\over4}\alpha^2\left[
{1\over\epsilon}+8{\zeta^{\prime}(-1)\over\zeta(-1)}-
{\zeta^{\prime}(-3)\over\zeta(-3)}+{91\over15}
\right]\left({\mu\over4\pi T}\right)^{6\epsilon}
\right.\\ &&
\left.
+{5\sqrt{6}\over2}\alpha^{5/2}\left[
{1\over\epsilon}+8-4\log2
\right]\left({\mu\over2m}\right)^{6\epsilon}
\right\}\;,\\
{\mathcal F}_{\rm 2c}&=&{\pi^2T^4\over90}\left[{5\sqrt{6}\over4}\alpha^{3/2}
-{15\over2}\alpha^2
\right] \;,\\
{\mathcal F}_{\rm 2d}&=&-{\pi^2T^4\over90}{5\sqrt{6}\over8}\alpha^{3/2} 
\label{3f4}
\;.
\eqa
\comment{jm}

\subsubsection{Pressure through $g^5$}

Combining the one- , two-, and three-loop contributions given
by Eqs.~(\ref{pf1}), (\ref{pf2}), and (\ref{pf3}), respectively,
gives the free energy through order $g^5$
\bqa\nonumber
{\mathcal F}&=&-{\pi^2T^4\over90}\left[
1-{5\over4}\alpha+{5\sqrt{6}\over3}\alpha^{3/2}
+{15\over4}\left(\log{\mu\over2\pi T}
-{59\over15}-3\log2+\gamma
\right.\right.\\ \nonumber&&
\hspace{15mm}
\left.\left.
+4{\zeta^{\prime}(-1)\over\zeta(-1)}
-2{\zeta^{\prime}(-3)\over\zeta(-3)}
\right)\alpha^2
-{15\sqrt{6}\over2}\left(\log{\mu\over2\pi T}
-{2\over3}\log\alpha
+{5\over6}
\right.\right.\\ &&
\hspace{20mm}
\left.\left.
-{5\over3}\log2
+{2\over3}\log3+{1\over3}\gamma
-{2\over3}{\zeta^{\prime}(-1)\over\zeta(-1)}
\right)
\alpha^{5/2}
\right]\;.
\label{fscaw}
\eqa
\comment{mj}
The pressure through order $g^5$ was first calculated using resummation 
by Parwani and Singh~\cite{singh} and later by Braaten and Nieto using
effective field theory \cite{ea1}.

The renormalization group equation for the coupling $g^2$ is
\bqa
\mu{d\alpha\over d\mu}&=&{3\alpha^2}\;.
\label{rsca}
\eqa
\comment{mj}
Using Eq.~(\ref{rsca}), one can verify that the free energy (\ref{fscaw}) 
is RG-invariant up to corrections of order $g^6\log g$.

\vspace{6mm}
\begin{figure}[htb]
\centerline{\psfig{file=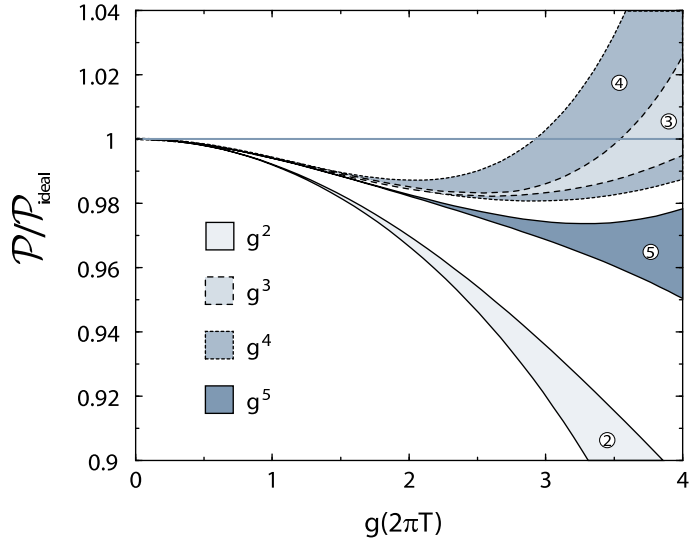,width=9cm}}
\vspace*{8pt}
\caption{Weak-coupling expansion for the pressure 
to orders $g^2$, $g^3$, $g^4$, and $g^5$
normalized to that of an ideal gas as a function
of $g(2\pi T)$.}
\label{fpert}
\end{figure}
In Fig.~\ref{fpert}, we show the successive perturbative approximations to the pressure
as a function of $g(2\pi T)$. The bands are obtained by varying the 
renormalization scale $\mu$ by a factor of two around the central value 
$\mu=2\pi T$.
The lack of convergence of the weak-coupling expansion 
is evident from this Figure.
The band obtained by varying $\mu$ by a factor of two is not 
necessarily a good measure of the error, 
but it is certainly a lower bound on the theoretical error.
Another indicator of the  theoretical error is the deviation 
between successive approximations.  We can infer from Fig.~\ref{fpert} 
that the error grows rapidly when $g(2 \pi T)$ exceeds 1.5.

\subsection{\it Gauge theories}
In this section, we discuss the application of resummation techniques to
QCD. The Euclidean 
Lagrangian for an $SU(N_c)$ gauge theory with $N_f$ fermions 
in the fundamental representation is
\bqa
{\mathcal L}&=&{1\over4}G_{\mu\nu}^aG_{\mu\nu}^a
+\bar{\psi}\gamma_{\mu}D_{\mu}\psi
\;,
\eqa
\comment{mj}
where $G_{\mu\nu}^a=\partial_{\mu}A_{\nu}^a-\partial_{\nu}A_{\mu}^a
+gf^{abc}A_{\mu}^bA_{\nu}^c$ is the
field strength, $g$ is the gauge coupling and $f^{abc}$ are the structure
constants. 
The covariant derivative is $D_{\mu}=\partial_{\mu}+igA_{\mu}^aT^a$, where
$T^a$ are the generators in the fundamental representation. 
In terms of the group-theory factors
$C_A$, $C_F$, $d_A$, $d_F$, and $T_F$, we have the relations
\bqa
f^{abc}f^{abd}&=&C_A\delta^{cd}\;, \nonumber \\
\delta^{aa} &=& d_A \;, \nonumber \\
\left(T^aT^a\right)_{ij}&=&C_F\delta_{ij}\;,\nonumber \\
{\rm Tr}\left(T^aT^b\right)&=&T_F\delta^{ab}\;,\nonumber \\
\delta^{ii} &=& d_F \;.
\eqa
\comment{mj}
Moreover, for $SU(N_c)$, 
we have $C_A=N_c$, $C_F=(N^2_c-1)/2N_c$, $T_F=N_f/2$
$d_A=N_c^2-1$, and $d_F=N_cN_f$.

If we are only interested in static quantities, we can apply the same
simplified resummation scheme also to gauge theories. Thus we are
interested in the static limit of the polarization tensor 
$\Pi_{\mu\nu}(\omega_n,{\bf p})$. In that limit $\Pi_T$ vanishes and 
$\Pi_L=m_D^2$.
In analogy with the scalar field theory, we rewrite the Lagrangian 
by adding and subtracting a mass term
${1\over2}m^2_DA_0^aA_0^a\delta_{p_0,0}$.
One of the mass terms is then absorbed into the propagator for the timelike
component of the gauge field $A_0$, while the other is treated as a 
perturbation. 

The free energy through $g^5$ requires the 
evaluation of diagrams up to three loops. The strategy is the same in the
scalar case, where one distinguishes between hard and soft loop momenta.
The result for QCD ($N_c=3$) is
\bqa\nonumber
{\mathcal F}&=&-{8\pi^2T^4\over45}\Bigg\{
1+{21\over32}N_f-{15\over4}\left(1+{5\over12}N_f\right)\!\!{\alpha_s\over\pi}
+30\left(1+{1\over6}N_f\right)^{3/2}\!\!\left({\alpha_s\over\pi}\right)^{3/2}
\\ \nonumber &&
\hspace{-5mm}
+\left[237.2+15.97N_f-0.413N_f^2
+{135\over2}\left(1+{1\over6}N_f\right)
\log\left[{\alpha\over\pi}\left(1+{1\over6}N_f\right)\right]
\right.
\\ \nonumber &&\left.
\hspace{-5mm}
-{165\over8}\left(1+{5\over12}N_f\right)\left(1-{2\over33}N_f\right)
\log{\mu\over2\pi T}\right]
\left({\alpha_s\over\pi}\right)^{2}
\\ \nonumber &&
\hspace{-5mm}
+\left(1+{1\over6}N_f\right)^{1/2}\Bigg[-799.2-21.96N_f-1.926N_f^2
\\ &&
\hspace{-5mm}
+{495\over2}\left(1+{1\over6}N_f\right)\left(1-{2\over33}N_f\right)
\log{\mu\over2\pi T}\Bigg]
\left({\alpha_s\over\pi}\right)^{5/2}
+{\mathcal O}\left(\alpha_s^3\log\alpha_s\right)
\Bigg\}.
\label{weakqcd}
\eqa
\comment{mj}

\begin{figure}[t]
\centerline{\psfig{file=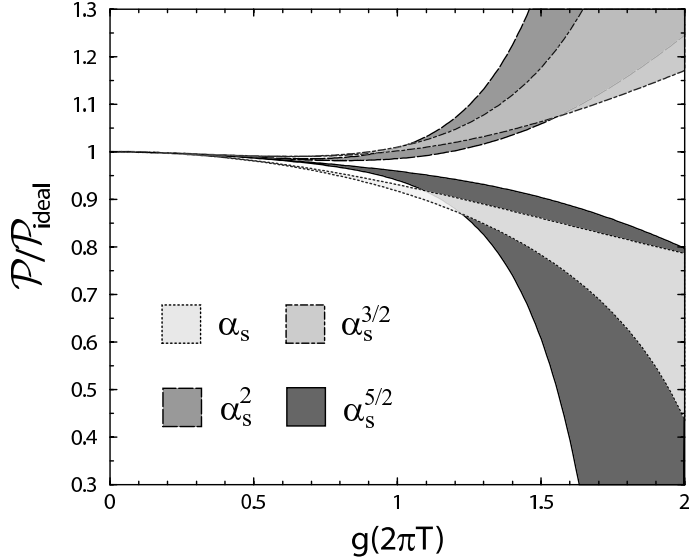,width=9cm}}
\vspace*{8pt}
\caption{Weak-coupling expansion for the pressure of 
$SU(3)$ gauge theory with $N_f=2$
to orders $g^2$, $g^3$, $g^4$, and $g^5$
normalized to that of an ideal gas as a function
of $g(2\pi T)$.}
\label{fpertqcd}
\end{figure}

The free energy for QCD through order $g^4$ was first derived by Arnold
and Zhai \cite{arnold1}. Later it was extended to order $g^5$
by Zhai and Kastening \cite{zhai} using the above resummation techniques, 
and by Braaten and Nieto using effective field theory \cite{ea2}.
We shall return to that in the next section.
The order-$g^5$ contribution is the last contribution that can be 
calculated using perturbation theory. At order $g^6$, perturbation theory
breaks down due to infrared divergences in the magnetic sector \cite{linde}.

In Fig.~\ref{fpertqcd}, we show the weak-coupling expansion for the pressure
of $SU(3)$ gauge theory with $N_f=2$
to orders $g^2$, $g^3$, $g^4$, and $g^5$
divided by that of an ideal gas as a function of $g(2\pi T)$.
The fluctuations of the predictions as one adds terms in the weak-coupling
expansion is evident from the Figure.  Additionally, if the 
series were well-behaved, the expectation would be that
the scale variation would be reduced as each additional order in $g$
was included; however, as can be seen from Fig.~\ref{fpertqcd}, for
low-temperatures, the variation of the pressure 
obtained by varying $\mu$ by a factor of two is becoming larger with 
each additional perturbative order.
This points to the necessity to resum higher logs in the final result.
As we will see in Sec.~\ref{dimred} this is possible using dimensional
reduction and effective field theory methods.

%% file: dimred.tex
\section{Dimensional Reduction}
\label{dimred}

In the imaginary time formalism, bosonic fields are periodic 
and fermionic fields are antiperiodic in the time direction
with period $\beta$. A generic field
can therefore be expanded in a Fourier series:
\bqa
\phi({\bf x},\tau)&=&\sqrt{\beta}
\sum_{n}\int_p
\phi_n(p)e^{i{\bf p}\cdot{\bf x}
+i\omega_n\tau}\;.
\eqa
\comment{jm}
where 
$\omega_n=2n\pi T$ for bosons and $\omega_n=(2n+1)\pi T$
for fermions. 
The free propagator of the $n$-th component of the field is given by
\bqa
\Delta(\omega_n,{\bf p})&=&{1\over p^2+\omega^2_n}\;.
\eqa
\comment{jm}

One can therefore view a thermal field theory in $d+1$
dimensions as a Euclidean field theory in $d$ dimensions with
infinitely many fields, where the Matsubara frequencies act as masses.
The nonzero bosonic modes and the fermionic modes have a mass of order $T$, 
while the zero-frequency mode is massless.
If we would use a resummed propagator 
the propagator
for the $n=0$ Matsubara mode would include an effective 
mass of order $gT$, while the 
masses for the 
nonzero bosonic modes and the fermionic modes would still be of order $T$.
If the coupling constant $g$ is small, the scales $T$ and $gT$ are
well separated and one can use effective field theory methods 
to organize the calculations of physical quantities such that one takes care
of one scale at a time. Not only can effective field theory methods be used
to unravel contributions from various momentum scales, they also simplify
practical calculations by treating one scale at a time. 
More specifically, 
one constructs a three-dimensional field theory for the
$n=0$ bosonic mode by integrating out all the nonzero bosonic modes as well
as all the fermionic modes. This process is called 
{\it dimensional reduction} (DR) and the nonzero Matsubara modes 
{\it decouple} \cite{gins,gpy,appel,nad,landsman89}.

The simplest way to construct the effective field theory for the 
zero-frequency bosonic modes is to use methods from effective 
field theory \cite{lepage}.
Having identified the degrees of freedom and their symmetries in the 
effective three-dimensional theory, one writes the most general Lagrangian
${\mathcal L}_{\rm eff}$ for the fields that satisfies those symmetries.
The effective Lagrangian has infinitely many interaction terms and is,
in general, 
nonrenormalizable. The terms in ${\mathcal L}_{\rm eff}$ are not arbitrary but
restricted by the symmetries of ${\mathcal L}_{\rm eff}$ and the 
coupling constants are determined by a 
{\it matching procedure}. One calculates static correlators in the
full and effective theory and demands that they be the same at long distances
scales $R\gg1/T$ \cite{ea1}.
Unfortunately, the matching procedure is complicated by the occurrence
of ultraviolet divergences. There are two sources of UV divergences. One
source is associated with the original four-dimensional field theory and
these divergences are removed by standard renormalization at zero 
temperature. The other source is associated with UV divergences in 
the effective theory in three dimensions and these are generated by 
integrating out the nonzero Matsubara modes. Thus the parameters of the
effective theory must depend on the ultraviolet cutoff $\Lambda$ that is used
to regulate the loop integrals in the effective theory in such a way that  
physical quantities do not depend on $\Lambda$. The ultraviolet 
cutoff $\Lambda$ plays the role of a factorization scale that separates
the momentum scale $T$ from the momentum scale $gT$ which can be described
by the effective three-dimensional theory.

Modern renormalization theory implies that the static correlators
at distances $R\gg1/T$ in the full theory can be reproduced to any
desired accuracy provided we include sufficiently many operators
in ${\mathcal L}_{\rm eff}$ and tune their coefficients as functions of
the coupling constants in the full theory, $T$, and $\Lambda$ \cite{ea1}.
The matching procedure is carried out using {\it strict perturbation theory},
which is 
identical to naive perturbation theory of Sec.~\ref{resum}.
This expansion is, of course, plagued with the infrared divergences
due to the long-range forces mediated by massless particles.
We know these divergences are screened, but taking screening into account
requires summing an infinite set of higher-order diagrams.
However, we can still use strict
perturbation theory to determine the coefficients of ${\mathcal L}_{\rm eff}$
because they encode the physics on the scale $T$ and are insensitive to the
physics on the scale $gT$. If we make the same incorrect 
assumptions about screening in the effective theory, the infrared divergences
will be the same in the correlators that are being calculated in the
two theories and they will cancel in the matching procedure \cite{ea1}.

\subsection{\it Scalar field theory}

In this subsection, we discuss dimensional reduction and calculations in the
effective theory for a thermal scalar field theory defined by the 
Lagrangian~(\ref{sl}). In particular, the weak-coupling expansion 
for the pressure~(\ref{fscaw}) is rederived.

\subsubsection{Effective Lagrangian}

The three-dimensional effective field theory obtained by dimensional reduction
describes a scalar field, $\phi({\bf x})$, that up to normalization can be
identified with the zero-frequency mode of the original field:
\bqa
\phi({\bf x})\approx\sqrt{T}\int_0^{\beta}d\tau\;\Phi({\bf x},\tau)\;.
\eqa
\comment{mj}
Based on the general discussion above, we can easily write down the
effective Lagrangian for a massless scalar field theory. 
There is a three-dimensional rotational symmetry as well
as the symmetry $\phi\longrightarrow-\phi$. One finds 
\bqa
{\mathcal L}_{\rm eff}&=&
{1\over2}(\nabla\phi)^2+{1\over2}m^2(\Lambda)\phi^2(\Lambda)+
{1\over24}\lambda(\Lambda)\phi^4(\Lambda)
+\delta{\mathcal L}_{\rm eff}\;,
\label{leffsca}
\eqa
\comment{mj}
where $\delta{\mathcal L}_{\rm eff}$ denotes all operators that are higher order
in the $\phi$ and $\nabla\phi$ and respect the symmetries. 
We have indicated that the parameters of the effective theory depend on the
ultraviolet cutoff $\Lambda$. Sometimes we will omit this dependence
for notational simplicity.
One term that was omitted in~(\ref{leffsca}) was $f(\Lambda)$, which 
is the coefficient of the unit operator. This parameter can be interpreted
as the contribution to the free energy 
from the scale $T$, or equivalently from the nonzero Matsubara frequencies.

Including $f$ in the effective Lagrangian, 
we can express the partition function as a path integral over
the three-dimensional field $\phi$:
\bqa
{\mathcal Z}&=&e^{-f(\Lambda)V}\int{\mathcal D}\phi
\exp\left[-\int d^3x\;{\mathcal L}_{\rm eff}\right]\;,
\label{zeff}
\eqa 
\comment{mj}
In the full theory, the partition function can also be 
written as a path integral:
\bqa
{\mathcal Z}&=&\int{\mathcal D}\phi
\exp{\left[-\int_0^{\beta}d\tau\int d^3x\;{\mathcal L}\right]}\;.
\label{zfull}
\eqa
\comment{mj}
Comparing~(\ref{zeff}) and~(\ref{zfull}), we obtain
\bqa
\log{\mathcal Z}&=&-{f(\Lambda)}V+{\log{\mathcal Z}_{\rm eff}}\;,
\label{fmatching}
\eqa
\comment{jm}
where ${\mathcal Z}_{\rm eff}$ is the partition function in the effective theory.
In order to carry out matching, we treat the mass term 
in the effective theory as a perturbation. The Lagrangian is then divided
according to
\bqa
\left({\mathcal L}_{\rm eff}\right)_{\rm free}&=& 
{1\over2}(\nabla\phi)^2\;,\\
\left({\mathcal L}_{\rm eff}\right)_{\rm int}&=&
{1\over2}(\nabla\phi)^2+{1\over2}m^2\phi^2+{\lambda\over24}\phi^4
+\delta{\mathcal L}_{\rm eff}\;.
\eqa
\comment{mj}
The mass parameter $m^2$ in~(\ref{leffsca}) can be interpreted as the
contribution to the Debye mass from the nonzero Matsubara modes 
or equivalently from the momentum scale $T$. The simplest way to determine
the mass parameter is to calculate the Debye mass in the full theory
and in the effective theory and demand that they be the same.
In  the original theory, 
the Debye mass is defined by the pole of the static propagator:
\bqa
p^2+\Pi(0,{\bf p})=0\hspace{1cm}p^2=m_D^2\;,
\eqa
\comment{mj}
where $\Pi(\omega_n,{\bf p})$ is the self-energy function. In the effective
theory, the Debye mass is given by 
\bqa
p^2+m^2+\Pi_{\rm eff}({\bf p})=0\hspace{1cm}p^2=m_D^2\;,
\eqa
\comment{mj}
where $\Pi_{\rm eff}({\bf p})$
is the self-energy in the effective theory.
The diagrams contributing to the self-energy in the full theory 
are shown in Fig.~\ref{thmass} and the resulting integral 
expression is given by
\bqa%
\nonumber
\Pi(P)&\approx&{1\over2}Z_{g^2}g^2\sumint_Q{1\over Q^2}
-{1\over4}g^4\sumint_{QR}{1\over Q^2R^4} \\
&& \hspace{3cm} -{1\over6}g^4\sumint_{QR}{1\over Q^2R^2(P+Q+R)^2}\;.
\label{appr}
\eqa
\comment{jm}

We have used the symbol $\approx$ in~(\ref{appr}) as a reminder that
the relation only holds in strict perturbation theory.
The last sum-integral depends on the external momentum $P$, but it can be 
simplified by expanding it around ${\bf p}=0$. This expansion can be justified
by noting that the leading-order self-energy is independent of the external
momentum and that the leading-order Debye mass therefore is of order
$g^2$. To order $g^4$, we can therefore set ${\bf p}=0$ in the last
two-loop sum-integrals.

In the effective theory, the self-energy 
to two loops is given by the diagrams in Fig.~\ref{thmass} and is
\bqa
\Pi_{\rm eff}({\bf p})&\approx&{1\over2}\lambda\int_q{1\over q^2}
-{1\over4}\lambda^2\int_{qr}{1\over q^2}{1\over r^4}
-{1\over6}\lambda^2\int_{qr}{1\over q^2}{1\over r^2}
{1\over({\bf p}+{\bf q}+{\bf r})^2}+\delta m^2\;,
\eqa
\comment{mj}
where $\delta m^2$ is a mass counterterm. 
The self-energy $\Pi_{\rm eff}({\bf p})$ 
can similarly be expanded in powers of the external momentum
${\bf p}$. To order $\lambda^2$, we can set ${\bf p}=0$. The external 
momentum ${\bf p}$ is the only scale in the loop integrals.
When ${\bf p}=0$, these integrals
vanish in dimensional regularization. Thus we can write
\bqa
\Pi_{\rm eff}(0)&\approx&\delta m^2\;,
\eqa
\comment{jm}
Matching the two expressions for
the Debye mass, we obtain
\bqa\nonumber
m^2&=&\Pi(0,0)-\delta m^2\\
&=&
{1\over2}Z_{g^2}g^2\sumint_Q{1\over Q^2}
-{1\over4}g^4\sumint_{QR}{1\over Q^2R^4} \nonumber \\
&& \hspace{2cm}-{1\over6}g^4\sumint_{QR}{1\over Q^2R^2(Q+R)^2}-\delta m^2\;.
\eqa
\comment{jm}

Renormalization of the coupling constant $g^2$ is carried out by 
the substitution
\bqa
Z_{g^2}&=&1+{3g^2\over2(4\pi)^2\epsilon}\;.
\label{scount}
\eqa
\comment{mj}
After coupling constant renormalization, there is still a remaining 
pole in $\epsilon$. This divergence is cancelled by a mass counterterm
which is
\bqa
\delta m^2&=&{g^4T^2\over24(4\pi)^2\epsilon}\;.
\label{deltamass4}
\eqa
\comment{mj}
After mass renormalization, the mass parameter becomes
\bqa
m^2&=&
{g^2T^2\over24}
\Bigg\{1
+\left[-3\log{\mu\over 4\pi T}
+4\log{\Lambda\over 4\pi T} \right. \nonumber \\
&& \hspace{2cm} \left. +2-\gamma+
2{\zeta^{\prime}(-1)\over\zeta(-1)}
\right]{g^2\over(4\pi)^2}
\Bigg\}
\;.
\label{masspara}
\eqa
\comment{mj}
The coupling constant $g^2$ satisfies a renormalization group equation:
\bqa
\mu{dg^2\over d\mu}&=&{3g^4\over(4\pi)^2}\;.
\label{rgsca}
\eqa
\comment{mj}
The logarithmic term $-3\log(\mu/4\pi T)$ in~(\ref{masspara}) 
cancels the $\mu$-dependence of
the $g^2$-term. The logarithmic term
$\log(\Lambda/4\pi T)$ shows that the mass term depends
explicitly on the ultraviolet cutoff $\Lambda$. This is necessary to cancel
the $\Lambda$-dependence that arises in calculations using
the effective theory.

In the following, we need the coupling constant $\lambda$ in the effective
theory only 
to leading order in the coupling constant $g^2$  in the full theory.
At tree level we can simply read off $\lambda$
from the Lagrangian~(\ref{sl}) of the full theory. Making the substitution
$\Phi(\tau,{\bf x})\longrightarrow\sqrt{T}\phi({\bf x})$
and comparing $\int_0^{\beta}d\tau{\mathcal L}$ with the effective Lagrangian 
in~(\ref{leffsca}), we obtain
\bqa
\lambda&=&g^2T\;.
\label{lambdadef}
\eqa
\comment{mj}
We next consider the coefficient of the unit operator $f$.
This parameter also has an expansion in powers of $g^2$ and 
can be interpreted as the contribution to the free energy $\log{\mathcal Z}$
from the distance scale $1/T$. It
can be determined by calculating the free energy in the two theories and 
demanding that they be the same.
In the full theory the Feynman diagrams contributing to $\log{\mathcal Z}$ 
through  three loops are shown in Fig.~\ref{scalarpertgraphs} (except
for those with a cross):
\bqa\nonumber
{T\log{\mathcal Z}\over V}&\approx&{1\over2}\sumint_P\log P^2-
{1\over8}Z_{g^2}g^2\left(
\sumint_P{1\over P^2}
\right)^2+{1\over16}g^4\left(\sumint_P{1\over P^2}\right)^2
\sumint_P{1\over P^4}
\\ &&
+{1\over48}g^4\sumint_{PQR}{1\over P^2Q^2R^2(P+Q+R)^2}\;.
\label{funit}
\eqa
\comment{mj}
The expression~(\ref{funit}) is ultraviolet divergent. Renormalization
is carried out by substituting the renormalization constant
$Z_{g^2}$~(\ref{scount}).
The sum-integrals are given in appendix A and we obtain 
\bqa\nonumber
{T\log{\mathcal Z}\over V}&\approx&{\pi^2T^4\over90}\Bigg\{
1-{5\over4}\alpha
+{15\over4}\left[\log{\mu\over4\pi T}+{31\over45}+{1\over3}\gamma
+{4\over3}{\zeta^{\prime}(-1)\over\zeta(-1)}
\right.\\ &&\left.
\hspace{3cm}
-{2\over3}{\zeta^{\prime}(-3)\over\zeta(-3)}
\right]\alpha^2
\Bigg\}\;.
\label{fmatch}
\eqa
\comment{mj}
In the effective theory, the diagrammatic expansion of the partition function
is given by the diagrams in Fig.~\ref{scalarpertgraphs}.
Since the mass term is treated as a perturbation, the
resulting loop integrals are massless and therefore vanish in dimensional
regularization.
Eq~(\ref{fmatching}) therefore reduces to
\bqa
{T\log{\mathcal Z}\over V}\approx-fT\;.
\eqa
\comment{mj}
$f(\Lambda)$ is then given by minus the left hand side of~(\ref{fmatch})
divided by $T$. Using the renormalization group equation~(\ref{rgsca})
it can be shown that $f(\Lambda)$ is independent of the ultraviolet cutoff
$\Lambda$ up to corrections of order $g^6$.

\subsubsection{Calculations in the effective theory}
\label{dimredsub}
Having determined the parameters in the effective Lagrangian, it can now
be used to calculate physical quantities using perturbation theory.
In order to take into account the physical effects of screening, we need to
include the mass term in free part of the Lagrangian. The effective
Lagrangian~(\ref{leffsca}) is then decomposed as
\bqa
\left({\mathcal L}_{\rm eff}\right)_{\rm free}&=& 
{1\over2}(\nabla\phi)^2+{1\over2}m^2\phi^2\;,\\
\left({\mathcal L}_{\rm eff}\right)_{\rm int}&=&
{\lambda^2\over24}\phi^4+\delta{\mathcal L}\;.
\eqa
\comment{jm}

Using this decomposition, we next consider the free energy.
The diagrams contributing to the free energy 
through  three loops are shown in Fig.~\ref{scalarpertgraphs} (except
for those with a cross):
\bqa\nonumber
{\log{\mathcal Z}_{\rm eff}\over V}&=&-
{1\over2}\int_{p}\log\left(p^2+m^2\right)
+{1\over2}\delta m^2\int_p{1\over p^2+m^2}
-{1\over8}\lambda\left(\int_{p}{1\over p^2+m^2}\right)^2
\\ \nonumber
&&+{1\over16}\lambda^2\left(\int_{p}{1\over p^2+m^2}\right)^2
\int_{p}{1\over(p^2+m^2)^2}
\\ 
&&+{1\over48}\lambda^2
\int_{pqr}{1\over p^2+m^2}{1\over q^2+m^2}{1\over r^2+m^2}
{1\over({\bf p}+{\bf q}+{\bf r})^2+m^2}\;,
\eqa
\comment{jm}
where $\delta m^2$ is the mass counterterm given in~(\ref{deltamass4}).
Using $g^2=\lambda T$ together with the value for the mass counterterm and 
the results for the integrals listed in the appendix, we obtain
\bqa
{\log{\mathcal Z}_{\rm eff}\over V}&=&
{1\over12\pi}m^3-{\lambda\over8(4\pi)^2}m^2
-{\lambda^2\over12(4\pi)^3}
\left[\log{\Lambda\over2m}+{9\over8}-\log2\right]m\;.
\label{feff}
\eqa
\comment{jm}

The free energy is the sum of~(\ref{funit}) and~(\ref{feff}).
Substituting the expression~(\ref{masspara}) for the mass parameter
and expanding~(\ref{feff}) in powers of $g$, 
we obtain
\bqa\nonumber
T{\log{\mathcal Z}_{\rm eff}\over V}&=&
{\pi^2T^4\over90}\left[
{5\sqrt{6}\over3}\alpha^{3/2}
-{15\over2}\alpha^2
-{15\sqrt{6}\over2}\left(\log{\mu\over2\pi T}
-{2\over3}\log\alpha
\right.\right.\\ &&\left.\left.
+{5\over6}-{5\over3}\log2
+{2\over3}\log3+{1\over3}\gamma
-{2\over3}{\zeta^{\prime}(-1)\over\zeta(-1)}
\right)
\alpha^{5/2}
\right]\;.
\label{fexp}
\eqa
\comment{jm}
Combining Eq.~(\ref{funit}) and Eq.~(\ref{fexp}), 
we recover the weak-coupling expansion
result~(\ref{fscaw}) through order $g^5$. This result is accurate up to
corrections of order $g^6\log g$.
Note that the explicit $\Lambda$-dependence of~(\ref{feff}) cancels against
the $\Lambda$-dependence of the mass parameter~(\ref{masspara})
to next-to-leading order in $g$.

We have seen that the parameters $f(\Lambda)$ and $\lambda(\Lambda)$
are independent of $\Lambda$, while $m^2(\Lambda)$ depends explicitly on 
the cutoff. The parameters of the effective theory satisfy some
evolution or renormalization group equations. These equations
follow from the condition that physical quantities such as the
Debye mass or the free energy are independent of the cutoff $\Lambda$.
They can be derived by calculating the logarithmic ultraviolet divergences
of the loop diagrams in the effective theory.
For example, the coefficient of the unit operator satisfies the 
evolution equation
\bqa
\Lambda{df\over d\Lambda}&=&-
{\pi^4\over12}{\lambda^3\over(4\pi)^6}\;.
\label{frun}
\eqa
\comment{jm}
It can be derived as follows. At the four-loop level there are two diagrams
contributing to the free energy that have logarithmic ultraviolet divergences.

The poles of the first diagram are cancelled by the poles of the 
two-loop diagram with one mass counterterm insertion. The $\Lambda$ dependence
of the second diagram is cancelled by the $\Lambda$ dependence of the
parameter $f$ and this yields the RG equation~(\ref{frun}).
Similarly, the mass parameter satisfies a renormalization group equation
which is derived in the same manner and reads
\bqa
\Lambda{dm^2\over d\Lambda}&=&{\lambda^2\over6(4\pi)^2}\;.
\label{massrgeq}
\eqa\comment{jm}
The solutions to~(\ref{frun}) and~(\ref{massrgeq}) are trivial
\bqa
f(\Lambda)&=&f(2\pi T)
-{\pi^4\over12}{\lambda^3\over(4\pi)^6}
\log{\Lambda\over2\pi T}\;,
\label{rgf}
\\
m^2(\Lambda)&=&m^2(2\pi T)+{\lambda^2\over6(4\pi)^2}\log{\Lambda\over2\pi T}\;.
\label{rgmass}
\eqa
\comment{jm}
The solutions to the evolution equations can be used to sum up 
leading logarithms of $g$ in physical quantities such as the free energy.
This is accomplished by using the expressions~(\ref{rgf})
for the 
short-distance part of the free energy and~(\ref{rgmass})
for the mass parameter of the effective theory:
\bqa\nonumber
{\mathcal F}&=&Tf(2\pi T)-{\pi^4\lambda^3T\over12(4\pi)^6}
\log{\Lambda\over2\pi T}
-{1\over12\pi}m^3T+{\lambda\over8(4\pi)^2}m^2 T
\\ &&
+{\lambda^2\over96(4\pi)^3}
\left[9-16\log2\right] m T
\;.
\label{above}
\eqa
\comment{mj}
The first two terms in~(\ref{above}) is the contribution to the free energy
from the scale $T$ and include a term $g^6\log g$ that arises from the
solution~(\ref{rgf}) to the evolution equation. The other contribution
to comes from expanding the $\lambda m^2$ term in~(\ref{above})
using~(\ref{rgmass}). 

\begin{figure}
\centerline{\psfig{file=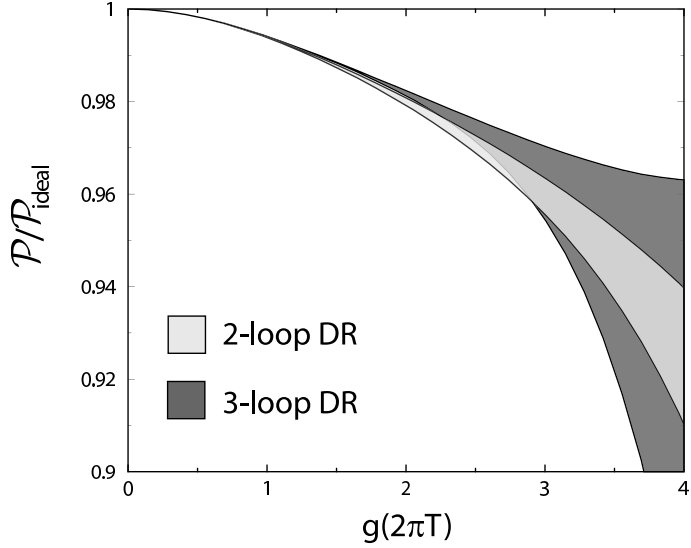,width=9cm}}
\vspace*{8pt}
\caption{The two- and three-loop approximations to the pressure
where the soft part is not expanded out in powers of $g$.}
\label{fdim}
\end{figure}

Fig.~\ref{fpert} clearly showed that the weak-coupling expansion is poorly
behaved at large coupling. 
This fact motivated Blaizot, Iancu and Rebhan~\cite{bir3} to
treat the soft sector differently. Instead of expanding the soft contribution
to the pressure of QCD
as a power series in $g$ and truncating at the appropriate
order, they calculated it in the loop expansion and kept the 
parameters of the effective Lagrangian as given by the 
dimensional reduction step. We apply the same strategy to the scalar field
theory and show the resulting successive approximations in Fig.~\ref{fdim};
at two loops we take the two-loop hard contribution and combine this with
the two-loop soft contribution using the one-loop mass parameter. At
three loops we take the three-loop hard contribution and combine this with
the three-loop soft contribution using the two-loop mass. The bands are
obtained by varying the renormalization scale $\mu=\Lambda$ by 
a factor of $2$ around $2\pi T$.

The three-loop pressure is then given by the sum of~(\ref{funit}) 
and~(\ref{feff}), where the parameters $m^2$ and $\lambda$ are given 
by~(\ref{masspara}) and~(\ref{lambdadef}). The result is shown in 
Fig.~\ref{fdim}. It is interesting to note that the two-loop band 
is almost inside the three-loop band.
The uncertainty in the three-loop approximation within this
scalar toy model is quite large.  Note that if the ${\mathcal O}(g^6 \log
g)$ term is included as in Eq.~(\ref{above}) then the variation of
the result with respect to the scale $\mu$ is less, namely, 
$0.93 < {\mathcal P}/{\mathcal P}_{\rm ideal} < 0.95$ at $g=4$.  However,
in presenting the result shown in Fig.~\ref{fdim} we have not included
this term because it formally contributes at four-loop order.  Also note
that within QCD the variation of the corresponding 
three-loop result is less owing
to the independence of $m_E$ on the electric factorization scale,
$\Lambda_E$ \cite{bir3}.

\subsection{QED}
In this subsection, we will discuss dimensional reduction and
the construction of a three-dimensional effective field theory
for QED.
We call this effective field theory {\it electrostatic QED} (EQED) in 
accordance with the terminology introduced by
Braaten and Nieto in the case of QCD \cite{ea2}.
The effective field theory is used to calculate the Debye mass and the
free energy through order $e^5$.
In massless QED, there are two momentum scales that contribute to 
equilibrium quantities such as the free energy. In analogy with the
scalar case, 
these are the scales $T$ and $eT$, where $e$ is the electromagnetic
gauge coupling.
These scales are associated with the typical momentum of a particle in the
plasma and the screening of static electric fields, respectively.

\subsubsection{Electrostatic QED}

The fields of the effective Lagrangian can, up to field redefinitions,
be identified with the zero-frequency modes of the fields in QED.
The timelike component of the gauge field, $A_0$, behaves as a real massive
interacting field. The fact that $A_0$ develops a thermal mass is a consequence
of the breakdown of Lorentz invariance at finite temperature.
Moreover, ${\mathcal L}_{\rm eff}$ is a gauge-invariant function of the 
spatial components $A_i$. There is three-dimensional rotational symmetry
and a discrete symmetry $A_0\longrightarrow-A_0$.
The effective Lagrangian can then be written as
\bqa
{\mathcal L}_{\rm eff}&=&{1\over2}\left(\nabla A_0\right)^2
+{1\over2}m^2_EA_0^2+{\lambda_E}A_0^4
+{1\over4}F_{ij}^2+\delta{\mathcal L}_{\rm eff}\;,
\label{eqedl}
\eqa
\comment{jm}
where $\delta{\mathcal L}_{\rm eff}$ again denotes higher order operators
that respect the symmetries.
The operator $A_0^4$ first contributes to the Debye mass at order
$e^5$ and to the free energy at order $e^6$.

The coefficient of the unit operator $f(\Lambda)$ was calculated 
in Ref. \cite{jensqed} through three loops:
\bqa\nonumber
f(\Lambda)&=&{11\pi^2T^4\over180}\Bigg\{1-{50\over11}\alpha
+{20\over11}\left[
-{20\over3}\log{\Lambda\over4\pi T}
+{8\over3}{\zeta^{\prime}(-3)\over\zeta(-3)}
-{16\over3}{\zeta^{\prime}(-1)\over\zeta(-1)}
\right.
\\&&\hspace{25mm}
\left.
-4\gamma
-{319\over12}+{208\over5}\log2
\right]\alpha^2\Bigg\}\;,
\label{qedf}
\eqa
\comment{jm}
where $\alpha = e^2/(4\pi)^2$.
The mass parameter has been calculated by various authors
through two loops
\cite{bip},\cite{jensqed}:
\bqa
m^2_E(\Lambda)&=&{16\pi^2\over3}T^2\alpha\left[
1-{8\over3}\left(\log{\mu\over4\pi T}
+\gamma+2\log2+{7\over4}
\right)\alpha
\right]\;.
\label{massparaqed}
\eqa
\comment{mj}
The mass parameter coincides with the Debye mass through order $\alpha^2$.
The reason is that there is no $\alpha^{3/2}$-term in weak-coupling expansion
for the Debye mass. This is due to the facts that there are only fermionic
propagators in the one-loop self-energy diagram in QED and that fermions need
no resummation.

The coupling constant $e_E$ is determined in the same manner
as $\lambda$ in $\phi^4$-theory. The result is~\cite{lands}
\bqa
e_E^2(\Lambda)&=&e^2T\;.
\eqa
\comment{jm}
The coupling constant $\lambda_E$ is calculated by evaluating the
four-point function for timelike photons at zero external momenta.~\cite{lands}
Taking the different normalizations of the fields in full QED and
EQED into account, one obtains
\bqa
\lambda_E(\Lambda)&=&-{e^4T\over12\pi^2}\;.
\eqa
\comment{jm}
Note that $\lambda_E$ requires no renormalization and vanishes at zero
temperature. This reflects the gauge symmetry of QED.

The renormalization group equation for the running gauge coupling is
\bqa
\mu{de^2\over d\mu}&=&{e^4\over6\pi^2}\;.
\label{runqed}
\eqa
\comment{jm}
From Eq.~(\ref{runqed}), it is easy to check that the parameters
$f$, $m_E^2$, $e_E^2$, and $\lambda_E$ are in fact renormalization
group invariant up to higher-order corrections.
\subsubsection{Calculations in EQED}
We next turn to calculations in the effective theory. In EQED, this is 
particularly simple since interactions first contribute to the
free energy at order $e^6$.
Thus through $e^5$, the contribution to the free energy is given by a simple
one-loop calculation:
\bqa\nonumber
\label{qedeff}
{\log{\mathcal Z}_{\rm eff}\over V}&=&-{1\over2}\int_p\log(p^2+m_E^2)\\
&=&{m_E^3\over12\pi}\;.
\label{free3d}
\eqa
\comment{jm}
Expanding~(\ref{free3d})
in powers of $e$ and truncating at order $e^5$, we obtain 
\bqa
T{\log{\mathcal Z}_{\rm eff}\over V}&=&{11\pi^2T^4\over18}\left[
{32\sqrt{3}\over33}\alpha^{3/2} 
\right. \nonumber \\
&& \left. \hspace{1.7cm} -{128\sqrt{3}\over33}
\left(\log{\mu\over 4\pi T}+\gamma+2\log2
+{7\over4}\right)\alpha^{5/2}
\right]\;.
\label{xxx}
\eqa
\comment{jm}
The free energy through $e^5$ 
is the sum of Eqs.~(\ref{qedf}) and~(\ref{xxx}). This result was first
obtained by Parwani using resummation.~\cite{parwani} In that calculation,
the order $e^5$ contribution is obtained by expanding 
two- and three-loop
diagrams as it was done in the scalar case in Sec.~\ref{weak}.
Here, the result is obtained by a straightforward one-loop calculation and
demonstrates that effective field theory significantly 
simplifies calculations of physical quantities.

Similarly, the contribution to the 
Debye mass $m_D^2$ from the scale $gT$
is also given by a simple one-loop diagram in the effective theory.
If the self energy in EQED is denoted by $\Pi_{\rm eff}(p,\Lambda)$, we have
\bqa\nonumber
m_D^2&=&m_E^2+\Pi_{\rm eff}(p,\Lambda)\\\nonumber
&=&m_E^2+12\lambda_E\int_k{1\over k^2+m_E^2}\\
&=&{16\pi^2\over3}T^2\alpha\left[
1-{8\over3}\left(\log{\mu\over4\pi T}
+\gamma+2\log2+{7\over4}
\right)\alpha
+16\sqrt{3}\alpha^{3/2}
\right].
\eqa
\comment{jm}
The Debye mass to order $e^5$ was first obtained by Blaizot, Iancu, and 
Parwani using resummation techniques \cite{bip}.

The resummation of logarithms of $e$
is straightforward. The evolution equations for $f(\Lambda)$ and 
$m_E^2(\Lambda)$ are derived in the same manner as in the 
scalar case. The logarithmic divergences in the effective theory
occur in the three-loop basketball diagram for $f$ and the two-loop
setting sun diagram for the mass parameter. These diagrams are both 
proportional to $\lambda_E^2$. Thus the leading logarithmic term in 
the free energy is proportional to
$e^9\log e$ and in the Debye mass is proportional to $e^8\log e$.
Thus the results for the free energy and Debye mass are correct through order
${\mathcal O}\left(\alpha^3\log\alpha\right)$.

\subsection{QCD}
In scalar field theory and QED, there are two momentum scales,
$T$ and $gT$,
contributing to equilibrium properties.
In nonabelian gauge theories, there are, in general, 
three momentum scales. They are

\vspace{3mm}
\begin{itemize}
\item{The scale $T$, which is the typical momentum of a particle in the 
plasma.}\\
\item{The scale $gT$, which is the scale associated with the screening of 
color-electric fields.}\\
\item{The scale $g^2T$, which is the scale associated with the screening of
color-magnetic fields.}
\end{itemize}
\vspace{3mm}

The fact that there is an additional scale $g^2T$ in nonabelian gauge
theories such as QCD, can be seen from their infrared properties.
Perturbative QCD at high temperature is plagued by infrared divergences.
The most severe divergences from a given infro
ared divergent
loop diagram comes from the region where all the Matsubara frequencies vanish
and the loop momenta go to zero. Thus the infrared properties are those
of pure-glue QCD in three spatial dimensions. This theory has a dimensionful
coupling constant $g^2T$ and this is the only scale in the theory.
It is a confining theory with a mass gap of order $g^2T$ that can only
be calculated nonperturbatively.~\cite{gpy}

This hierarchy of momentum scales suggests that one constructs a sequence
of two effective field theories that take care of the scales $gT$ and
$g^2T$. This strategy was first proposed by Braaten and Nieto \cite{ea2}.
These effective fields theories are called {\it electrostatic QCD} (EQCD) and
{\it magnetostatic QCD} (MQCD), respectively. 
The contribution to the free energy from the scale $T$ is given by the
coefficient of the unit operator $f_{E}(\Lambda_E)$ in EQCD. 
$f_{E}(\Lambda_E)$ has an expansion in powers of $g^2$ and starts to contribute
to the free energy
at order $g^0$. The contribution from the scale $gT$ is given by the
coefficient of the unit operator $f_M(\Lambda_E\Lambda_M)$ in 
MQCD.~\footnote{The arbitrary factorization scales $\Lambda_E$ and 
$\Lambda_M$ separate the momentum scales $T$ from $gT$, and $gT$ from 
$g^2T$, respectively.} 
It has an expansion
in powers of $g$ and starts to contribute to the free energy at order $g^3$.
Finally, the contribution to the free energy from the scale $g^2T$ is
given by the partition function in MQCD. It has an expansion in powers
of $g^2$, but first starts to contribute to the free energy
at order $g^6$.

\subsubsection{Electrostatic QCD}
EQCD contains an electrostatic field $A_0^a({\bf x})$ and a magnetostatic
field $A_i({\bf x})$. The field $A_0^a({\bf x})$ behaves as a scalar field
in the adjoint representation.
The effective Lagrangian is
\bqa
{\mathcal L}_{\rm EQCD}&=&{1\over4}G_{ij}^aG_{ij}^a
+{1\over2}({\mathcal D}_{i}A_0^a)^2
+{1\over2}m^2_EA_0^aA_0^a+{1\over8}\lambda_E(A_0^a)^2
+\delta{\mathcal L}_{\rm EQCD}\;,
\label{eqcdlag}
\eqa
\comment{jm}
where $G_{ij}=\partial_iA_j^a-\partial_jA_i^a+g_Ef^{abc}A_i^aA_j^a$ is the
magnetostatic field strength and $D_i A_0^a=(\partial_i+g_E {\epsilon}^{abc} 
A_i^b A_0^c)A_0^a$ is the
covariant derivative. $\delta{\mathcal L}_{\rm EQCD}$ consists of all higher-order
gauge invariant operators than can be constructed out of
the scalar field $A_0$ and the gauge field $A_i$. In the matching procedure,
the mass term in EQCD is again treated as a perturbation.

The free energy for QCD can be written as 
\bqa
{\mathcal F}&=&Tf_E(\Lambda_E)-{\log{\mathcal Z}_{EQCD}\over V}
\eqa
\comment{jm}
where 
${\mathcal Z}_{EQCD}$ is the partition function for EQCD.
The coefficient of the unit operator through three loops
was calculated by Braaten and
Nieto~\cite{ea2}
\bqa
\nonumber
f_E(\Lambda_E)&=&-d_A{\pi^2\over45}T^3\Bigg\{1
+{7\over4}{d_f\over d_A}
-5\left(C_A+{5\over2}T_F\right){\alpha\over4\pi}
+5\left[C_A^2\left(48\log{\Lambda_E\over4\pi T}
\right.\right.
\\ \nonumber&&
\left.
\hspace{-8mm} -{22\over3}\log{\mu\over4\pi T}+{116\over5}+4\gamma
+{148\over3}{\zeta^{\prime}(-1)\over\zeta(-1)}
-{38\over3}{\zeta^{\prime}(-3)\over\zeta(-3)}
\right)
\\\nonumber &&
\hspace{-8mm} +C_AT_F\left(
48\log{\Lambda_E\over4\pi T}-{47\over3}\log{\mu\over4\pi T}
+{401\over60}-{37\over5}\log2+8\gamma
\right.\\ \nonumber&&
\left.
\hspace{-8mm} +{74\over3}{\zeta^{\prime}(-1)\over\zeta(-1)}
-{1\over3}{\zeta^{\prime}(-3)\over\zeta(-3)}
\right)
+T_F^2\left(
{20\over3}\log{\mu\over4\pi T}+{1\over3}-{88\over5}\log2+4\gamma
\right.
\\ &&
\left.\left.
\hspace{-8mm} +{16\over3}{\zeta^{\prime}(-1)\over\zeta(-1)}
-{8\over3}{\zeta^{\prime}(-3)\over\zeta(-3)}
\right)
+C_FT_F\left({105\over4}-24\log2\right)\right]\left(\alpha\over4\pi\right)^2
\Bigg\}\,,
\label{fqcdf}
\eqa
\comment{jm}
where $\alpha = g^2/(4 \pi)$.
A few comments are in order. After calculating the diagrams that contribute
to $f_E$ up to three-loop order and renormalizing the coupling constant
by the substitution $g^2\longrightarrow Z_g^2$, 
where 
\bqa
Z_g^2&=&1-{(11C_A-4T_F)g^2\over3(4\pi)^2\epsilon}\;,
\eqa
\comment{jm}
there is still a remaining pole in $\epsilon$. This pole is cancelled by
the counterterm $\delta f_E$:
\bqa
\delta f_E&=&-{d_AC_A\over16\pi}\alpha m_E^2{1\over\epsilon}\;.
\label{dfqcd}
\eqa
\comment{jm}
This counterterm is analogous to the mass counterterm that was needed in the
case of the scalar theory.
The explicit $\Lambda$-dependence of $f(\Lambda_E)$ is necessary to cancel
logarithms of $\Lambda/m_E$ arising in calculations of vacuum diagrams
in EQCD. The coefficient of the unit operator therefore satisfies the
evolution equation
\bqa
\Lambda_E{d\over d\Lambda_E}f_E&=&
-{d_AC_A\over4\pi}\alpha m_E^2\;.
\eqa
\comment{jm}
Note also that we need the order $\epsilon$ of the mass parameter
$m_E^2$ since it multiplies the pole in $\epsilon$ in~(\ref{dfqcd}).
The mass parameter $m_E$ is calculated by matching the Debye mass
in QCD and EQCD using strict perturbation theory.~\footnote{Note that the
Debye mass cannot be defined perturbatively beyond leading order in $g$
due to infrared divergences associated with the magnetostatic scale 
$g^2T$~\cite{reb}
and therefore requires a nonperturbative definition \cite{ay}.
See also Sec.~\ref{phi}.}  
The result through next-to-leading order has been calculated 
by Braaten and Nieto~\cite{ea2} through order $\epsilon$.
The order $\epsilon^0$ term is  
\bqa\nonumber
m^2_E&=&{4\pi\over3}\alpha T^2\Bigg\{
C_A+T_F+\left[
C_A^2\left({5\over3}+{22\over3}\gamma+{22\over3}\log{\Lambda_E\over4\pi T}\right)
\right.\\ \nonumber&& \left.
+C_AT_F\left(
3-{16\over3}\log2+{14\over3}\gamma+{14\over3}\log{\Lambda_E\over4\pi T}
\right)
\right.\\ && \left.
+T_F^2\left({4\over3}-{16\over3}\log2-{8\over3}\gamma-{8\over3}
\log{\Lambda_E\over4\pi T}\right)-6C_FT_F
\right]{\alpha\over4\pi}
\Bigg\}\;.
\label{debyeqcd}
\eqa
\comment{jm}
The order-$\epsilon$ term is
\bqa\nonumber
{\partial m_E^2\over\partial\epsilon}&=&
{4\pi\over3}\alpha T^2\Bigg\{
C_A\left(
2{\zeta^{\prime}(-1)\over\zeta(-1)}+2\log{\Lambda_E\over4\pi T}
\right)
\\ && \hspace{1cm}
+T_F\left(1-2\log2
+2{\zeta^{\prime}(-1)\over\zeta(-1)}+2\log{\Lambda_E\over4\pi T}
\right)
\Bigg\}\;.
\eqa
\comment{jm}
The coupling constants $g_E$ and $\lambda_E$ are calculated in the same
manner as in QED. The result is
\bqa
g_E^2&=&g^2T\;,
\label{ge}
\\
\lambda_E&=&{(9-N_f)g^4\over12\pi^2}T\;.
\eqa
\comment{jm}
The gauge coupling $g^2$ satisfies the RG-equation
\bqa
\mu{dg^2\over d\mu}&=&{22C_A-8T_F\over3}{g^4\over(4\pi)^2}\;.
\label{dmu}
\eqa
\comment{jm}
Using~(\ref{dmu}), it can be shown that 
the mass parameter $m_E^2$ as well as the coupling constants $g_E^2$ and
$\lambda_E$ are independent of the factorization scale $\Lambda_E$.

Having determined the parameters in EQCD, one can calculate the
contribution to various physical quantities from the scale $gT$.
In order to do so, we again include the effects of the 
mass term for $A_0$ into the free part of the Lagrangian of EQCD. 
The contribution to the free energy from the scale $gT$ is given by 
$f$ which is equal to the logarithm of the partition function
of EQCD:
\bqa
Tf_M&=&-{T\log{\mathcal Z}_{\rm EQCD}\over V}\;.
\eqa
\comment{jm}
The calculation of ${\mathcal Z}_{\rm EQCD}$ through three loops was 
done by Braaten and Nieto.~\cite{ea2} The result is
\bqa\nonumber
f_M(\Lambda_E,\Lambda_M)&=&-{d_Am_E^3\over12\pi}\Bigg\{
1-\left[{9\over4}+3\log{\Lambda_E\over2m_E}\right]{C_Ag_E^2\over4\pi m_E}
\\ && \hspace{5mm}
-\left[{89\over8}-{11\over2}\log2+{1\over2}\pi^2\right]
\left({C_Ag_E^2\over4\pi m_E}\right)^2
\Bigg\}\;.
\label{eqcdf2}
\eqa
\comment{jm}
By expanding~(\ref{eqcdf2}) in powers of $g_E$, substituting~(\ref{debyeqcd})
and~(\ref{ge}),
we obtain the contribution to the free energy from the scale $gT$ through
$g^5$:
\bqa\nonumber
f_M(\Lambda_E,\Lambda_M)&=&
-d_A{\pi^2T^3\over45}\Bigg\{
30 \bar\alpha_s^{3/2} \\
&& + {135 \over 2} \left( {C_A \over C_A + T_F} \right) 
\left[ \log\bar\alpha_s - 2 \log{\Lambda_E\over 4\pi T} - {3\over2}\right] \bar\alpha_s^2 \nonumber \\
&& 
+ {45 \over 2 (C_A + T_F)^2} \Biggl[
C_A^2
\left( 11 \log{e^\gamma  \Lambda_E \over4\pi T} + 
{33\over2}\log2 - {3 \pi^2\over2}-{247\over8}\right)
\nonumber \\
&&
+ C_A T_F \left(7 \log{e^\gamma \Lambda_E\over4\pi T} - 8\log2 +{9\over2}\right)
- 9 C_F T_F 
\nonumber \\
&&- T_F^2 \left(4\log{e^\gamma \Lambda_E\over4\pi T}+8\log2-2\right)\Biggr]
\bar\alpha_s^{5/2}
\Bigg\}\;,
\label{fqcdff}
\eqa
\comment{mj}
where $\bar\alpha_s = (C_A + T_F) \alpha_s/3 \pi$.
The free energy through $g^5$ is the sum of~(\ref{fqcdf}) 
and~(\ref{fqcdff}) and is given by~(\ref{weakqcd}) for $C_A=N_c=3$.
Note that the dependence on factorization scale $\Lambda_E$ cancels between
$f_E$ and $f_M$ up to corrections of order $g^6$.

\subsubsection{Magnetostatic QCD}
The effective Lagrangian of magnetostatic QCD is
\bqa
{\mathcal L}_{\rm MQCD}&=&
{1\over4}G_{ij}^aG_{ij}^a+\delta{\mathcal L}_{\rm MQCD}\;,
\label{mqcd}
\eqa
\comment{jm}
where $G_{ij}=\partial_iA_j^a-\partial_jA_i^a+g_Mf^{abc}A_i^aA_j^b$
is the magnetostatic field strength and $g_M$ is the gauge coupling.
The term $\delta{\mathcal L}_{\rm MQCD}$ consists of all possible gauge-invariant
terms that can be constructed out of the spatial gauge field $A_i$.
In Ref.~\cite{ea2} it was shown that the operators in
$\delta {\mathcal L}$ first contribute to the free energy at order $g^{12}$.
In the remainder of this section, we therefore omit $\delta{\mathcal L}$
and only consider the minimal gauge theory with action
$\int d^3x\;{1\over4}G_{ij}^aG_{ij}^a$.

For the present discussion, the only parameter  we need to determine is the 
gauge coupling. To leading order in $g_E$, one 
finds~\cite{ea2}
\bqa
g_M&=&g_E\;.
\eqa
\comment{jm}
To this order in $g$, the parameter $g_M$ is independent of the 
factorization scale $\Lambda_M$.
The free energy of QCD can be written as
\bqa
{\mathcal F}&=&Tf_E(\Lambda_E)
+Tf_M(\Lambda_E,\Lambda_M)-{T\log{\mathcal Z}_{MQCD}\over V}
\;,
\eqa
\comment{jm}
where 
$f_M(\Lambda_E,\Lambda_M)$ is the unit operator of MQCD
and $\log{\mathcal Z}_{MQCD}$ is the partition function in magnetostatic QCD.

Perturbation theory using the Lagrangian~(\ref{mqcd})
is plagued with infrared divergences. In the case of the free energy,
these infrared divergences first enter at the four-loop 
level.
Thus perturbation theory can not be used to calculate the contribution to the
free energy from the scale $g^2T$. However, for dimensional reasons,
the nonperturbative contribution has to be of the form: 
\bqa
-{\log{\mathcal Z}_{MQCD}\over V}&=&
\left(a+b\log{\Lambda_M\over g_M^2}\right)g_M^6\;,
\label{fg}
\eqa
\comment{jm}
where $a$ and $b$ are pure numbers. The coefficient $a$ can be determined by
using nonperturbative methods such as lattice simulations. The coefficient
$b$ can be calculated by extracting the logarithmic ultraviolet
divergences in the four-loop diagrams contributing to the free energy in MQCD.
This was recently done in a calculation by 
Kajantie, Laine, Rummukainen and Schr\"oder \cite{kajaloop,kaja}:
\bqa
b&=&d_AC_A^3{g_M^6\over(4\pi)^4}
\left(
{43\over12}-{157\pi^2\over768}
\right)\;.
\label{bcoeff}
\eqa
\comment{jm}
The free energy to order $g^5$ has been calculated both by resummation 
and effective field theory methods. Using effective field theory, one can
also determine the coefficient of $g^6\log g$. There are two contributions
to this coefficient. The first contribution comes from the scale $gT$
and can be determined by perturbative calculations of the logarithm
of the partition function in EQCD.
It involves the evaluation the of four-loop vacuum diagrams and has recently
been determined in a very impressive calculation by
Kajantie, Laine, Rummukainen and Schr\"oder:~\cite{kajaloop,kaja}
\bqa
f_{M}^{(4)}(\Lambda_E,\Lambda_M)
&=&
d_AC_A^3{g_M^6\over(4\pi)^4}
\left(
{43\over4}-{491\pi^2\over768}\right)\log{\Lambda_E\over m_E}\;,
\eqa
\comment{jm}
where the superscript indicates the number of loops.
The second contribution comes from the scale $g^2T$ and is given by
the coefficient $b$ in Eq.~(\ref{bcoeff}). 


%% file: spt.tex
 \section{Screened Perturbation Theory}
\label{spt}

In this section, we discuss one possibility to reorganize perturbation
theory, which is {\it screened perturbation theory} (SPT) which was
introduced by Karsch, Patk\'os and Petreczky \cite{K-P-P-97}.
It can be made more systematic by using a framework called
``optimized perturbation theory'' that Chiku and Hatsuda~\cite{C-K-98}
have applied to a spontaneously broken scalar field theory.
In this approach, one introduces a single variational parameter $m$, which 
has a simple interpretation as a thermal mass. The advantage of SPT is that 
it is relatively easy to apply and that higher-order corrections can be 
calculated so that one can study convergence properties.

In SPT a mass term is added and subtracted to the scalar Lagrangian 
with the added bit kept as part of the free Lagrangian and the 
subtracted bit associated  with the interactions.  This can be 
accomplished by introducing a
parameter $\delta$ in the spirit of the linear delta expansion \cite{deltaexp}.
The resulting Euclidean Lagrangian is
\bqa
{\mathcal L}&=&{1\over2}(\partial_{\mu}\phi)^2
+(1-\delta)m^2\phi^2
+{g^2\delta\over24}\phi^4
\;,
\label{sptlag}
\eqa
\comment{jm}
where we have also introduced a factor of $\delta$ in the $\phi^4$
term.  In the limit $\delta \rightarrow 1$ the SPT Lagrangian reduces 
to the standard scalar Lagrangian.

The Lagrangian (\ref{sptlag}) is then expanded into free and interacting parts
with all interaction terms proportional to $\delta$
\bqa
{\mathcal L}_{\rm free}&=&{1\over2}(\partial_{\mu}\phi)^2+{1\over2}m^2\phi^2\,\\
{\mathcal L}_{\rm int}&=&\delta\left({g^2\over24}\phi^4-{1\over2}m^2\phi^2\right) 
\;.
\eqa
\comment{jm}

\begin{figure}[t]
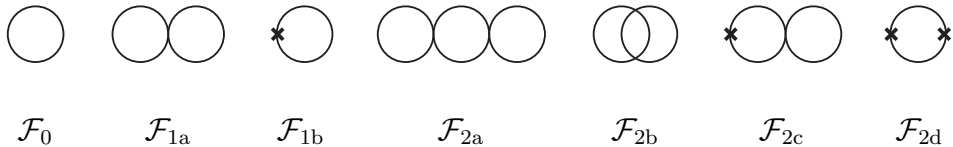


\begin{tabular}{ccccccc}
$\oneloop$ &
$\figureeight$ &
$\oneloopX$ &
$\triplebubble$ &
$\basketball$ &
$\figureeightX$ &
$\oneloopXX$
\vspace{5mm}
\\
${\mathcal F}_{\rm 0}$ &
${\mathcal F}_{\rm 1a}$ &
${\mathcal F}_{\rm 1b}$ &
${\mathcal F}_{\rm 2a}$ &
${\mathcal F}_{\rm 2b}$ &
${\mathcal F}_{\rm 2c}$ &
${\mathcal F}_{\rm 2d}$ 
\end{tabular}

\caption[a]{ 
Diagrams which contribute up to three-loop order in screened perturbation theory.  
A boldfaced $\times$ indicates an insertion of $m^2$.
}  
\label{fig:SPTgraphs}

\end{figure}

The loop expansion then simply becomes an expansion in $\delta$.  
Note also that the counterterms that are necessary to renormalize
the theory also have expansions in powers of $\delta$.
After expansion of all of the contributions 
to the appropriate order in $\delta$ we set $\delta=1$.  
If we were able to calculate the result to all orders in $\delta$
the result would independent of the mass parameter $m$; however,
at any finite order in the expansion there will be a residual 
dependence on $m$.  We will discuss three possible ways
to fix the parameter $m$ including a variational prescription. 
The diagrams which contribute up to the free energy up to three-loop order 
in SPT are shown in Fig.~\ref{fig:SPTgraphs}.

\subsection{One-loop contribution}

The one-loop free energy is
\bqa
\label{freebos}
{\mathcal F}_0 &=& {\mathcal F}_{\rm 0a} + \Delta_0{\mathcal E}_0 \; .
\eqa
\comment{jm}

Using the results from Ref.~\cite{spt} for the sum-integrals
and counterterm $\Delta_0{\mathcal E}_0$, we obtain the final result for 
the one-loop free energy
\bqa
\label{sptf1loop}
(4\pi)^2{\mathcal F}_0&=&-{1\over8}(2L+3)m^4
-{1\over2}J_0T^4\;,
\eqa
\comment{jm}
where $J_n$ are functions of $m/T$ defined in (\ref{jndef})
evaluated at $\epsilon=0$ and $L=\log(\mu^2/m^2)$.

\subsection{Two-loop contribution}

\noindent
The contribution to the free energy of order $\delta$ is
\bqa
\label{twobare}
{\mathcal F}_1 &=& {\mathcal F}_{\rm 1a} +
{\mathcal F}_{\rm 1b}+ \Delta_1{\mathcal E}_0 +
{\partial {\mathcal F}_{\rm 0a} \over \partial m^2} \Delta_1 m^2\;,
\label{f2}
\eqa
\comment{jm}
where $\Delta_1{\mathcal E}_0$ and $\Delta_1 m^2$ are 
the order-$\delta$ vacuum and
mass counterterms, respectively.  Using the results from 
Ref.~\cite{spt} for the sum-integrals and counterterms, the 
two-loop contribution to the free energy becomes
\bqa
(4\pi)^2{\mathcal F}_1&=&
{1\over2}\left[(L+1)m^2-J_1T^2\right]m^2
+{1\over8}\alpha\left[
(L+1)m^2-J_1T^2
\right]^2\;.
\label{sptf2loop}
\eqa
\comment{jm}
\subsection{Three-Loop contribution}

The contribution to the free energy of order $\delta^2$ is
\bqa
{\mathcal F}_2 &=& {\mathcal F}_{\rm 2a} +
{\mathcal F}_{\rm 2b} +{\mathcal F}_{\rm 2c} + {\mathcal F}_{\rm 2d}
+ \Delta_2{\mathcal E}_0 +
{\partial {\mathcal F}_{\rm 0a} \over \partial m^2} \Delta_2 m^2
+{1\over2}
{\partial^2{\mathcal F}_{\rm 0a}\over (\partial m^2)^2}
\left(\Delta_1 m^2\right)^2
\nonumber \\
&&
\hspace{5mm}
\label{3unren}
+
\left({\,\partial {\mathcal F}_{\rm 1a} \over \partial m^2}
+{\,\partial {\mathcal F}_{\rm 1b} \over \partial m^2}
\right) \Delta_1 m^2 +
{{\mathcal F}_{\rm 1a} \over g^2} \Delta_1 g^2
\;,
\label{f3}
\eqa
\comment{jm}
where we have included all of the order-$\delta^2$ 
counterterms.  Using the 
results from Ref.~\cite{spt} for the integrals and counterterms 
the three-loop contribution to the free energy becomes
\bqa\nonumber
(4\pi)^2{\mathcal F}_2&=&
-{1\over4}\left(L+J_2\right)m^4_1
-{\alpha\over4}
(L+J_2)\left[
(L+1)m^2-J_1T^2
\right]m^2\\ \nonumber
&&\hspace{-2cm}-
{1\over48}\alpha^2
\left[\left(
5L^3+17L^2+\mbox{$41\over2$}L-23-\mbox{$23\over12$}\pi^2
        -\psi^{\prime\prime}(1)+C_0+3(L+1)^2J_2
\right)m^4\right.
\\ \nonumber
&&
\left.
\hspace{-1cm}-\left(12L^2+28L-12-\pi^2-4C_1
+6(L+1)J_2\right)J_1m^2T^2
\right. \\
&&\hspace{1cm}
\left.
+\left(3(3L+4)J_1^2+3J_1^2J_2+6K_2+4K_3\right)
T^4
\right]\;,
\label{sptf3loop}
\eqa
\comment{jm}
where $C_0=39.429$, $C_1=-9.8424$, and $K_2$ and $K_3$ are functions of 
$m/T$ given in Ref.~\cite{massive}.

\subsection{Pressure to three loops}

The contributions to the free energy of zeroth, first, and second order in
$\delta$ are given in~(\ref{sptf1loop}),~(\ref{sptf2loop}),
and~(\ref{sptf3loop}), respectively. Adding them we obtain the successive
approximations to the free energy in screened perturbation theory.
Using the fact that ${\mathcal P} = - {\mathcal F}$ we obtain
the one-loop approximation to the pressure:
\bqa
\label{spt1loop}
(4\pi)^2{\mathcal P}_0={1\over8}\left[
4J_0T^4+\left(2L+3\right)m^4
\right]\;.
\eqa
\comment{jm}
The two-loop approximation is obtained by
including~(\ref{sptf2loop}):
\bqa\nonumber
(4\pi)^2{\mathcal P}_{0+1}&=&{1\over8}\left[
4J_0T^4+4J_1m^2T^2-\left(2L+1\right)m^4
\right] \\ && \hspace{15mm}
-{1\over8}\alpha\left[
J_1T^2-(L+1)m^2
\right]^2\;.
\label{ambi}
\eqa
\comment{jm}
The three-loop approximation is obtained by including~(\ref{sptf3loop})
to obtain:
\bqa\nonumber
(4\pi)^2{\mathcal P}_{0+1+2}&=&{1\over8}\left[
4J_0T^4+4J_1m^2T^2+2J_2m^4-m^4
\right]
\nonumber\\ \nonumber
&&
-{1\over8}\alpha\left[J_1T^2-(L+1)m^2\right]\left[J_1T^2
+2J_2m^2+(L-1)m^2\right]
\nonumber
\\ \nonumber
&&
+{1\over48}\alpha^2
\left[
3J_2\left(J_1T^2-(L+1)m^2\right)^2
\right.
\nonumber \\
&& \hspace{5mm} + \left.\left(3(3L+4)J_1^2+6K_2+4K_3\right)T^4\right.
\nonumber\\ \nonumber&& \hspace{5mm} 
-\left(12L^2+28L-12-\pi^2-4C_1\right)J_1m^2T^2
\nonumber
\\ \nonumber
&&
\hspace{5mm} +\left( 5L^3+17L^2+\mbox{${41\over2}$}L-23-\mbox{${23\over12}$}\pi^2
        -\psi^{\prime\prime}(1)+ C_0 \right)m^4
\label{sptfreefin}
\Big]\;,
\\ &&
\eqa
\comment{jm}
The only remaining task is to determine the mass parameter $m$ which
we will discuss next.

\subsection{Mass prescriptions}
The mass parameter $m$ in screened perturbation theory is completely arbitrary.
In order to complete a calculation using SPT, we need a prescription for
the mass parameter $m = m_*(T)$.
The prescription of Karsch, Patk\'os, and Petreczky for $m_*(T)$ is the solution to the
one-loop gap equation:
\bqa
\label{pet}
m_{*}^2={1\over2}\alpha(\mu_*)\left[
J_1(m_*/T)T^2-\left(2\log{\mu_*\over m_*}+1\right)m_*^2
\right]
\;,
\eqa
\comment{jm}
where the function $J_1$ is defined in~(\ref{jndef}).
Their choice for the scale was $\mu_*=T$.
In the weak-coupling limit, the solution to~(\ref{pet}) is 
$m_*=g(\mu_*)T/\sqrt{24}$. 
There are many possibilities for generalizing~(\ref{pet}) to higher orders
in $g$. We will consider three different possibilities in the following.

\subsubsection{Debye mass}

One class of possibilities 
is to identify $m_*$
with some physical mass in the system.  The simplest choice is the 
Debye mass $m_D$ defined by the location of the pole in the static
propagator:
\bqa
p^2+m^2+\Sigma(0,p)&=&0\;,\hspace{1cm}p^2=-m_D^2\;.
\eqa
\comment{jm}
The Debye mass is a well defined quantity in scalar field theory and
abelian gauge theories at any order in perturbation theory. However,
in nonabelian gauge theories, it is plagued by infrared divergences beyond
leading order.~\cite{reb} 

\subsubsection{Tadpole mass}

The {\it tadpole mass} is another generalization of Eq.~(\ref{pet})
to higher loops.
The tadpole mass is defined by 
\bqa
m_t^2&=&g^2\langle\phi^2\rangle\;.
\eqa
\comment{jm}
The tadpole mass is well defined at all orders in scalar field theory, but
the generalization to gauge theories is problematic. The natural replacement
of $\langle\phi^2\rangle$ would be $\langle A_{\mu}A_{\mu}\rangle$, which
is a gauge-variant quantity.

\subsubsection{Variational mass}

There is another class of prescriptions that is variational in spirit.
The results of SPT would be independent of $m$
if they were calculated to all orders. This suggests choosing $m$ to
minimize the dependence of some physical quantity on $m$.
The {\it variational mass} is defined by minimizing the free energy;
\bqa
{\partial {\mathcal F}\over \partial m^2}&=&0\;.
\eqa
\comment{jm}
The variational mass has the benefit that it is well defined at 
all orders in perturbation theory and can easily be generalized to 
gauge theories.

\begin{figure}
\centerline{\psfig{file=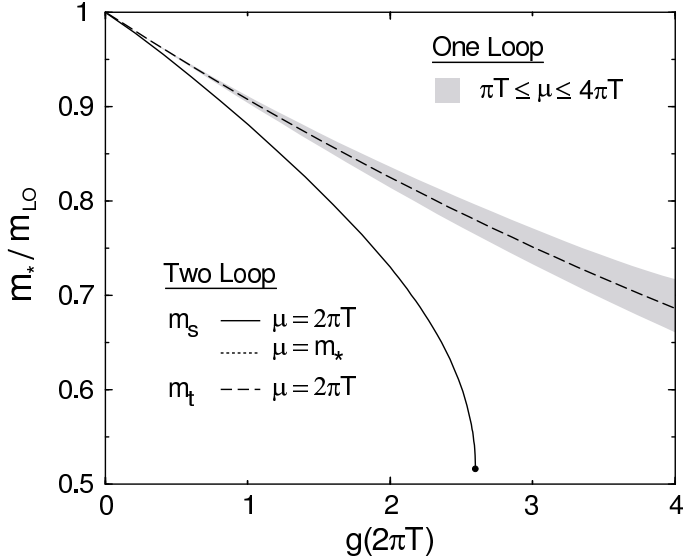,width=9cm}}
\vspace*{8pt}
\caption{Solutions $m_*(T)$ to the SPT one-loop gap equation (shaded bands)
        and two-loop gap equations (lines) as functions of $g(2\pi T)$.}
\label{sptmasses}
\end{figure}

\subsubsection{Comparison}

At one loop, the three different prescriptions give the same 
gap equation, Eq.~(\ref{pet}). Moreover, it turns out that the two-loop
tadpole mass coincides with the one-loop tadpole mass.  However, at 
two-loops the screening and variational masses are 
ill-behaved \cite{spt}.  The
screening mass solution ceases to exist beyond $g \sim 2.6$ and the
variational gap equation has no solution for finite $g$.  
In Fig.~\ref{sptmasses}
we show the various solutions $m_*(T)$ to the SPT one-loop gap equation 
(shaded band) and two-loop gap equations (lines) as functions of $g(2\pi T)$.
From this we conclude that the best prescription is to use the tadpole mass
which results in the same gap equation at one- and two-loops given by
Eq.~(\ref{pet}).

\subsection{Pressure}

\begin{figure}
\centerline{\psfig{file=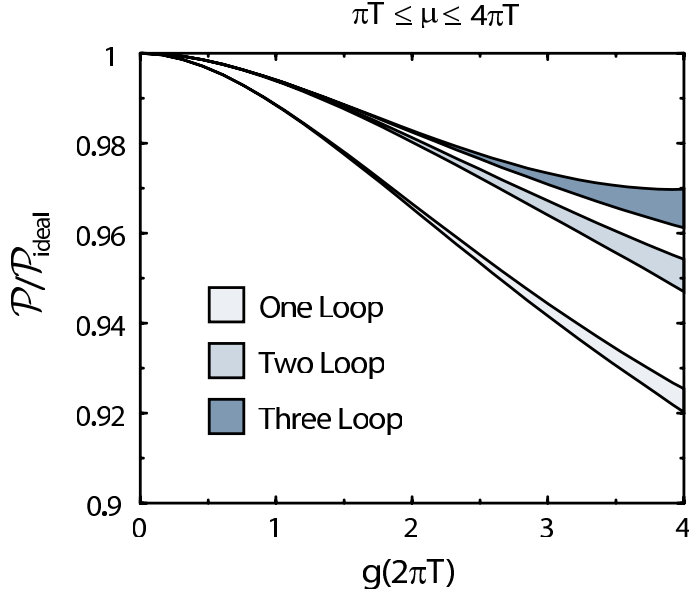,width=9cm}}
\vspace*{8pt}
\caption{One-, two-, and three-loop SPT-improved pressure
        as a function of $g(2\pi T)$.  Bands correspond to
        a variation of the renormalization scale 
        $\pi T < \mu < 4 \pi T$.}
\label{sptpressure}
\end{figure}

In Fig.~\ref{sptpressure}, we show the one-, two-, and three-loop SPT-improved
approximations to the pressure using the tadpole gap
equation.
The bands are obtained by varying $\mu$ by a factor of two around
the central values $\mu=2\pi T$.
The one-loop bands in Fig.~\ref{sptpressure} lie below the other bands;
however, the two- and three-loop bands both lie within the $g^5$ band of the
weak-coupling expansion in Fig.~\ref{fpert}.
The one-, two-, and three-loop approximations to the pressure are
perturbatively correct up to order $g^1$, $g^3$, and $g^5$, respectively;
however, we see a dramatic improvement in the apparent convergence
compared to the weak-coupling expansion.
Improved convergence properties were also found for the entropy and the Debye
mass \cite{spt}.

This result demonstrates the effectiveness of SPT
in providing stable and apparently converging predictions for the thermodynamic
functions of a massless scalar field theory.  An essential ingredient of this
approach is using the solution to a gap equation as the prescription for
the mass parameter $m$.  In the following section we will see how this
idea can be extended to gauge theories.  

%% file: htl.tex
\section{Hard-thermal-loop Perturbation Theory}
\label{htl}
In this section, we discuss the generalization of SPT to gauge theories.
This framework is called hard-thermal-loop perturbation theory (HTLPT). 
The HTLPT framework is attractive because it allows for a systematic analytic 
reorganization of perturbative series based on the HTL effective
action.  Additionally, it is manifestly gauge invariant and is applicable to
calculating 
both static and dynamical quantities.  For simplicity, here we restrict the 
discussion to pure-glue QCD. Calculations including fermions can be found 
in Refs.~\cite{fermions,aps1}.

The Minkowski-space Lagrangian density that generates the perturbative 
expansion for
pure-glue QCD can be expressed in the form
\bqa
{\mathcal L}_{\rm QCD}=-{1\over2}{\rm Tr}\left(G_{\mu\nu}G^{\mu\nu}\right)
+{\mathcal L}_{\rm gf}+{\mathcal L}_{\rm ghost}+\Delta{\mathcal L}_{\rm QCD},
\label{L-QCD}
\eqa
\comment{mj}
where $G_{\mu\nu}=\partial_{\mu}A_{\nu}-\partial_{\nu}A_{\mu}
	-ig[A_{\mu},A_{\nu}]$ is the gluon field strength
and $A_{\mu}$ is the gluon field expressed
as a matrix in the $SU(N_c)$ algebra.
The ghost term ${\mathcal L}_{\rm ghost}$ depends on the choice of
the gauge-fixing term ${\mathcal L}_{\rm gf}$.
In HTLPT the Lagrangian density is written as
\bqa
{\mathcal L}= \left({\mathcal L}_{\rm QCD}
+ {\mathcal L}_{\rm HTL} \right) \Big|_{g \to \sqrt{\delta} g}
+ \Delta{\mathcal L}_{\rm HTL} \; .
\label{L-HTLQCD}
\eqa
\comment{mj}
The HTL improvement term appearing above is $(1-\delta)$ times 
the isotropic HTL effective action which generates all HTL $n$-point
functions\cite{bpaction}
\bqa
\label{L-HTL}
{\mathcal L}_{\rm HTL}=-{1\over2}(1-\delta)m_D^2 {\rm Tr}
\left(G_{\mu\alpha}\left\langle {y^{\alpha}y^{\beta}\over(y\cdot D)^2}
	\right\rangle_{\!\!y}G^{\mu}_{\;\;\beta}\right) \; ,
\eqa
\comment{mj}
where
$D_{\mu}$ is the covariant derivative in the adjoint representation,
$y^{\mu}=(1,\hat{{\bf y}})$ is a light-like four-vector,
and $\langle\ldots\rangle_{ y}$
represents the average over the directions
of $\hat{{\bf y}}$.
The free Lagrangian in general covariant gauge
is obtained by setting $\delta=0$ in~(\ref{L-HTLQCD}):
\bqa
\nonumber
{\mathcal L}_{\rm free}&=&-{\rm Tr}\left(\partial_{\mu}A_{\nu}
\partial^{\mu}A^{\nu}-\partial_{\mu}A_{\nu}\partial^{\nu}A^{\mu}\right)
-{1\over\xi}{\rm Tr}\left[\left(\partial^{\mu}A_{\mu}\right)^2\right] \\
&&-{1\over2}m_D^2\mbox{Tr}
\left[(\partial_{\mu}A_{\alpha}-\partial_{\alpha}A_{\mu})
\left\langle{y^{\alpha} y^{\beta}\over(y\cdot\partial)^2}\right\rangle_{\!\!y}
(\partial^{\mu}A_{\beta}-\partial_{\beta}A^{\mu})\right].
\label{L-free}
\eqa
\comment{mj}
The resulting propagator is the HTL gluon propagator and
the remaining terms in (\ref{L-HTLQCD}) are treated as perturbations.
The propagator can be decomposed into transverse and longitudinal
pieces which in Minkowski space are given by
\bqa
\Delta_T(p)&=&{1 \over p^2 - \Pi_T(p)}\;,
\label{Delta-T}
\\
\Delta_L(p)&=&{1 \over -n_p^2 p^2+\Pi_L(p)} \;,
\label{Delta-L}
\eqa
\comment{mj}
where $\Pi_T$ and $\Pi_L$ are the transverse and longitudinal self-energies,
respectively, and the four-vector $n_p^\mu$ is
\bqa
n_p^\mu = n^\mu - {n\!\cdot\!p \over p^2} p^\mu \; .
\eqa
\comment{mj}
The four-vector $n^\mu$ specifies the thermal rest frame 
(canonically $n=(1,{\bf 0})$).

In terms of $\Pi_T$ and $\Pi_L$ the full self-energy tensor is given by
\bqa
\label{pi-def}
\Pi^{\mu\nu}(p) \;=\; - \Pi_T(p) T_p^{\mu\nu}
- {1\over n_p^2} \Pi_L(p) L_p^{\mu\nu}\;,
\eqa
\comment{mj}
where the tensors $T_p$ and $L_p$ are
\bqa
T_p^{\mu\nu}&=&g^{\mu\nu} - {p^{\mu}p^{\nu} \over p^2}
-{n_p^{\mu}n_p^{\nu}\over n_p^2}\;,\\
L_p^{\mu\nu}&=&{n_p^{\mu}n_p^{\nu} \over n_p^2}\;.
\eqa
\comment{mj}

The HTL gluon self-energy tensor for a gluon of momentum $p$ is given in
Eq.~(\ref{scomp}) with $m_D^2$ given by Eq.~(\ref{qcdmd})
and in the limit that $d \rightarrow 3$ the resulting 
expressions for the transverse and longitudinal gluon self-energies 
reduce to the expressions given in Eqs.~(\ref{redt}) and (\ref{redl})
with $m_D^2$ given by Eq.~(\ref{qcdmd3}).  
Note that there are also HTL vertex corrections which are given by similar but
somewhat more complicated expressions which can be found in Refs.~\cite{er,htl2,aps1}.

As mentioned above, HTLPT is a systematic framework for performing
calculations in thermal gauge theories which is gauge invariant by construction. 
It systematically
includes several physical effects of the plasma such as the propagation
of massive particles, screening of interactions, and Landau damping.
We briefly discuss these next.

\begin{figure}
\centerline{\psfig{file=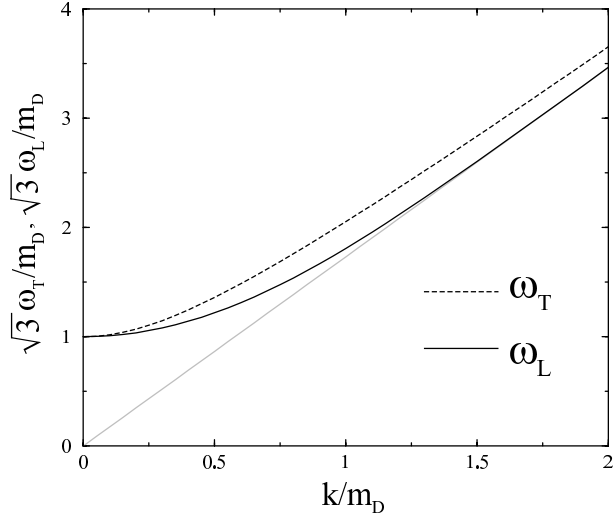,width=8cm}}
\vspace*{8pt}
\caption{Longitudinal and transverse dispersion relations.}
\label{gdisp}
\end{figure}

\vspace{0.2cm}

\noindent
{\it Massive quasi-particles:}

\vspace{0.1cm}

The HTL self-energies are included in the zeroth order 
propagators which results in the resummed HTL propagator for gluons.
The dispersion relations for the transverse and longitudinal 
gluonic degrees of freedom are determined by locating the zeros
of the inverse propagator which gives the following two equations
\bqa
\omega_T^2-k^2-\Pi_T(\omega_T,k)&=&0\;,\\
k^2+\Pi_L(\omega_L,k)&=&0\;.
\eqa
\comment{mj}
The dispersion relations
for transverse and longitudinal gluons are shown in 
Fig.~\ref{gdisp}.  As can been seen from this Figure both modes
approach a constant in the limit of small momentum and
approach the light-cone in the limit of large momentum.

\vspace{0.2cm}

\noindent 
{\it Screening:}

\vspace{0.1cm}

HTLPT also includes screening of interactions which can be seen by
examining the static limit of the HTL propagators. For instance, the inclusion
of the longitudinal self-energy changes the Coulomb potential of two
static charges in the plasma to a Yukawa potential:
\bqa
\lim_{\omega\rightarrow0}
\Delta_L(\omega,k)
&=&{1 \over k^2+m_D^2} \;,
\eqa
\comment{mj}
This result shows that chromoelectric fields are screened on a
scale $r \sim m_D^{-1}$.
Likewise, the screening 
of long wavelength chromomagnetic fields is determined by the
transverse propagator for small frequencies
\bqa
\Delta_T(\omega,k)\sim
{1\over k^2 +  i {\pi\over 4} m_D^2 \omega/k}\;.
\label{ms}
\eqa
\comment{mj}
This shows that there is no screening of static magnetic fields meaning
that the magnetic mass problem of nonabelian gauge theories at high
temperature is not solved by HTL resummation.  However, HTL resummation 
does give {\it dynamical} screening at a scale 
$r\sim\left(m^2_D\omega\right)^{-{1 \over 3}}$.
Note that
the divergences associated with the absence of static magnetic screening do not
pose a problem until four-loop order; however, at four loops the 
lack of static magnetic screening gives
rise to infrared divergences that cause perturbation theory to break down.

\vspace{0.2cm}

\noindent 
{\it Landau damping:}

\vspace{0.1cm}

The transverse and longitudinal HTL self-energies also contain the physics
of Landau damping.  Landau damping represents a transfer of energy from
the soft modes to the hard modes for spacelike momentum.  This can be seen
from the analytic structure of the self-energies given by 
Eqs.~(\ref{redt}) and (\ref{redl}).  Because of the logarithms appearing
in these functions there is an imaginary contribution to the self-energies
for $-k<\omega<k$ which gives the rate of energy transfer from the
soft to hard modes.  Note that ignoring this contribution leads to gauge
variant and unrenormalizable results.

\subsection{Calculation of the free energy using HTLPT}

\begin{figure}[htb]
\centerline{\psfig{file=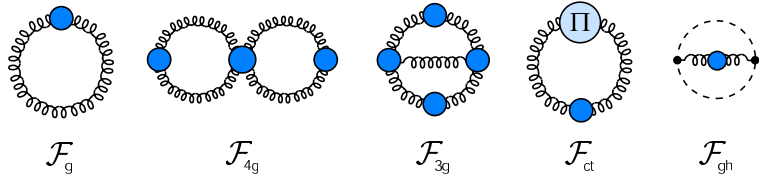,width=10cm}}
\vspace*{8pt}
\caption[a]{Diagrams contributing through NLO in HTLPT.  Shaded circles
indicate dressed HTL propagators and vertices.}
\label{htl2feynman}
\end{figure}

We will now present results of a two-loop calculation of the free energy
of pure-glue QCD within HTLPT.  The relevant one- and two-loop diagrams 
are shown in Fig.~\ref{htl2feynman} with shaded circles indicating dressing
of propagators/vertices with HTL resummed propagators/vertices.

\subsubsection{One-loop contribution}

The thermodynamic potential at leading order in HTLPT is
\begin{equation}
\Omega_{\rm LO} \;=\; (N_c^2-1) {\mathcal F}_g
	\;+\; \Delta_0{\mathcal E}_0 \;,
\label{Omega-LO:def}
\end{equation}
\comment{mj}
where $\Delta_0{\mathcal E}_0$ is a vacuum counterterm and
${\mathcal F}_g$ is shown in Fig.~\ref{htl2feynman}. 
Using the results from Ref.~\cite{htl2} for the sum-integrals
and counterterms the one-loop free energy becomes
\bqa
{\Omega_{\rm LO}\over {\mathcal F}_{\rm ideal}}
&=& 
1 - {15 \over 2} \hat m_D^2
+ 30 \hat m_D^3
\nonumber \\ && \hspace{1cm}
+ {45 \over 8}
\left( {1\over \epsilon} + 2 \log {\hat \mu \over 2}
	- 7 + 2 \gamma 
+ {2 \pi^2\over 3} \right)
	\hat m_D^4  
\;,
\label{Omega-1loop}
\eqa
\comment{mj}
where $\hat{m}=m_D/2\pi T$ and $\hat{\mu}=\mu/2\pi T$.

\subsubsection{Two-loop contribution}
The diagrams that contribute to the two-loop free energy 
are shown in Fig.~\ref{htl2feynman}. 
The thermodynamic potential at next-to-leading order in HTLPT
can be written as
\bqa\nonumber
\Omega_{\rm NLO}&=& \Omega_{\rm LO} \;+\; (N_c^2-1)
\left[{\mathcal F}_{3g}+{\mathcal F}_{4g}+{\mathcal F}_{gh}+{\mathcal F}_{\rm ct} \right]
+ \Delta_1{\mathcal E}_0
\\ &&
+ \Delta_1m_D^2 {\partial \ \ \over \partial m_D^2} \Omega_{\rm LO} \;,
\label{F1}
\eqa
\comment{mj}
where $\Delta_1{\mathcal E}_0$ and $\Delta_1m_D^2$ are the terms of order
$\delta$ in the vacuum energy density and mass counterterms, respectively.
Again using the results from Ref.~\cite{htl2} for the sum-integrals
and counterterms the two-loop contribution to the free energy can
be computed.  Combining this with the one-loop contribution gives the
total next-to-leading order result for the thermodynamic potential
\bqa
{\Omega_{\rm NLO}\over {\mathcal F}_{\rm ideal}}
&=& 
1 - 15 \hat m_D^3
- {45 \over 4} \left( \log{\hat \mu \over 2}
		- {7\over 2} + \gamma + {\pi^2\over3} \right)
		\hat m_D^4
\nonumber
\\
&&  \hspace{2mm}
+ {N_c \alpha_s \over 3 \pi}
\Bigg[ - {15 \over 4} + 45 \hat m_D
 - {165 \over 4}
\left( \log{\hat \mu \over 2 }
	- {36 \over 11} \log \hat m_D - 2.001 \right) \hat m_D^2
\nonumber
\\
&& \hspace{3.5cm}
+ {495\over 2}
\left( \log{\hat \mu \over 2} + {5\over22} + \gamma \right) \hat m_D^3 \Bigg]
\;.
\label{Omega-NLO}
\eqa
\comment{mj}

\begin{figure}
\centerline{\psfig{file=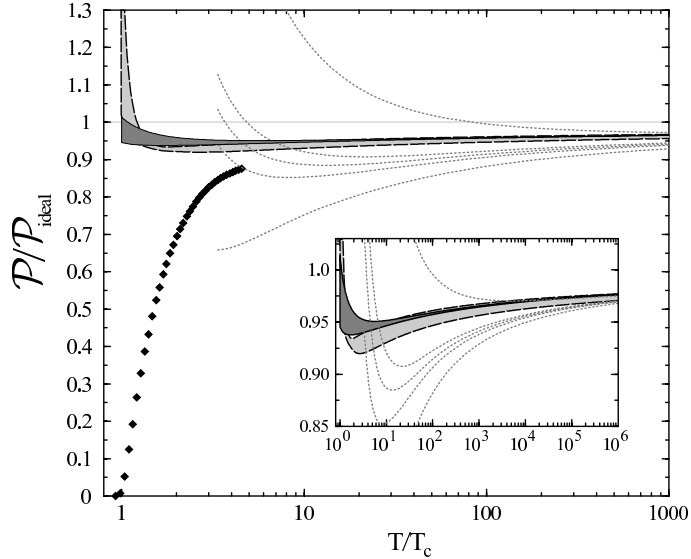,width=9cm}}
\vspace*{8pt}
\caption{
The LO and NLO results for the pressure in HTLPT
compared with 4-d lattice results (diamonds)
and 3-d lattice results (dotted lines) for various values of an unknown
coefficient in the 3-d effective Lagrangian.
The LO HTLPT result is shown as a light-shaded band outlined by a dashed line.
The NLO HTLPT result is shown as a dark-shaded band outlined by a solid line.
The shaded bands correspond to variations
of the renormalization scale $\mu$ by a factor of two around $\mu=2\pi T$.
}
\label{NLOfig}
\end{figure}

In Fig.~\ref{NLOfig},
we have plotted the LO and NLO HTLPT predictions for
the pressure of pure-glue QCD
as a function of $T/T_c$, where $T_c$ is 
the critical temperature for 
the deconfinement phase transition.
We have also included the 4-dimensional lattice gauge theory
results of Boyd et al. \cite{lattice-0} and the 3-dimensional
lattice gauge theory results of Kajantie et al. \cite{KLRS}
in this Figure.

To translate $\alpha_s(2 \pi T)$ into a value of $T/T_c$,
we use the two-loop running formula for pure-glue QCD with
$\Lambda_{\rm \overline{MS}} = 1.04\,T_c$.
Thus $\alpha_s(2 \pi T)= 0.06$ and 0.2 translate into
$T/T_c \sim 400$ and $T/T_c \sim 1$, respectively.
The LO and NLO HTLPT results are shown in Fig.~\ref{NLOfig} as
a light-shaded band outlined by a dashed line and
a dark-shaded band outlined by a solid line, respectively.
The LO and NLO bands overlap all the way down to $T=T_c$,
and the bands are very narrow compared to the corresponding bands
for the weak-coupling predictions in Fig.~\ref{fpertqcd}.
Thus the convergence of HTLPT
seems to be dramatically improved over naive perturbation theory
and the final result is extremely insensitive to the scale $\mu$.

Despite the apparently improved convergence, the predictions of
HTLPT seem to lie significantly above the available 4d lattice
results available for $T < 5 T_c$.  This could be due to at least three
possibilities:  
(a) the expansion in the variational mass performed is
poorly convergent;
(b) by including the HTL resummation also for the
hard modes HTLPT treats the hard modes improperly suggesting that
a method which explicitly treats these separately would be better;
or (c) HTLPT is missing crucial physics related to confinement near
$T\sim T_c$.  Out of these three possibilities the last two
are the most plausible explanations given the success of such
mass expansions applied to scalar theories \cite{AS-01}.  We
will discuss an alternative method of reorganization which addresses
the second possibility in Secs.~\ref{drspt} and~\ref{drphisec}.
The third possibility could, in principle, be taken into account
by using an effective theory which is written in terms of 
Polyakov loops \cite{polrob1}.  For a particularly nice introduction
see Ref.~\cite{polrob2}.  Unfortunately, work
on this is still unable to produce a method for calculating
the coefficients in the Polyakov loop effective Lagrangian near $T_c$ where
it would be needed.  Note though that there have been recent
advances in the calculation of the one-loop Polyakov effective
action \cite{diaos,megias} and 
attempts to extract renormalized Polyakov loops from the 
lattice \cite{polrob3}.  Also there has been an attempt to introduce
models which include a non-trivial Polyakov loop expectation
value in a quasiparticle-based model \cite{polmei}.

The reader may also notice that the HTLPT results also seem to
disagree with the 3d lattice calculations shown in Fig.~\ref{NLOfig}.
In this context it is important to note that the lattice results
from Ref.~\cite{KLRS} are now known to be incorrect due to
the assumption that one of the coefficients in the 3d
effective Lagrangian, namely $\beta_{\rm E1}$, was zero.  It
is now known to be non-zero and this is expected to modify
the result significantly at low temperatures \cite{kaja}.

%% file: phi.tex
\section{$\Phi$-derivable Approach}
\label{phi}
In the previous sections, we have discussed SPT and HTLPT, which are
approaches that involve the variational determination of a mass parameter
$m^2$.
In this section, we will discuss the $\Phi$-derivable approach, which is
another approach that is variational of nature. In this approach, one uses
the exact propagator as a variational function.
Its formulation 
was first constructed by Luttinger and Ward~\cite{lw} and by Baym \cite{baym}.
Later it was generalized to relativistic field theories by Cornwall, Jackiw
and Tomboulis \cite{CJT-74}. The approach is based on the fact that the
thermodynamic potential can be expressed 
in terms of the 2 particle-irreducible (2PI)
effective action which has a diagrammatic expansion 
involving the 2PI skeleton graphs. Although here we focus
on equilbrium physics we also note that the 2PI formalism and its generalizations are 
also very useful when studying non-equilibrium real-time physics
\cite{aarts1,aarts2,berges}.

Applying the $\Phi$-derivable approach to quantum field theories, one is facing
two nontrivial issues. The first issue is the question of renormalization.
The three-loop
calculation by Braaten and Petitgirard~\cite{ep1} for scalar field
theory shows that there are ultraviolet divergences at order $g^6$ that cannot
be eliminated by renormalization. This explicit calculation seems to
be in contradiction with the papers by van Hees and Knoll~\cite{knoll}
and by Blaizot, Iancu and Reinosa~\cite{urko}, which show that 
$\Phi$-derivable approximations can be systematically renormalized.

The second issue is that of gauge dependence. While the exact 2PI effective
action is gauge independent at the stationary point, this property may be
lost in approximations. This problem has recently been studied by 
Arrizabalaga and Smit \cite{arri}, where they showed that the 
$n$-loop$~\Phi$-derivable
approximation $\Phi_n$, defined as the truncation of the action functional 
after $n$ loops, has a controlled gauge dependence.
The gauge dependence of $\Phi_n$ shows up at order $g^{2n}$, where $g$ is
the gauge coupling. Moreover, if the $n$th order solution to the gap
equation is used to evaluate the complete effective action, the 
gauge dependence first shows up at order $g^{4n}$.
 
The $\Phi$-derivable approach has 
several attractive features. First, it respects the global symmetries of the
theory. Thus it is consistent with
the conservation laws that follow from the Noether's theorem.
Second, when evaluated at the stationary point, one is guaranteed 
thermodynamic consistency.~\cite{baym}
Moreover, it turns out that 
the two-loop $\Phi$-derivable approximation has an additional property.
The entropy reduces to the one-loop expression at the variational point.
This property was first shown for QED by Vanderheyden and Baym \cite{van}, and
later generalized to QCD by Blaizot, Iancu and Rebhan \cite{rebb}.

Since the $\Phi$-derivable approach 
involves the propagator as a variational function and not simply
a mass parameter, it is also difficult to solve.  
Braaten and Petitgirard~\cite{ep1} 
have therefore developed a strategy to systematically
solve the $n$-loop $\Phi$-derivable approximation to the thermodynamics of
massless $\phi^4$ field theory. 
The method involves expanding sum-integrals in powers of $m/T$ and $g^2$,
where $m$ is a variational mass parameter.
Blaizot, Iancu and Rebhan~\cite{rebb}
have solved approximately the two-loop $\Phi$-derivable approximation in QCD
by exploiting the fact that the hard-thermal-loop self-energies are
the solutions to the gap equations to leading order in $g$ for soft 
external momenta. We discuss these approximations next.

\subsection{Scalar field theory}
We start our discussion by again considering a massless thermal scalar field theory
with a Lagrangian given by~(\ref{sl}). The thermodynamic potential is
\bqa
\Omega[D]&=&
{1\over2}{\rm Tr}\log D^{-1}
-{1\over2}{\rm Tr}\,\Pi D
+\Phi[D]\;, 
\label{phieq}
\eqa
\comment{jm}where $D$ is the exact propagator,
$\Pi$ is the exact self-energy, and 
$\Phi[D]$ is the sum of all two-particle irreducible diagrams.
The two-particle irreducible diagrams are 
shown diagrammatically in Fig.~\ref{s} up to three-loop order.
\begin{figure}[htb]
\begin{eqnarray*}
-\Phi[D] =  {1\over8} \figureeight + {1\over48} \basketball +...
\end{eqnarray*}
\caption[a]{
$\Phi$-derivable two- and three-loop skeleton graphs.
}
\label{s}
\end{figure}
The exact propagator $D(P)$ satisfies the variational equation
\bqa
\delta\Omega[D]\over\delta D&=&0\;.
\label{vareq}
\eqa
\comment{jm}Using~(\ref{phieq}) 
the variational equation~(\ref{vareq})
can be written as
\bqa
\Pi(P)&=&2{\delta\Phi[D]\over\delta D(P)}\;.
\label{dpi}
\eqa
\comment{jm}In the case of a thermal scalar field, the 
exact propagator is
\bqa
D(P)&=&{1\over P^2+\Pi(P)}\;.
\label{eprop}
\eqa
\comment{jm}Substituting~(\ref{eprop}) into~(\ref{phieq}), 
the thermodynamic potential $\Omega[D]$ becomes
\bqa
\Omega[D]&=&{1\over2}\sumint_P\log\left(P^2+\Pi(P)\right)
-{1\over2}\sumint_P\Pi(P){1\over P^2+\Pi(P)}
+\Phi[D]\;.
\label{termpot}
\eqa
\comment{jm}If we truncate $\Phi$ at the $n$th order in the loop expansion, one refers
to the corresponding variational approximation $\Omega_n$ to the exact 
thermodynamic potential $\Omega$ 
as the {$n$-loop $\Phi$-derivable approximation}. 
The $n$-loop $\Phi$-derivable approximations, where $n=2,3,...$ define
a systematically improvable sequence of approximations to the exact
thermodynamic potential.

\subsubsection{Two-loop $\Phi$-derivable approximation}
In the two-loop $\Phi$-derivable approximation there is only a single
diagram contributing to $\Phi[D]$ which is the first diagram 
on the right-hand side
in Fig.~\ref{s}.
The thermodynamic potential $\Omega_2$ is
\bqa\nonumber
\Omega_2[D]&=&
{1\over2}\sumint_P\log\left(P^2+\Pi(P)\right)
-{1\over2}\sumint_P{\Pi(P)\over P^2+\Pi(P)}
\\ &&
+{1\over8}g^2\left(\sumint_P{1\over P^2+\Pi(P)}\right)^2\;.
\label{tlphi}
\eqa
\comment{jm}It follows from Eq.~(\ref{dpi}) that the self-energy is independent of the
momentum and we shall denote the exact self-energy 
$\Pi$ in this approximation by
$m^2$.
The one-loop gap equation becomes
\bqa
m^2&=&{1\over2}g^2\sumint_P{1\over P^2+m^2}\;.
\label{1lgapphi}
\eqa
\comment{jm}One can use the one-loop gap equation to simplify the 
expression for the
thermodynamic potential:
\bqa
\Omega_2&=&{1\over2}\sumint_P\log\left(P^2+m^2\right)
-{1\over4}m^2\sumint_P{1\over P^2+m^2}\;.
\label{renphi}
\eqa
\comment{jm}This eliminates all reference to terms which 
are explicitly of order $\alpha$.

The two sum-integrals are given in (\ref{s1}) and (\ref{s2}). 
They are separately 
ultraviolet divergent, but the sum is finite. We can take the limit
$\epsilon\rightarrow0$ and obtain
\bqa
\Omega_2&=&-{1\over8(4\pi)^2}\left[
4J_0T^4+2J_1m^2T^2+m^4
\right]\;.
\label{2lt}
\eqa
\comment{mj}
This result is identical to the free energy in SPT at two loops.
The gap equation~(\ref{1lgapphi}) 
is ultraviolet divergent and requires renormalization.
We write the renormalization constant $Z_{g^2}$ as 
\bqa
Z_{g^2}g^2&=&g^2+\Delta_1g^2+\Delta_2g^2+...\;,
\label{rc}
\eqa
\comment{mj}
where the subscript indicates the loop order.
Making the substitution $g^2\rightarrow Z_{g^2}g^2$, the gap 
equation~(\ref{1lgapphi}) can be
written as 
\bqa
m^2\left(1-{\Delta_1g^2\over g^2}\right)&=&
{1\over2}\sumint{1\over P^2+m^2}\;.
\eqa
\comment{mj}
Using Eq.~(A.9) in the appendix and 
\bqa
\Delta_1g^2={g^4\over2(4\pi)^2\epsilon}\;,
\label{d1g2}
\eqa
\comment{mj}
the gap equation becomes finite:
\bqa
m^2&=&{1\over2}\alpha\left[J_1T^2-(L+1)m^2\right]\;.
\label{2lgap}
\eqa
\comment{jm}Here $L=\log(\mu^2/m^2)$.

It is interesting to note that both the renormalized thermodynamic potential 
and renormalized gap equation 
are identical to the expressions in
SPT, Eq.~(\ref{sptf2loop}) and~(\ref{pet}). 
However, the coupling constant in the two-loop $\Phi$-derivable approximation
runs differently with the renormalization
scale $\mu$:
\bqa
\mu{dg^2\over d\mu}&=&{g^4\over(4\pi)^2}\;.
\eqa
\comment{jm}Hence, the $\beta$-function differs from the perturbative 
one~(\ref{rsca}) by a factor of three.

The two-loop $\Phi$-derivable approximation to the entropy can be found by 
differentiating the thermodynamic potential~(\ref{tlphi}) 
with respect to the temperature and using the gap equation to eliminate
the ultraviolet divergences.
This gives
\bqa
T{\mathcal S}&=&{1\over(4\pi)^2}\left[
2J_0T^4+J_1m^2T^2\right]\;.
\label{phis}
\eqa
\comment{jm}This expression is equal to the entropy of an ideal gas of massive bosons.

The fact that the two-loop expression for the entropy reduces to the
one-loop expression does not depend on the fact that the self-energy
is a constant mass term. This simplification at the two-loop level motivated
Blaizot, Iancu and Rebhan~\cite{rebb} to use the expression for the entropy
as the starting point for various approximations to the thermodynamic
functions. 
In the case of a scalar field theory, we have
\bqa
{\mathcal S}&=&-{\partial\over\partial T}\sumint_P\log\left(
P^2+m^2\right)\;,
\label{sbir}
\eqa
\comment{jm}
where $m^2$ again is a thermal mass. If we use the weak-coupling result for
$m^2$, Eq.~(\ref{sbir}) reproduces the weak-coupling result for the 
entropy through order $g^2$. However, it underestimates the order-$g^3$
correction to the entropy by a factor of four \cite{rebb}. The reason is that
the weak-coupling result for $m^2$ is the solution to the gap equation
only to order $g^2$. Blaizot, Iancu and Rebhan corrected for that by 
using the order-$g^3$ solution to the gap equation:
\bqa
m^2={2\pi^2\over3}\alpha T^2\left[1-\sqrt{6}\alpha^{1/2}\right]\;.
\label{birm}
\eqa
\comment{jm}

The approximation using Eq.~(\ref{birm}) for the mass obviously breaks
down for sufficiently high values of $g$, since the mass turns negative.
The authors of Ref.~\cite{rebb} then introduced 
the mass 
\bqa
m^2={2\pi^2\over3}\alpha T^2\left[1+\sqrt{6}\alpha^{1/2}\right]^{-1}\;.
\label{mbir}
\eqa
\comment{jm}
in their approximation to entropy. This approximation gives the same 
parametric improvement as~(\ref{birm}). Finally, they introduced what they
called the next-to-leading order HTL approximation by truncating the
gap equation after terms that are third order in $g$ and $m/T$ and solving
the resulting quadratic equation with respect to $m$. This amounts
to using the mass parameter
\bqa
m&=&\left[{\sqrt{6}\over3}\pi\alpha^{1/2}\sqrt{1+3\alpha/32\pi^2}
-{1\over4}\alpha\right]T\;.
\label{mnlo}
\eqa
\comment{jm}
The entropy obtained by using the
mass prescriptions~(\ref{mbir}) or~(\ref{mnlo}) is nonperturbative
in the sense that there are contributions from all orders in $g$.

\subsubsection{Three-loop $\Phi$-derivable approximation}
The three-loop $\Phi$-derivable  approximation to the free
energy is
\bqa\nonumber
\Omega_3[D]&=&
{1\over2}\sumint_P\log\left(P^2+\Pi(P)\right)
-{1\over2}\sumint_P{\Pi(P)\over P^2+\Pi(P)}
\\ && \nonumber
+{1\over8}g^2\left(\sumint_P{1\over P^2+\Pi(P)}\right)^2
\\ && 
-{1\over48}g^4\sumint_{PQR}{1\over P^2+\Pi(P)}
{1\over Q^2+\Pi(Q)}{1\over R^2+\Pi(R)}{1\over S^2+\Pi(S)}
\;,
\label{3lphi}
\eqa
\comment{jm}
where $S=-(P+Q+R)$. The corresponding gap equation that follows from the
variation of~(\ref{3lphi}) with respect to $\Pi(P)$ is
\bqa\nonumber
\Pi(P)&=&{1\over2}g^2\sumint_Q{1\over Q^2+\Pi(Q)}
-{1\over6}g^4\sumint_{QR}{1\over Q^2+\Pi(Q)}{1\over R^2+\Pi(R)}
{1\over S^2+\Pi(S)}\;.
\\&& 
\label{3lgap}
\eqa
\comment{jm}
It follows from the gap equation~(\ref{3lgap}) that the self-energy 
$\Pi(P)$ is a nontrivial function of the external momentum $P$.

In the previous subsection, we saw that we could use
the gap equation to eliminate the ultraviolet divergences in 
two-loop approximation to the thermodynamic potential.
We will follow that strategy again and use~(\ref{3lgap}) to
eliminate some of the divergences in~(\ref{3lphi}).
We then obtain
\bqa\nonumber
\Omega_3&=&{1\over2}\sumint_P\log\left(P^2+\Pi(P)\right)
-{1\over4}\sumint_P{\Pi(P)\over P^2+\Pi(P)}
\\ &&
+{1\over48}g^4\sumint_{PQR}{1\over P^2+\Pi(P)}
{1\over Q^2+\Pi(Q)}{1\over R^2+\Pi(R)}{1\over S^2+\Pi(S)}
\;.
\label{1omega}
\eqa
\comment{mj}
In section~\ref{spt}, 
we saw that a reasonable approximation in SPT is to expand
the sum-integrals in powers of $m/T$ treating $m$ to be of order $gT$.
A similar strategy was developed 
and applied to the three-loop $\Phi$-derivable approximation by 
Braaten and Petitgirard \cite{ep1}. The idea is to introduce a mass variable
which in the weak-coupling limit is of order $gT$, and then expand the
sum-integrals as double expansions in $g^2$ and $m/T$. 
Braaten and Petitgirard chose to use the Debye mass as a mass variable.
In the three-loop $\Phi$-derivable approximation, the Debye mass
is the solution to the gap equation
\bqa
m^2&=&{1\over2}g^2\sumint_{Q}{1\over Q^2+\Pi(Q)}
-{1\over6}g^4{\mathcal I}_{\rm Debye}\;,
\eqa
\comment{jm}
where
\bqa
{\mathcal I}_{\rm Debye}&=&\sumint_{QR}{1\over Q^2+\Pi(Q)}
{1\over R^2+\Pi(R)}{1\over S^2+\Pi(S)}\bigg|_{p=im}\;.
\eqa
\comment{jm}

In the three-loop $\Phi$-derivable approximation, 
the variational equation~(\ref{3lgap}) can be rewritten as 
\bqa
\Pi(P)&=&m^2-{1\over6}g^4\left[{\mathcal I}_{\rm sun}
-{\mathcal I}_{\rm Debye}\right]\;.
\label{piexp}
\eqa
\comment{jm}
where
\bqa
{\mathcal I}_{\rm sun}(P)&=&\sumint_{QR}{1\over Q^2+\Pi(Q)}
{1\over R^2+\Pi(R)}{1\over S^2+\Pi(S)}\;.
\eqa
\comment{jm}

Since we are assuming that the scales $m\sim gT$ and $2\pi T$
can be separated, we are allowed to expand $\Pi(P)$ in 
powers of $g$ and $m/T$. The variational gap equation will be solved 
in the two momentum regions separately.
For hard momenta, the self-energy is expanded as 
\bqa
\Pi(P)&=&m^2+g^4
\left[\Pi_{4,0}(P)+\Pi_{4,1}(P)+...\right]+\dots \;,
\label{hexp}
\eqa
\comment{jm}
where the functions $\Pi_{n,k}(P)$ are of $T^2(m/T)^k$ when the external 
momentum $P$ is hard.
For soft momentum $P=(0,{\bf p})$, 
the self-energy is written as 
$\Pi(0,{\bf p})=m^2+\sigma(p)$, where 
\bqa
\sigma(p)&=&g^4
\left[\sigma_{4,-2}(p)+\sigma_{4,0}(p)+\dots\right]\;.
\label{sexp}
\eqa
\comment{jm}

Here, the functions $\sigma_{n,k}(p)$ are 
of order $m^2(m/T)^k$ when $p$ is soft.
By inserting the expansions for $\Pi(P)$ and $\sigma(p)$ into the gap
equation and expanding systematically in $g$ and $m/T$, 
we can find expressions for $\Pi_{n,k}(p)$ and $\sigma_{n,k}(p)$
by matching coefficients of order $g^n$ on both sides and solving
it recursively.
In  this manner, we find
\bqa
\Pi_{4,0}(P)&=&-{1\over6}\sumint_{QR}{1\over Q^2R^2(P+Q+R)^2}+
{1\over6}T^2I_{\rm Debye}\;, \\ 
\Pi_{4,1}(P)&=&-{1\over2}T\int_q{1\over q^2+m^2}\sumint_R\left(
{1\over Q^2(P+Q)^2}-{1\over Q^4}
\right)\;,\\
\sigma_{4,-2}(p)&=&-{1\over6} T^2
\left[I_{\rm sun}-I_{\rm Debye}\right]\;,
\label{onlyf}
\eqa
\comment{jm}
where $I_{\rm sun}$ and $I_{\rm Debye}$ are the three-dimensional
counterparts of ${\mathcal I}_{\rm sun}$ and ${\mathcal I}_{\rm Debye}$.
These are the only functions we need in the expansion for $\Pi(P)$.

In appendix C, we expand the necessary sum-integrals and integrals
in powers of $m/T$.
Substituting these expressions,
keeping all the terms through order $g^5$, the gap equation becomes
\bqa
m^2&=&{1\over2}g^2\sumint_P\left[{1\over P^2}-{m^2\over P^4}\right]
+{1\over2}g^2T\int_p{1\over p^2+m^2}
-{1\over6}g^4T^2I_{\rm Debye}
\nonumber \\  &&
-{1\over2}g^4T\int_p{1\over p^2+m^2}\sumint_P{1\over P^4}
-{1\over2}g^6T\int_p{\sigma_{4,-2}(p)\over(p^2+m^2)^2}\;.
\label{massexp}
\eqa
\comment{jm}
The gap equation has poles in $\epsilon$ and requires renormalization.
Making the substitution $g^2\rightarrow Z_{g^2}g^2$, the gap 
equation~(\ref{massexp}) can be written as 
\bqa\nonumber
m^2\left(1-{\Delta_1g^2\over g^2}\right)&=&
-{1\over2}g^2m^2\sumint_P{1\over P^4}
-{1\over6}g^4T^2I_{\rm Debye}
\\
\nonumber && +{1\over2}\left(g^2+\Delta_2g^2\right)\left(
\sumint_P{1\over P^2}+T\int_p{1\over p^2+m^2}
\right)
\\ && \nonumber 
-{1\over2}g^4T\int_p{1\over p^2+m^2}\sumint_P{1\over P^4}
-{1\over2}g^6T\int_p{\sigma_{4,-2}(p)\over(p^2+m^2)^2}\;.
\\ &&
\eqa
\comment{mj}
Using the expressions for the sum-integrals and integrals
in appendix A and B and
\bqa
\Delta_2g^2&=&{g^4\over(4\pi)^2\epsilon}\;,
\eqa
\comment{jm}
the gap equation~(\ref{massexp}) becomes finite:
\bqa\nonumber
m^2&=&{2\pi^2\over3}\alpha T^2\left[
1-12\left({m\over4\pi T}\right)-24\left({m\over4\pi T}\right)^2\left(
\gamma+\log{\mu\over4\pi T}
\right)
\right.\\ &&\left.\nonumber
+2\left(2 \log{m\over4\pi T} - \log{\mu\over4\pi T}
-2+6\log2
+{\zeta^{\prime}(-1)\over\zeta(-1)}
\right)\alpha
\right.\\ &&\left. 
+24\left({m\over4\pi T}\right)\left(\gamma+\log{\mu\over4\pi T}\right) \alpha
-\left({m\over4\pi T}\right)^{-1}\left(1-\log2\right)\alpha^2
\right]\,.
\eqa
\comment{mj}
Note that the running of the coupling in the three-loop $\Phi$-derivable
approximation 
agrees with the perturbative one~(\ref{rsca}) through order $g^4$.

Substituting the expansions of the sum-integrals and integrals
into~(\ref{1omega}) and
keeping all terms through order $g^5$, the thermodynamic potential becomes
\bqa\nonumber
\Omega_3&=&{1\over2}\sumint_P\log P^2+{1\over4}\sumint_P{m^2\over P^2}
+{1\over2}T\int_p\log(p^2+m^2)-{1\over4}T\int_p{m^2\over p^2+m^2}
\\ &&\nonumber
-{1\over48}g^4\sumint_{PQR}{1\over P^2Q^2R^2S^2}
+{1\over24}g^4T^2\sumint_P{1\over P^2}I_{\rm Debye}
\\ &&\nonumber
+{1\over24}g^4T^3\int{1\over p^2+m^2}I_{\rm Debye}
\\ &&\nonumber
-{1\over48}g^4T^3\int_{pqr}{1\over p^2+m^2}{1\over q^2+m^2}{1\over r^2+m^2}
{1\over({\bf p}+{\bf q}+{\bf r})^2+m^2}
\\ &&
+{1\over8}g^4T\int_p{1\over p^2+m^2}\sumint_{QR}{1\over Q^4R^2}
+{1\over4}m^2g^4T\int{\sigma_{4,-2}(p)\over(p^2+m^2)^2}
\;.
\eqa
\comment{jm}
Using the expressions for the integrals and sum-integrals in the 
appendix, we obtain
\bqa\nonumber
\Omega_3&=&-{\pi^2T^4\over90}\Bigg\{1
-30\left({m\over4\pi T}\right)^2
+120\left({m\over4\pi T}\right)^3
\\ && \hspace{1cm} \nonumber
+5\left(\log{m\over2\pi T}
-{29\over60}+2\log2
+{3\over2}{\zeta^{\prime}(-1)\over\zeta(-1)}
-{1\over2}{\zeta^{\prime}(-3)\over\zeta(-3)}
\right) \alpha^2
\\ && \hspace{1cm} 
+60\left({m\over4\pi T}\right)\left(
\log{m\over2\pi T}
+{1\over2}\gamma
-{3\over2}+{1\over2}\log2\right)
\alpha^2\Bigg\}. \nonumber
\\&&
\eqa
\comment{m}
The divergences through order $g^5$ have been eliminated
by using the gap equation. 
In the calculations by Braaten and Petitgirard, they kept
divergent terms through order $g^7$ in the gap equation and the
thermodynamic potential. The single poles in $\epsilon$ 
of order $g^6$ could be eliminated
by defining a renormalized coupling constant, but the double poles could not.
The appearance of these double poles was identified with the truncation
of interaction functional at three loops by Braaten and Petitgirard.
However, it seems that it is possible to eliminate poles at all
orders in the $\Phi$-derivable approach if the renormalization is
performed properly \cite{urko}.  Note that at the order presented here, 
namely $g^5$, the difference between the two renormalization schemes does not
enter since it only affects the treatment of terms of order $g^6$ and 
higher in the pressure.

\begin{figure}
\centerline{\psfig{file=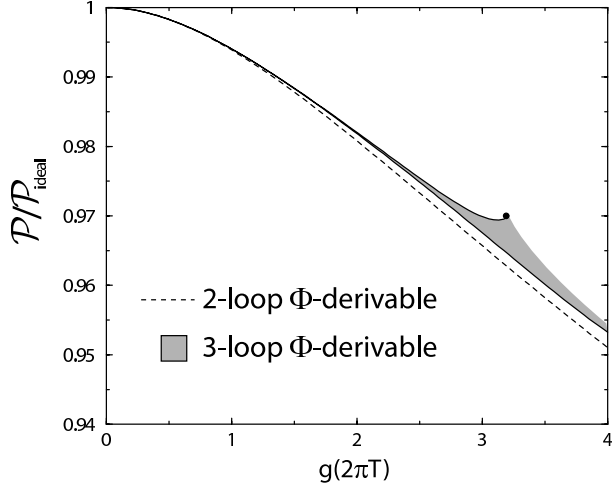,width=8cm}}
\vspace*{8pt}
\caption{Two- and three-loop $\Phi$-derivable approximations to the
 pressure normalized to that of an ideal gas as a function of $g(2\pi T)$.
Three-loop band is obtained by varying the renormalization scale, $\mu$,
by a factor of two around $\mu= 2 \pi T$.  Note that using the running
given by Eq.~(\ref{d1g2}) at two-loops gives a result which is independent
of $\mu$.
}
\label{newphi}
\end{figure}

In Fig.~\ref{newphi}, the two- and three-loop $\Phi$-derivable approximations
to the pressure divided by that of an ideal gas are shown.
Here we encounter the same problem as in SPT, namely that one cannot 
extend the solution to the gap equation
past a critical value $g^*$ which we have indicated by a dot in the Figure. 
From the Figure it is evident that there is a significant improvement in the
convergence compared to the weak-coupling expansion.
Improved convergence properties were also found for the entropy and the Debye
mass \cite{ep1}.

\subsection{Gauge theories}

In  this subsection, we review gauge theories within the 
$\Phi$-derivable approach.  This approach was first applied 
to gauge theories
by Blaizot, Iancu and Rebhan~\cite{bir1,bir2,rebb} and 
by Peshier.~\cite{Peshier-00}
For simplicity, we restrict the discussion to pure-glue QCD and take $N_c=3$
in which case the thermodynamic potential $\Omega$ is given by
\bqa\nonumber
\Omega[D,D_{\rm gh}]&=&
{1\over2}{\rm Tr}\log D^{-1}-{\rm Tr}\log D^{-1}_{\rm gh} \nonumber \\
&& \hspace{1cm} -{1\over2}{\rm Tr}\,\Pi D +{\rm Tr}\,\Pi_{\rm gh}D_{\rm gh}+\Phi[D,D_{\rm gh}]\;, 
\label{tdgauge}
\eqa
\comment{jm}
where the trace is over color and Lorentz indices. $D(P)$ and 
$D_{\rm gh}(P)$
are the exact gluon and ghost propagators, while $\Pi(P)$ and $\Pi_{\rm gh}(P)$
are the corresponding exact self-energies. $\Phi[D,D_{\rm gh}]$ is the sum of
all two-particle irreducible vacuum diagrams.

The self-energy tensor $\Pi_{\mu\nu}(P)$ in Abelian gauge theories
is transverse with respect to the four-momentum 
$P=(\omega,{\bf p})$ and is also gauge-fixing independent.
In nonabelian gauge theories, the self-energy tensor $\Pi_{\mu\nu}(P)$
is generally not four-dimensionally 
transverse, but it can be decomposed into four different
structure functions.~\cite{gpy} The axial gauges which are defined by
$n_{\nu}A^{\mu\;a}=0$, where $n_{\mu}$ is a constant vector
represent a class of gauges in which the self-energy tensor
is transverse with respect to the four-momentum. However, an approximation 
does not necessarily respect transversality due to the fact that 
there are no vertex corrections.~\cite{rebb} In the temporal axial 
gauge~\footnote{At finite temperature, there are problems with this gauge
related to the fact it is incompatible with the periodicity of 
the field.~\cite{bellac} There are other problems as well, 
see e.g. Ref.~\cite{landshoff}.}
$A_0=0$, this presents no problem since the self-energy tensor can be
expressed exclusively in terms of $\Pi_T$ and $\Pi_L$.
Moreover, in this gauge, the ghost 
self-energy
$\Pi_{\rm gh}$ vanishes and the ghost field decouples. Thus the skeleton
expansion of $\Phi$ simplifies significantly.

\begin{figure}[htb]
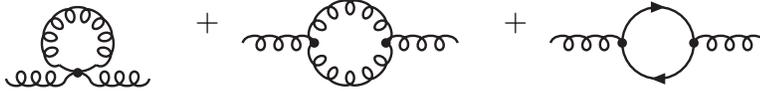

\begin{center}
$$
 \oneloopQCDgsea \: + \: \oneloopQCDgseb \: + \: \oneloopQCDfse
$$
\end{center}
\caption[a]{One-loop QCD self-energy graphs.}
\label{fig:qcdselfenergy}
\end{figure}

The gap equations that are obtained by varying the thermodynamic potential
with respect to $D_T(P)$ and $D_L(P)$ are 
shown in Fig.~\ref{fig:qcdselfenergy}:
\bqa
\Delta_T^{-1}&=&\omega^2-k^2-\Pi_T\;,
\label{gap1}\\
\Delta_L^{-1}&=&k^2+\Pi_L\;,
\label{gap2}
\eqa
\comment{jm}
where the transverse and longitudinal propagators are defined in
Eqs.~(\ref{Delta-T})--(\ref{Delta-L}). The 
solutions to the nonlocal equations~(\ref{gap1})--(\ref{gap2}) are very 
difficult. However, to leading order in $g$ and for soft momenta, the 
solutions are simply given by the hard-thermal loop
self-energies~(\ref{redt})--(\ref{redl}) \cite{rebb}.
For hard transverse modes, Blaizot, Iancu and Rebhan solved the gap 
equations to next-to-leading order in $g$. The solution is
\bqa
\Pi_T&=&\Pi_T^{(2)}+\delta\Pi_T\;,
\label{pdpi}
\eqa
\comment{jm}
where $\Pi_T^{(2)}$ is given by the one-loop self-energy diagrams with
bare propagators, and $\delta\Pi_T$ is given by the same diagrams with one line
being soft and one line being hard. The calculation of the
function $\delta\Pi_T$ was performed in Ref.~\cite{rebb} where the
explicit form can be found.

Having found the approximate solutions to the gap equation, one can 
substitute them into to expression~(\ref{tdgauge}) to obtain the 
pressure. In the imaginary time formalism, the only soft mode is the
zero-frequency mode. In that case, the longitudinal self-energy
$\Pi_L$ reduces to a mass term $m_D^2$, while $\Pi_T$ vanishes.
For the nonzero Matsubara modes, it suffices to use bare propagators.
The thermodynamic  potential then reduces to
\bqa
\Omega&=&4(d-1)\sumint_P^{\prime}\log P^2
+6(d-1)^2g^2\left(\sumint_P^{\prime}{1\over P^2}\right)^2
\nonumber \\
&& \hspace{1cm} +4T\int_p\log\left(p^2+m_D^2\right)
\;.
\label{gaugephi}
\eqa
\comment{jm}
Using the expressions for the sum-integrals and the integrals
in the appendices, Eq.~(\ref{gaugephi}) reduces to 
\bqa
\Omega&=&-
{8\pi^2T^4\over45}\bigg[1-{15\over4}{\alpha\over\pi}
+{15\over4\pi^3}(\beta m_D)^3
\bigg]\;.
\label{p3}
\eqa
\comment{mj}
Using the weak-coupling expression for the Debye mass, it is easy to show that
Eq.~(\ref{p3}) agrees with the weak-coupling result for the pressure through
order $g^3$. The entropy, which follows from differentiation of~(\ref{p3})
with respect to $T$ becomes
\bqa
T{\mathcal S}&=&
{32\pi^2T^4\over45}\bigg[1-{15\over4}{\alpha\over\pi}
+{15\over16\pi^3}(\beta m_D)^3
\bigg]\;.
\label{ss3}
\eqa
\comment{mj}
The expression for the entropy~(\ref{ss3}) also agrees with the weak-coupling
expansion result to order $g^3$.

Again Blaizot, Iancu and Rebhan~\cite{rebb} exploited the fact that the
two-loop expression for the entropy
reduces to the one-loop expression. In the case of gauge
theories, one can write~\cite{rebb}
\bqa
{\mathcal S}&=&{\mathcal S}_L+{\mathcal S}_T\;,
\eqa
\comment{jm}
where
\bqa
{\mathcal S}_{L}&=&-8\int{d\omega\over2\pi}\int_k
{\partial n(\omega)\over\partial T}\left[{\rm Im}
\log\left(k^2+\Pi_L\right)+{\rm Im}\,\Pi_L\,{\rm Re}\,D_L\right]
\;,
\label{sll}
\\ \nonumber
{\mathcal S}_{T}&=&-16
\int{d\omega\over2\pi}\int_k
{\partial n(\omega)\over\partial T}\left[{\rm Im}
\log\left(-\omega+k^2+\Pi_T\right)-{\rm Im}\,\Pi_T \,{\rm Re}\,D_T\right]\;.
\\ &&
\label{st}
\eqa
\comment{jm}

The Stefan-Boltzmann result for the entropy is obtained by using bare 
transverse and longitudinal propagators in Eqs.~(\ref{sll})--(\ref{st}).
The order-$g^2$ correction to the entropy is obtained by using the
leading-order solution to the gap equation $\Pi_T^{(2)}$ in
Eq.~(\ref{st}). Note that the longitudinal modes do not contribute at this
order in $g$.  The order-$g^3$ correction receives contribution from both
the soft and hard modes.  The soft contribution can obtained from
Eqs.~(\ref{sll})--(\ref{st}) by using the hard-thermal-loop self-energies
and expanding the Bose-Einstein distribution function, 
$n(\omega)\simeq T/\omega$. 
For example, the contribution from soft longitudinal
soft gluons is
\bqa
{\mathcal S}_{L}&=&-8\int{d\omega\over2\pi}\int_k
{1\over\omega}\bigg[{\rm Im}
\log\left(k^2+\Pi_L\right)+{\rm Im}\,\Pi_L\,{\rm Re}\,D_L\bigg]\;.
\eqa
\comment{jm}

The hard contributions are found by evaluating
Eq.~(\ref{st}) using the next-to-leading order solution to the gap equation
for transverse modes $\delta\Pi_T(\omega,{\bf k})$ on the light cone, i.e.
$\omega=k$:
\bqa
S_3^{\rm hard}&=&-8
\int_k{1\over k}
{\partial n(k)\over\partial T}{\rm Re}\,\delta\Pi_T(\omega=k)\;.
\eqa
\comment{jm}

Blaizot, Iancu and Rebhan define the HTL entropy by using the 
hard-thermal-loop self energies in Eqs.~(\ref{sll})--(\ref{st})
for both soft and hard momenta. While this is not the leading-order solution
to the gap equation for the hard modes, it nevertheless reproduces the 
order-$g^2$ term in the entropy (since this is obtained by using the 
one-loop self-energies with bare propagators in the expression for the
entropy and is exclusively given by the hard modes). 
The full numerical evaluation of the HTL entropy is of course also 
nonperturbative since its expansion contains all orders in the coupling $g$.

The ultraviolet divergences that appear in the approximate solutions to the
one-loop gap equations can be eliminated by wave function renormalization
at zero temperature. However, they drop out of the expression for the entropy.
Thus consideration of the entropy gives no 
renormalization group equation describing the running
of the coupling constant. In their numerical treatment, Blaizot, Iancu and
Rebhan took the standard two-loop running of the coupling constant.
The HTL resummation in Refs.~\cite{bir1,bir2,rebb} is specific
to the entropy. In order to obtain the pressure, one must integrate the entropy
with respect to the temperature. This introduces an unknown integration 
constant which the authors fixed by requiring
that the pressure vanish at $T_c$.

Peshier~\cite{Peshier-00} considered another approximation to the two-loop
$\Phi$-derivable approximation for gauge theories 
motivated by scalar field theory. 
The two-loop $\Phi$-derivable approximation 
is equivalent to making the substitution
\bqa
\Phi_2[D]\rightarrow{1\over4}\sumint_P{m^2\over P^2+m^2}\;,
\label{pes1}
\eqa
\comment{jm}
where $m$ is the solution to the one-loop gap equation~(\ref{1lgapphi}).
If one replaces the mass $m$ by the hard-thermal-loop
approximation to the self-energy, one reproduces the pressure
correctly to order $g^2$. The order-$g^3$ contribution is, however,
overestimated by a factor of four.
%
Peshier suggested to
approximate the two-loop approximation to
the functional $\Phi[D]$ in QCD by the analog to Eq.~(\ref{pes1}):
\bqa
\Phi[D^*]&=&{1\over4}{\rm Tr}D^*\,\Pi^*\;,
\label{pes3}
\eqa
\comment{jm}
where the $*$ indicates that the self-energies and propagators are 
taken in the HTL approximation.
By construction, this approximation to the two-loop $\Phi$-derivable 
thermodynamic potential
reproduces the order-$g^2$ correction to the pressure. The 
order-$g^3$ is again incorrect by a factor of four. 
The entropy of Peshier is calculated by differentiating~(\ref{pes3})
with respect to the temperature. The resulting entropy does not coincide
with that of Blaizot, Iancu and Rebhan.

Finally, we would like to discuss
the strategy developed by Braaten and Petitgirard to solve the 
$n$-loop $\Phi$-derivable approximation in QCD. We know that the static
infrared limit of $\Pi_{00}(\omega,{\bf p})$ is the order-$g^2$ 
result for the Debye mass $m_D^2$, while all other components vanish.
For soft momenta, we therefore expand
\bqa
\Pi_{00}(0,p)&=&m_D^2+g^4\left[\sigma^{4,-2}_{00}(p)
+...\right]\;,
\eqa
\comment{jm}
where $m_D$ is the Debye mass is defined by the pole position of the 
static propagator:
\bqa
m^2_D&=&\Pi_{00}(\omega_n=0,{p})\big|_{p=im_D}\;.
\label{gapqcd}
\eqa
\comment{jm}
For hard momenta, we write the expansion of
$\Pi_{00}$ in the form
\bqa
\Pi_{00}(P)&=&g^2\left[\Pi^{2,0}_{00}(p)+\Pi^{4,0}_{00}(p)
+...\right]\;.
\eqa
\comment{jm}

In order to solve the gap equation~(\ref{gapqcd}) beyond leading order,
we must incorporate the self-energies in the propagators. 
In a next-to-leading-order calculation of the pressure, 
it suffices to use bare propagators for hard loop momenta:
\bqa
\Delta_{\mu\nu}(P)&=&\left[{\delta_{\mu\nu}\over P^2}
+(1-\xi){P_{\mu}P_{\nu}\over P^4}\right]\;,
\label{hardprop}
\eqa
\comment{jm}
where $\xi$ is the gauge parameter in the class of covariant gauges.
For soft momenta, we include the leading term in the 
expansion for $\Pi_{00}$. We must then use the dressed propagator:
\bqa
\Delta_{00}(0,p)&=&{1\over p^2+m_D^2}\;.
\eqa
\comment{jm}
For soft momenta, the other components of the propagator are still given
by~(\ref{hardprop}).
The gap equation then becomes:
\bqa\nonumber
m^2_D&=&3(d-1)^2g^2\sumint_K{1\over K^2}
+3 g^2 T \int_k\left[
{d-2\over k^2}+{2(m_D^2-p^2)\over k^2[({\bf p}+{\bf k})^2+m_D^2]}
\right.\\ &&
\left.
\hspace{1mm}+{1\over k^2+m_D^2}
+(\xi-1)\left(p^2+m_D^2\right)
{{\bf k}\cdot({\bf p}+{\bf k})\over k^4[({\bf p}+{\bf k})^2+m_D^2]}
\right]\bigg|_{p^2=-m_D^2}\!.
\label{debyenlo}
\eqa
\comment{mj}

The first term arises when all internal lines are hard and is
independent of the gauge parameter $\xi$.
The remaining terms
arise when all lines are soft. 
The case where one line is soft and one
line is hard precisely corresponds to $\delta\Pi_T$ in~(\ref{pdpi})
and is neglected in a NLO calculation.
The contribution from the soft sector was first calculated by Rebhan
when he calculated the order $g^3$-term in the weak-coupling expansion
for the Debye mass using resummed perturbation theory \cite{reb}.
When this contribution is evaluated on shell, the 
gauge dependent term vanishes algebraically before we have integrated over
$k$. However, the integral is linearly infrared divergent. Introducing
an infrared cutoff, performing the integral and taking the limit 
$p^2\rightarrow -m_D^2$ before removing the cutoff, it can be shown that this
term drops out \cite{rebbleik,reb,shif}.

The third term is logarithmically
infrared divergent and this singularity is caused by a massless
static gluon propagator. This is another example of the breakdown
of perturbation theory in the static magnetic sector of QCD and prevents
a perturbative definition of the Debye mass beyond leading order in $g^2$.
A magnetic mass $m_m$ of 
order $g^2T$ is expected to be generated nonperturbatively in QCD. 
Rebhan~\cite{reb} 
suggested to obtain an estimate of the NLO correction by 
incorporating $m_m$ 
by the replacement $1/p^2\rightarrow1/(p^2+m^2_m)$
in the transverse part of the propagator.
The integrals can then be evaluated analytically and the
gap equation~(\ref{debyenlo}) then becomes
\bqa
m^2_D&=& g^2T^2+
{3 g^2 m_D T \over 2 \pi} \left[
\left(1-{m_m^2\over4m_D^2}\right)\log{2m_D+m_m\over m_m}-{1\over2}-{m_m\over2m_D}
\right]\;,
\label{solmass}
\eqa
\comment{mj}
where the first term comes from hard loop momenta and the remaining terms
from the soft sector. The pressure is still given by Eq.~(\ref{p3}),
where the mass $m_D$ is the solution to~(\ref{solmass}) instead of the
weak-coupling expression that was used in Ref.~\cite{bir3}.
By introducing a magnetic mass, we have included
nonperturbative physics in an ad hoc manner. One might abandon the
NLO solution to the gap equation altogether and simply use the
leading-order result. Our approximate solution to the two-loop
$\Phi$-derivable approach then reduces to the leading-order solution of Blaizot,
Iancu and Rebhan \cite{rebb}.

%% file: drspt.tex
\section{Dimensionally Reduced SPT}
\label{drspt}

In Secs.~\ref{spt}--\ref{htl}, we  have seen that SPT and HTLPT
are ways of reorganizing perturbation theory so that the convergence
properties of the perturbative approximants is dramatically improved.
However, we have also seen that the two-loop result for the HTL-improved
pressure in pure-glue QCD fails to match onto that of
lattice gauge theory at temperatures for which they are available.
It has been suggested that the disagreement is due
to the fact that the HTL approximation breaks down for hard momenta. 
One should therefore use the HTL approximation only for soft momenta
and treat hard momenta in a different manner. 
Such a strategy has recently
been proposed by Blaizot, Iancu and Rebhan \cite{bir3}, where one uses
strict perturbation for the hard modes and applies screened perturbation
theory only to the dimensionally reduced theory for the soft modes.

This approach is called dimensionally reduced screened perturbation theory
(DRSPT). In the case of gauge theories, this approach is gauge invariant
order by order in the loop expansion. 
Of course, for nonabelian gauge theories,
it breaks down at four loops due to infrared divergences in the 
magnetic sector. This is reflected in Eq.~(\ref{fg}) where the constant
$a$ can only be determined nonperturbatively. 
Blaizot, Iancu, and Rebhan~\cite{bir3} 
applied DRSPT to EQCD and calculated the pressure through four loops
up to this unknown constant.
It seems that the agreement with available lattice data
is improved significantly by treating the hard modes perturbatively
but not insisting
on expanding the soft contributions in powers of the coupling, instead
keeping the full $g$-dependence of these contributions.

In this section, we apply DRSPT
to a dimensionally reduced scalar field theory.  
The effective Lagrangian is then written as
\bqa
{\mathcal L}&=&{1\over2}(\nabla\phi)^2+{1\over2}(m_E^2+\delta m^2)\phi^2
+{\lambda\over24}\phi^4-{1\over2}\delta m^2\phi^2\;,
\eqa
\comment{mj}
where $m_E$ is the mass parameter determined perturbatively in the 
dimensional reduction step and $\delta m$ is to be determined variationally:
\bqa
{\partial \over \partial (\delta m^2)}f&=&0\;.
\eqa
\comment{mj}
In the four-loop calculation below, we also need the coupling constant
$\lambda$ to next-to-leading order in $g^4$:
\bqa
\lambda&=&g^2-{3g^4T\over(4\pi)^2}\left(\log{\mu\over4\pi T}+\gamma\right)\;.
\label{lnlo}
\eqa
\comment{mj}
The diagrams through four loops are shown in Fig.~\ref{fig:drsptgraphs}.

\begin{figure}[t]

\begin{eqnarray*}
 &\mbox{1-loop}:\hspace{3mm}&
 \sy{}12\oneloop \nn
 &\mbox{2-loop}:\hspace{3mm}&
 \sy{}18 \figureeight
 \sy{-}12 \oneloopX \nn
 &\mbox{3-loop}:\hspace{3mm}&
 \sy{-}1{48}\basketball
 \sy{-}1{16}\triplebubble
 \sy{+}1{4}\figureeightX
 \sy{-}1{4}\oneloopXX
 \nn
 &\mbox{4-loop}:\hspace{3mm}&
 \sy{}1{48}\ctriangle
 \sy{+}1{24}\basketballL
 \sy{+}1{48}\threering
 \!\!\!\!\!\sy{+}1{32}\fourbubbles \nn
 &&\sy{-}1{8}\triplebubbleXl
 \,\sy{-}1{8}\triplebubbleXm
 \sy{-}1{12}\basketballX \nn
 &&\sy{+}1{8}\figureeightXoX
 \,\,\sy{+}1{4}\figureeightXXo
 \,\sy{-}1{6}\oneloopXXX
\end{eqnarray*}

\caption[a]{
Diagrams which contribute up to four-loop order in DRSPT.  A boldfaced
${\boldmath \times}$ indicates an insertion of $\delta m^2$.}
\label{fig:drsptgraphs}

\end{figure}

\subsection{One-loop order}

The one-loop contribution to the free energy is
\bqa\nonumber
f_0&=&{1\over2}\int_p\log\left(p^2+m^2\right)\\
&=&-{1\over12\pi}m^3\;.
\label{1fj}
\eqa
\comment{mj}
with $m^2=m_E^2+\delta m^2$.  At this order the only solution to the 
variational condition $\partial f_0/\partial (\delta m^2)=0$ is 
the trivial solution $\delta m^2=0$ or $m^2 = m_E^2$.

\subsection{Two-loop order}
The two-loop contribution to the free energy is
\bqa
f_1&=&{1\over8}\lambda\left(\int_p{1\over p^2+m^2}\right)^2
-{1\over2}\delta m^2\int_p{1\over p^2+m^2}\;, \nonumber \\
&=& {1\over8} \alpha m^2 T + {m\over 8 \pi} \delta m^2\;,
\label{2fj}
\eqa
\comment{mj}
where we have used the integrals contained in appendix B and
again $m^2=m_E^2+\delta m^2$.
Combining the one- and two-loop contributions and requiring that their sum
satisfy the variational condition,
gives the two-loop gap equation
\bqa
m^2 - m_E^2 &=&- 2 \pi \alpha m T \;.
\label{gapoir}
\eqa
\comment{mj}

The two-loop DRSPT approximation to the free energy is given by the
sum of~(\ref{fmatch}) truncated at order $\alpha$,~(\ref{1fj}) 
and~(\ref{2fj}). Using the
leading-order solution to the gap equation~(\ref{gapoir}), it is easy to
show that this approximation agrees with the weak-coupling expansion through
order $g^3$.

\subsection{Three-loop order}
The three-loop contribution to the free energy is
\bqa\nonumber
f_2&=&
- {1\over48}
\lambda^2\int_{pqr}
{1\over p^2+m^2}{1\over q^2+m^2}{1\over r^2+m^2}
{1\over({\bf p}+{\bf q}+{\bf r})^2+m^2}
\\ &&\nonumber
-{1\over16}\lambda^2\left(\int_p{1\over p^2+m^2}\right)^2\int_q{1\over (q^2+m^2)^2}
+{1\over4}\lambda\delta m^2\int_{pq}{1\over(p^2+m^2)^2}{1\over q^2+m^2}
\\ &&
-{1\over4}\left(\delta m^2\right)^2\int_p{1\over(p^2+m^2)^2} 
+ {\partial f_0 \over \partial m^2} \Delta_2 m^2 \;,
\label{3fj}
\eqa
\comment{mj}
where the mass counterterm is given by 
\bqa
\Delta_2 m^2 = {2 \over 3} \left( {\lambda \over 16 \pi} \right)^2 {1 \over \epsilon} \; .
\label{del2m2}
\eqa
\comment{mj}
Using the integrals listed in Appendix B, Eq.~(\ref{3fj}) reduces to
\bqa
f_2 = {1 \over3} \pi \alpha^2 m T^2 
\left[\log{\Lambda\over 2m}+{9\over8}-\log2\right]	
- {1 \over 8} \alpha T \delta m^2 - {1 \over 32 \pi m} (\delta m^2)^2 \;.
\label{3fjr}
\eqa
\comment{mj}
Combining this result with the one- and two-loop 
contributions~(\ref{1fj})--(\ref{2fj}), and imposing the 
variational constraint gives the three-loop gap 
equation
\bqa
\left(m^2-m_E^2\right)^2 &=&-{32\pi^2\over 3} \alpha^2 m^2 T^2 
\left[\log{\Lambda\over 2m}-\log2+{1\over8}\right]\;.
\label{dm4}
\eqa
The three-loop DRSPT approximation to the free energy is given by the
sum of~(\ref{fmatch}),~(\ref{1fj}),~(\ref{2fj}).  
and~(\ref{3fjr}). Upon substituting the expression for the mass 
parameter~(\ref{masspara}) and using~(\ref{dm4}), the $\Lambda$-dependence
in the three-loop DRSPT approximation to the free energy cancels to 
next-to-leading order in $g$. This approximation agrees with the 
weak-coupling expansion to order $g^4$.
\subsection{Four-loop order}

\begin{figure}
\centerline{\psfig{file=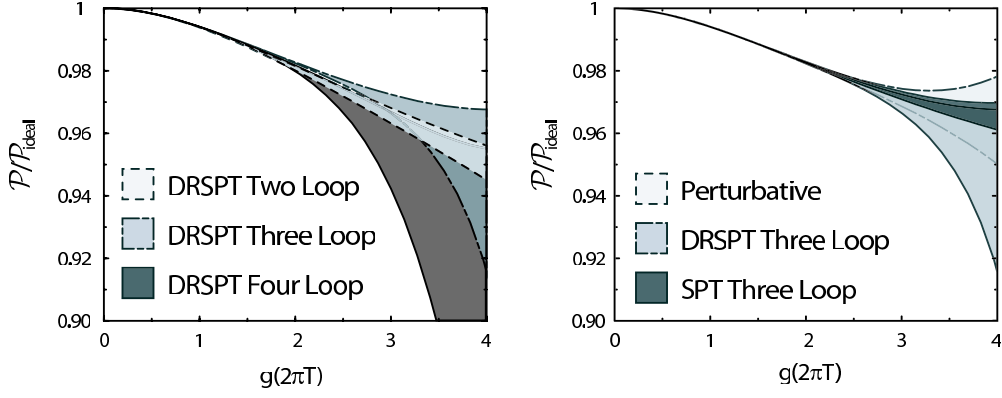,width=13.2cm}}
\vspace*{8pt}
\caption{(a) Two-, three-, and four-loop DRSPT results for the 
pressure as a function of $g(2\pi T)$. 
(b) Three-loop DRSPT and SPT results are compared with 
the perturbative result accurate to order $g^5$.}
\label{4f4}
\end{figure}

The four-loop contribution to the free energy is
\bqa\nonumber
f_3&=&
{\lambda^3\over48}\int_{pqrs}
{1\over p^2+m^2}{1\over q^2+m^2}{1\over r^2+m^2}{1\over s^2+m^2}
{1\over ({\bf p}+{\bf q}+{\bf r})^2+m^2}
\\ && \nonumber \hspace{7cm} \times
{1\over ({\bf p}+{\bf r}+{\bf s})^2+m^2} \\
&&\nonumber\hspace{-1cm}
+{\lambda^3\over24}\int_{pqr} 
{1\over p^2+m^2}{1\over q^2+m^2}{1\over r^2+m^2}
{1\over\left[({\bf p}+{\bf q}+{\bf r})^2+m^2\right]^2} 
\int_s {1\over s^2+m^2} \\
\nonumber
&&\hspace{-1cm}
+{\lambda^3\over48} \left(\int_{p} {1\over p^2+m^2}\right)^3 
\int_q {1\over (q^2+m^2)^3}
+{\lambda^3\over32} \left(\int_{p} {1\over p^2+m^2}\right)^2 
\left(\int_q {1\over (q^2+m^2)^2}\right)^2
\\ \nonumber
&&\hspace{-1cm}
-{\lambda^2\over8} \delta m^2 \left[\left(\int_{p} 
{1\over (p^2+m^2)^2}\right)^2 \int_q {1\over q^2+m^2}
+\left(\int_{p} {1\over p^2+m^2}\right)^2 \int_q{1\over (q^2+m^2)^3}\right]
\\ \nonumber
&&\hspace{-1cm}
-{\lambda^2\over12} \delta m^2 \int_{pqr} 
{1\over p^2+m^2}{1\over q^2+m^2}{1\over r^2+m^2} 
{1\over\left[({\bf p}+{\bf q}+{\bf r})^2+m^2\right]^2} 
\\ \nonumber
&&\hspace{-1cm}
+{\lambda\over4}\left(\delta m^2\right)^2
\int_{pq}{1\over (p^2+m^2)^3}{1\over q^2+m^2}
+{\lambda\over8}\left(\delta m^2\right)^2 
\left(\int_{p} {1\over (p^2+m^2)^2}\right)^2
\\ 
&&\hspace{-1cm}
-{1\over6}\left(\delta m^2\right)^3\int_p{1\over (p^2+m^2)^3}
+ {\partial f_1 \over \partial m^2} \Delta_2 m^2 
+ \delta f_3
\;,
\eqa
\comment{mj}
where $\Delta_2 m^2$ is given in Eq. (\ref{del2m2}) and $\delta f_3$ 
is a vacuum counterterm
given by
\bqa
\delta f_3&=&-{\pi^4\over96}{\lambda^3\over(4\pi)^6}{1\over\epsilon}\;.
\label{vc}
\eqa
\comment{mj}
The logarithmic ultraviolet divergence
that is cancelled by the vacuum counterterm~(\ref{vc}) is arising from 
the first four-loop diagram in Fig.~\ref{fig:drsptgraphs}. It is the same
divergence that we discussed in Sec.~\ref{dimredsub} that leads to the
running of $f(\Lambda)$ at four loops. Thus $f_3$ depends explicitly
on $\Lambda$ at order $\lambda^3$ and this dependence is cancelled only by
adding the hard contribution to the free energy through order $g^6$.

Using the integrals in Appendix B, we obtain the four-loop 
contribution to the free energy
\bqa
f_3&=& {\pi^2(\pi^2-4)\over12}\alpha^3T^3\left[\log{\Lambda\over2m}-0.845213 
\right]
\nonumber \\ && 
-{\pi\over6m}\alpha^2T^2\delta m^2\left[\log{\Lambda\over2m}-\log2
+{1\over8} \right]
-{1\over192 \pi m^3} \left(\delta m^2\right)^3\;.
\label{4fj}
\eqa\nonumber
\comment{mj}
Note that the contribution proportional to $\alpha \left(\delta m^2\right)^2$ 
vanishes.
Combining this with the lower order results and imposing the variational 
condition gives the four-loop gap equation
\bqa
\left(m^2-m_E^2\right)^3 &=& - {32\pi^2\over3}\alpha^2 m^2 T^2 (m^2 - m_E^2)
\left[\log{\Lambda\over 2m}-\log2+{9\over8}\right] \nonumber \\
&&\hspace{-4mm}+{16 \pi^3 \over 3} \alpha^3 m^3 T^3 \left(\pi^2 - 4 \right) 
+ 48 \pi \alpha^2 m^5 T \left( \log{\mu\over 4 \pi T} + \gamma \right)
\,.
\label{drspt4lg}
\eqa
\comment{mj}
Note that Eq.~(\ref{drspt4lg}) contains a contribution from 
the two-loop contribution (\ref{2fj}) 
evaluated using the NLO expression for $\lambda$ given by Eq.~(\ref{lnlo}). 
The soft part of the four-loop DRSPT approximation to the free energy is given
by the sum of~(\ref{1fj}),~(\ref{2fj}),~(\ref{3fjr}) and~(\ref{4fj}), again
with (\ref{2fj}) evaluated using the NLO expression for $\lambda$ 
given by Eq.~(\ref{lnlo}).

In Fig.~\ref{4f4} we have plotted the 
two-, three-, and four-loop DRSPT results for the 
pressure as a function of $g(2\pi T)$.
The fact that the four-loop  DRSPT approximation deviates from the 
two and three-loop DRSPT approximations is related to the 
fact that we have taken the hard contribution to the free energy
of order $\alpha^3$ to be identically zero which is highly
improbable; however, absent of a calculation of this coefficient
we prefer to not play games. Unfortunately, the calculation of the 
contributing four-loop diagrams is a highly nontrivial task.
Note that to obtain the three-loop approximation we used the gap
equation obtained at four-loops which is independent of this unknown
constant.   We use this gap equation because the three-loop
gap equation does not have a solution for $g\neq0$.~\footnote{
This is 
similar to the behavior found in the three-loop SPT gap equation \cite{spt} and 
the three-loop gap equation in DRSPT applied to QCD.~\cite{bir3}}
In Fig.~\ref{4f4}(b) we compare the three-loop DRSPT result with
the $g^5$ perturbative result and the three-loop SPT result.  As can
be seen from this Figure, although DRSPT improves the situation in terms
of the convergence of successive approximations, the result is still 
very sensitive to the renormalization scale having a larger variation than
even the perturbative result.  In the next section we
will demonstrate an alternative method for reorganizing the dimensionally-reduced
calculation.


%% file: pdr.tex
\section{Dimensionally Reduced $\Phi$-derivable Approximation}
\label{drphisec}

In this section, we apply the $\Phi$-derivable approach to the 
dimensionally reduced (DR) scalar field theory defined by~(\ref{leffsca}).
The total free energy is given by the sum of the hard and the 
soft contributions. The hard contribution is given by strict perturbation
theory and is an expansion in powers of $\alpha$. Through order
$\alpha^2$, the expression is given by~(\ref{fmatch}). The soft contribution
is given by calculating the free energy for the 
effective theory~(\ref{leffsca}) in the three dimensions
using the $\Phi$-derivable approach.
We will use the strategy developed by 
Braaten and Petitgirard in Ref.~\cite{ep1}
and explained in Sec.~\ref{phi}.

\subsection{Two-loop DR $\Phi$-derivable approximation}
The thermodynamic potential $\Omega_2$ in 
two-loop DR $\Phi$-derivable approximation is given
by
\bqa\nonumber
\Omega_2&=&
{1\over2}\int_p\log\left(p^2+m^2_E+\Pi(p)\right)
-{1\over2}\int_p{\Pi(p)\over p^2+m^2_E+\Pi(p)}
\\ &&
+{1\over8}\lambda\left(\int_p{1\over p^2+m^2_E+\Pi(p)}\right)^2\;.
\label{tp3d}
\eqa
\comment{mj}

Again the self-energy $\Pi(p)$ is momentum independent. If we write
$m^2=m^2_E+\Pi$, the gap 
equation that follows from varying~(\ref{tp3d}) is
\bqa
m^2&=&m^2_E+{1\over2}\lambda\int_p{1\over p^2+m^2}\;.
\eqa
\comment{mj}
Both the thermodynamic potential and the gap equations are
finite with dimensional regularization. Using~(\ref{1l0}) and~(\ref{1l1}) 
in appendix B, we obtain
\bqa
\Omega_2&=&-{m^3\over12\pi}-{1\over8}{m^2T}\alpha
\;,\\
m^2&=&m^2_E-2\pi\alpha mT\;.
\eqa
\comment{mj}
Note that the gap equation and the
thermodynamic potential are the same as 
those obtained in the two-loop
DRSPT approximation. The two-loop 
DR $\Phi$-derivable approximation then agrees with the weak-coupling
expansion through order $g^3$.

\subsection{Three-loop DR $\Phi$-derivable approximation}

The three-loop DR $\Phi$-derivable thermodynamic potential $\Omega_3$ is
given by
\bqa\nonumber
\Omega_3&=&{1\over2}\int_p\log\left(p^2+m^2_E+\Pi(p)\right)
-{1\over2}\int_p{\Pi(p)\over p^2+m^2_E+\Pi(p)}
\\ && \nonumber
\hspace{-8mm}+{1\over8}\lambda\left(\int_p{1\over p^2+m^2_E+\Pi(p)}\right)^2
-{1\over48}\lambda^2\int_{pqr}{1\over p^2+m^2_E+\Pi(p)}
{1\over q^2+m^2_E+\Pi(q)}
\\ && \hspace{-8mm} \times
{1\over r^2+m^2_E+\Pi(r)}
{1\over s^2+m^2_E+\Pi(s)}
+{1\over2}\Delta_2m^2_E\int_p{1\over p^2+m^2_E+\Pi(p)}\;.
\label{3drphi}
\eqa
\comment{mj}
where $\Delta_2m^2_E$ is the mass counterterm
and ${\bf s}=-({\bf p}+{\bf q}+{\bf r})$. 
\bqa
\Delta_2m^2_E&=&{g^4T^2\over24(4\pi)^2\epsilon}\;.
\label{cmdrp}
\eqa
\comment{mj}

This mass counterterm is identical to the one used in the dimensional reduction
step and in DRSPT.
The gap equation obtained by varying~(\ref{3drphi}) with respect to $\Pi(p)$ is
\bqa\nonumber
\Pi(p)&=&{1\over2}\lambda\int_q{1\over q^2+m^2_E+\Pi(q)}
\\ &&
\hspace{-13mm}
-{1\over6}\lambda^2\int_{qr}
{1\over q^2+m^2_E+\Pi(q)}{1\over r^2+m^2_E+\Pi(r)}
{1\over s^2+m^2_E+\Pi(s)}
+\Delta_2m^2_E\,.
\label{newgap}
\eqa
\comment{jm}
$\Omega_3$ can be simplified by substituting the gap equation~(\ref{newgap})
into~(\ref{cmdrp}):
\bqa\nonumber
\Omega_3&=&
{1\over2}\int_p\log\left(p^2+m^2_E+\Pi(p)\right)
-{1\over4}\int_p{\Pi(p)\over p^2+m^2_E+\Pi(p)}
\\ && \nonumber \hspace{-5mm}
+{1\over48}\lambda^2\int_{pqr}{1\over p^2+m^2_E+\Pi(p)}
{1\over q^2+m^2_E+\Pi(q)}
\\ && \hspace{-5mm} \times
{1\over r^2+m^2_E+\Pi(r)}
{1\over s^2+m^2_E+\Pi(s)}
+{1\over4}\Delta_2m^2_E\int_p{1\over p^2+m^2_E+\Pi(p)}\,.
\eqa
\comment{mj}
In the three-loop DR $\Phi$-derivable approximation, the gap equation is no
longer independent of the external momentum and we shall carry out the same
mass expansion as we did in Sec.~\ref{phi}. The Debye mass now satisfies
\bqa
m^2&=&m_E^2+{1\over2}\lambda\int_q{1\over q^2+m^2_E+\Pi(q)}
-{1\over6}\lambda^2I_{\rm Debye}
+\Delta_2m^2_E\;.
\eqa
\comment{mj}
The variational equation~(\ref{newgap}) then reduces to
\bqa
\Pi(p)&=&m^2-m_E^2+{1\over6}\lambda^2\left[I_{\rm Debye}-I_{\rm sun}\right]\;.
\eqa
\comment{mj}
We distinguish between hard and soft
loop momenta and the expansions for the self-energies are, respectively,
\bqa
\Pi(p)&=&m^2-m^2_E+\lambda^2
\left[\Pi_{4,0}(p)+\Pi_{4,1}(p)+...\right]+...\;,\;\;[{\rm p\;\;hard}]
\;,\label{hp}\\
\Pi(p)&=&m^2-m^2_E+\lambda^2\left[\sigma_{4,-2}(p)+\sigma_{4,0}(p)+...\right]
+...\;,\;\;[{\rm p\;\;soft}]\;.
\eqa
\comment{mj}
It should be noted that the expansion~(\ref{hp}) for hard momenta
is never needed. The reason is that if one momentum $p$ in a loop integral is
hard, one can expand the self-energy in powers of $\Pi(p)/p^2$.
The resulting integrals have no scale and vanish in 
dimensional regularization.~\footnote{This is in contrast to the
$\Phi$-derivable approximation in 3+1 dimensions, 
where the nonzero Matsubara frequencies set 
the scale.}

By inserting the expansion for $\sigma(p)$ into the gap equation, we 
can determine the functions $\sigma_{n,k}(p)$ by matching coefficients
of order $\lambda^n$ on both sides. The only function we need to carry out
calculations to order $g^5$, is $\sigma_{4,-2}(p)$ which is given
in Eq.~(\ref{onlyf}).
The gap equation reduces to
\bqa\nonumber
m^2-m^2_E&=&{1\over2}\lambda\int_p{1\over p^2+m^2}
-{1\over6}\lambda^2I_{\rm Debye}
\\ &&
-{1\over2}\lambda^3\int_p{\sigma_{4,-2}(p)\over(p^2+m^2)^2}
+\Delta_2 m^2_E\;.
\eqa
\comment{mj}
Using the expressions for the integrals in appendix B
and for the mass counterterm~(\ref{cmdrp}), the 
renormalized gap equation reduces to
\bqa\nonumber
m^2-m^2_E&=&-2\pi\alpha mT
-{8\over3}\pi^2\alpha^2T^2\left[\left(\log{\Lambda\over2m}
+{3\over2}-2\log2\right)
\right. \\ &&\left.
\hspace{1cm}
+\pi\alpha{T\over m}
\left(1-\log2\right)
\right]
\;.
\label{drphimass}
\eqa
\comment{mj}
After substituting the expression~(\ref{masspara}) for the mass parameter,
the $\Lambda$-dependence in~(\ref{drphimass}) cancels to next-to-leading
order in $g$. 

Again, using the expansions of the integrals in appendix C, 
the dimensionally reduced three-loop $\Phi$-derivable approximation,
$\Omega_3$,
to the thermodynamic potential through order $g^5$ 
reduces to
\bqa\nonumber
\Omega_3&=&{1\over2}\int_p\log(p^2+m^2)
-{1\over4}\int_p{m^2-m_E^2\over p^2+m^2}
+{1\over24}\lambda^2I_{\rm Debye}\int_p{1\over p^2+m^2}
\\ && \nonumber
-{1\over48}\lambda^2\int_{pqr}{1\over p^2+m^2}
{1\over q^2+m^2}{1\over r^2+m^2}
{1\over({\bf p}+{\bf q}+{\bf r})^2+m^2}
\\ &&
+{1\over4} \lambda^2 m^2 \int_p {\sigma_{4,-2} \over (p^2+m^2)^2}
+{1\over4}\Delta_2m^2_E\int_p{1\over p^2+m^2}
\;.
\label{omega3d}
\eqa
\comment{jm}
Using the expressions for the integrals in appendix B
and for the mass counterterm~(\ref{cmdrp}), we obtain 
the soft contribution to the renormalized three-loop free energy 
%
\bqa
\Omega_3&=&-{m^3\over12\pi}
-{m\over16\pi}\!\left(m^2-m^2_E\right)
+{1\over6}\pi mT^2\alpha^2\!\left(\log{\Lambda\over2m}
+2-{1\over2}\log 2\right)\,.
\label{3ldr}
\eqa
\comment{jm}
The complete three-loop DR $\Phi$-derivable approximation to the 
free energy is given by adding~(\ref{fmatch}) and~(\ref{3ldr}).
Upon substituting the expression for the mass parameter~(\ref{masspara})
and using the fact that $m^2$ is independent of $\Lambda$
to next-to-leading order, it is easy to
show that $\Lambda$-dependence in the soft contribution~(\ref{3ldr}) 
cancels to next-to-leading order in $g$. Using Eq.~(\ref{drphimass}), 
one can show that
the complete three-loop DR $\Phi$-derivable approximation to the 
free energy agrees with the weak-coupling expansion through order $g^5$.

\begin{figure}
\centerline{\psfig{file=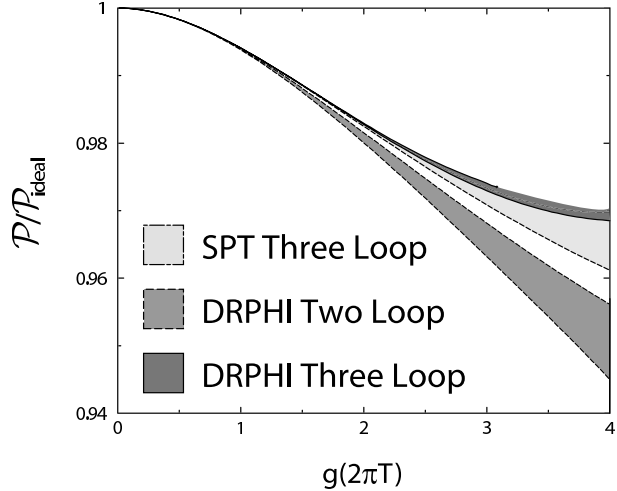,width=8cm}}
\caption{DR $\Phi$-derivable approximations to the pressure divided by that
of an ideal gas. 
For comparison we have shown the
three-loop SPT result for the pressure.}
\label{drphi}
\end{figure}

In Fig.~\ref{drphi}, we show the two- and three-loop DR $\Phi$-derivable
approximations to the thermodynamic pressure.
The bands are obtained by varying the renormalization scale by a factor of two
around the central value $\mu=2\pi T$.
From the Figure, we see that DRPHI also has better convergence than the 
perturbative expansion~(\ref{fscaw}).
Note that for $\mu=4\pi T$ the DRPHI three-loop gap equation has no
solution beyond $g\sim3.2$ which we have indicated by not continuing
the topmost DRPHI line in Fig.~\ref{drphi}.
For comparison, we have also shown the three-loop SPT approximation to the
pressure. The agreement between the two approaches is very good and the
final scale variation of the DRPHI result is smaller than the SPT result
which is most likely due to the fact that with three-loop SPT we are forced
to use a two-loop gap equation owing to the sickness of the three-loop 
variational gap equation.
The dimensionally reduced $\Phi$-derivable approach therefore 
offers an economical alternative to the full treatment in four dimensions. 
However, as in the
case of DRSPT, it is limited to calculating static quantities.

%% file: concl.tex
\section{Conclusions}
\label{conclude}

In this paper we have reviewed some of the progress that 
has been made concerning our understanding of ultrarelativistic hot 
field theories over the past 10-15 years concentrating on the 
systematic computation of thermodynamic observables.  We have 
presented results using multiple methods of computation and
perturbative reorganization.  In particular we have discussed
standard perturbative techniques, dimensional reduction, 
effective field theory methods, screened perturbation theory (SPT),
hard-thermal-loop perturbation theory (HTLPT),
the $\Phi$-derivable approximation, dimensionally reduced (DR) SPT,
and the DR $\Phi$-derivable approximation.

We began by showing that the perturbative expansions of the free
energy for both simple scalar field theories and QCD have a 
very small, if not vanishing, radius of convergence.  In the case of 
QCD this makes
the resulting expansions useless at phenomenologically relevant 
temperatures where $\alpha_s\sim0.2$ ($g_s \sim 2$).  We then showed that 
dimensional reduction and effective field theory methods can provide 
an efficacious way of organizing weak-coupling calculations.  By 
combining them with nonperturbative methods such as lattice gauge 
theory, the long-standing problem of calculating the order-$\alpha_s^3$ 
contribution to the QCD pressure can be solved.  

Another advantage
of using dimensional reduction and effective field theory methods
is that the explicit separation of the hard and soft scales allows for
a different treatment of the physics at these scales.  We showed,
for example, that if the soft-sector is treated non-perturbatively, 
the successive approximations to the thermodynamic functions are
much better behaved, giving reasonably constrained results for the
pressure in scalar field theories to $g \sim 4$. We showed that
two reorganizations of the soft-sector computation, the
DRSPT and DR $\Phi$-derivable approach, both gave better 
convergence of the successive approximations to the pressure at 
large coupling.

While the results are impressive, the methods are limited because
they can only be applied to
the computation of static quantities such as the pressure.  It
would be preferable to have a reorganization of the soft-sector
physics which could be applied to the calculation of real-time quantities
as well.
This desire has motivated the development of the $\Phi$-derivable approach
and the SPT and HTLPT reorganizations. The $\Phi$-derivable approach has 
only been taken to three-loops for
a scalar field theory where it improves covergence and scale variation of
the approximations to the pressure.  It would be nice to systematically extend this approach
also to three loops for QCD.  It has been demonstrated that when applied to the 
calculation of the
pressure in QCD the HTLPT reorganization yields impressive results
when one considers the convergence of the successive approximations.
Additionally, results are explicitly gauge-invariant within HTLPT
as opposed to $\Phi$-derivable approximations where gauge invariance
can only be guaranteed to a fixed order in the coupling constant.

Despite these theoretical benefits, when 
compared to existing lattice data for the pressure in 
the range $T_c < T < 5\,T_c$, the NLO 
HTLPT prediction for the pressure seems to underestimate 
the correction to ideal gas behavior.  
The failure of the method
in this region could come from the fact that within HTLPT the
requirement of gauge invariance causes a simultaneous modification of both
soft and hard modes.  Perhaps a method that makes an explicit
separation between these two scales is more appropriate.  
Another possible explanation
is that HTLPT, as implemented, does not properly take into account
the proper scaling relations for quantities near the phase transition
which is governed (approximately) by the center group of SU(3), Z(3).
However, deciding which of these two possibilities is the culprit
is not possible at this stage.  Results from the 
approximately self-consistent $\Phi$-derivable approach of Blaizot, Iancu,
and Rebhan seem to point to the first possibility, but are by no
means conclusive.  The only thing that can resolve this is 
further study.  We also mention in closing that given the success
of the DR $\Phi$-derivable approximation discussed in Sec.~\ref{drphisec}
it would be interesting to apply this method also to the QCD pressure.

%% file: app.tex
\appendix

\section{Sum-integrals}
\label{app1}
In the imaginary-time formalism for thermal field theory, 
the 4-momentum $P=(P_0,{\bf p})$is Euclidean with $P^2=P_0^2+{\bf p}^2$. 
The Euclidean energy $p_0$ has discrete values:
$P_0=2n\pi T$ for bosons and $P_0=(2n+1)\pi T$ for fermions,
where $n$ is an integer. 
Loop diagrams involve sums over $P_0$ and integrals over ${\bf p}$. 
With dimensional regularization, the integral is generalized
to $d = 3-2 \epsilon$ spatial dimensions.
We define the dimensionally regularized sum-integral by
\bqa
\label{sumint-def}
  \hbox{$\sum$}\!\!\!\!\!\!\int_{P}& \;\equiv\; &
  \left(\frac{e^\gamma\mu^2}{4\pi}\right)^\epsilon\;
  T\sum_{P_0=2n\pi T}\:\int {d^{3-2\epsilon}p \over (2 \pi)^{3-2\epsilon}}\;,\\ 
  \hbox{$\sum$}\!\!\!\!\!\!\int_{\{P\}}& \;\equiv\; &
  \left(\frac{e^\gamma\mu^2}{4\pi}\right)^\epsilon\;
  T\sum_{P_0=(2n+1)\pi T}\:\int {d^{3-2\epsilon}p \over (2 \pi)^{3-2\epsilon}}\;,
\label{sumint-def2}
\eqa 
\comment{mj}
where $d=3-2\epsilon$ is the dimension of space
and $\mu$ is an arbitrary
momentum scale. The factor $(e^\gamma/4\pi)^\epsilon$
is introduced so that, after minimal subtraction
of the poles in $\epsilon$
due to ultraviolet divergences, $\mu$ coincides
with the renormalization
scale of the $\overline{\rm MS}$ renormalization scheme.

\subsection{One-loop sum-integrals}
We need several simple massless one-loop sum-integrals
\bqa
\sumint_P \log P^2 &=& -{\pi^2 \over 45} T^4  + {\mathcal O}(\epsilon)\,,
\\
\sumint_P {1 \over P^2} \hspace{0.5cm} &=&
T^2 \left({\mu\over4\pi T}\right)^{2\epsilon}
{1 \over 12} \left[ 1
	+ \left( 2 + 2{\zeta'(-1) \over \zeta(-1)} \right) \epsilon
        + {\mathcal O}(\epsilon^2)
         \right] \,,
\label{sumint:2}
\\
\sumint_P {1 \over (P^2)^2} &=&
{1 \over (4\pi)^2} \left({\mu\over4\pi T}\right)^{2\epsilon}
\left[ {1 \over \epsilon} + 2 \gamma
	 + {\mathcal O}(\epsilon)
	 \right] \,,
\\
\sumint_{\{P\}}{1\over P^2}
&=&-{T^2\over24}
        + {\mathcal O}(\epsilon)
\;,
\label{simple1}
\eqa
\comment{jm}
The calculation of these sum-integrals is standard.
%
We also need some massive one-loop sum-integrals
\bqa
\label{s1}
  \sumint_P\log\left(P^2+m^2\right)&=&
{1\over(4\pi)^2}\left({\mu\over m}\right)^{2\epsilon}
\left[-{e^{\gamma\epsilon}\Gamma(1+\epsilon)
	\over\epsilon(1-\epsilon)(2-\epsilon)}m^4-J_0T^4
\right]\;,\\
\label{s2}
\sumint_P\frac{1}{P^2+m^2}&=&
{1\over(4\pi)^2}\left({\mu\over m}\right)^{2\epsilon}
\left[-{e^{\gamma\epsilon}\Gamma(1+\epsilon)\over\epsilon(1-\epsilon)}m^2
+J_1T^2\right]\;,
\\
\label{s3}
\sumint_P\frac{1}{(P^2+m^2)^2}&=&
{1\over(4\pi)^2}\left({\mu\over m}\right)^{2\epsilon}
\left[{e^{\gamma\epsilon}\Gamma(1+\epsilon)\over\epsilon}
+J_2\right]\;.
\eqa
\comment{mj}
The thermal terms can be expressed as integrals involving the Bose-Einstein
distribution function:
\bqa\nonumber
J_n(\beta m)&=&
{4e^{\gamma\epsilon}\Gamma({1\over2})\over\Gamma({5\over2}-n-\epsilon)}
\beta^{4-2n}m^{2\epsilon}
\int_0^{\infty}dk\;{k^{4-2n-2\epsilon}\over\left(k^2+m^2\right)^{1/2}}
{1\over e^{\beta\left(k^2+m^2\right)^{1/2}}-1}\;. 
\\ &&
\label{jndef}
\eqa
\comment{mj}These integrals satisfy the recursion relation
\bqa
\label{rec}
xJ_n^{\prime}(x)=2\epsilon J_n(x)-2x^2J_{n+1}(x)\;.
\eqa
\comment{mj}
In the limit $\beta m\longrightarrow 0$, these
integrals reduce to
\bqa
J_0 &\longrightarrow& {16\pi^4\over45}-{4\pi^2\over3}(\beta m)^2
+{8\pi\over3}\left(\beta m\right)^3
	\nonumber \\
	&& \hspace{4cm} \;+\; \left( \log{\beta m\over4\pi}-{3\over4}+\gamma\right) 
		(\beta m)^4 \;,\\
\label{j1}
J_1 &\longrightarrow& {4\pi^2\over3}
	\;-\; 4\pi \beta m
	\;-\; 2 \left( \log{\beta m\over4\pi}-{1\over2}+\gamma\right) 
		(\beta m)^2 \;,\\
J_2 &\longrightarrow& {2\pi\over\beta m}
	\;+\; 2 \left( \log{\beta m\over4\pi}+\gamma\right)\;,
\label{j2exp}
\eqa
\comment{mj}

\subsection{Two-loop sum-integrals}
We need a single massless two-loop sum-integral which happens
to vanish:
\bqa
\sumint_{PQ}{1\over P^2Q^2(P+Q)^2}&=&0\;.
\label{tl}
\eqa
\comment{mj}


\subsection{Three-loop sum-integrals}

We need a single massless three-loop sum-integral:
\bqa\nonumber
\sumint_{PQR}{1\over P^2Q^2R^2(P+Q+R)^2}&=&
{T^4\over24(4\pi)^2}\left[{1\over\epsilon}+6\log{\mu\over4\pi T}
+8{\zeta^{\prime}(-1)\over\zeta(-1)}
\right.\\ &&\left.
\hspace{15mm}
-2{\zeta^{\prime}(-3)\over\zeta(-3)}
+{91\over15}
+{\mathcal O}(\epsilon)\right]\;.
\eqa
\comment{mj}

\section{Three-dimensional integrals}
Dimensional regularization can be used to
regularize both the ultraviolet divergences and infrared divergences
in 3-dimensional integrals over momenta.
The spatial dimension is generalized to  $d = 3-2\epsilon$ dimensions.
Integrals are evaluated at a value of $d$ for which they converge and then
analytically continued to $d=3$.
We use the integration measure
\begin{equation}
 \int_p\;\equiv\;
  \left(\frac{e^\gamma\mu^2}{4\pi}\right)^\epsilon\;
\:\int {d^{3-2\epsilon}p \over (2 \pi)^{3-2\epsilon}}\,,
\label{int-def}
\end{equation}
\comment{mj}
We require several integrals that do not involve the
Bose-Einstein distribution function.
The momentum scale in these integrals is set by the mass
parameter $m$ or $m_D$.

\subsection{One-loop integrals}
The one-loop integrals are
\bqa
\int_p\log\left(p^2+m^2\right)&=&
-{m^3\over6\pi}\left[1+\left(
2\log{\mu\over2m}+{8\over3}\right)\epsilon+{\mathcal O}(\epsilon^2)\right]
\;,\label{1l0}\\
\int_p{1\over p^2+m^2}&=&
-{m\over4\pi}\left[1+\left(
2\log{\mu\over2m}+2\right)\epsilon+{\mathcal O}(\epsilon^2)\right]
\label{1l1}
\;,\\
\int_p{1\over\left(p^2+m^2\right)^2}&=&
{1\over8\pi m}\left[1+\left(
2\log{\mu\over2m}\right)\epsilon+{\mathcal O}(\epsilon^2)\right]
\label{1l2}
\;,\\
\int_p{1\over\left(p^2+m^2\right)^3}&=&
{1\over32\pi m^3}\left[1+\left(
2\log{\mu\over2m}+2\right)\epsilon+{\mathcal O}(\epsilon^2)\right]
\;, \\
\int_p{\sigma_{4,-2}(p)\over(p^2+m^2)^2}&=&
{T^2\over12(4\pi)^3m}\left(1-\log2\right)\;.
\label{new}
\eqa
\comment{mj}The calculation of~(\ref{1l0})--(\ref{1l2}) are standard. The integral
in~(\ref{new}) was calculated by Braaten and Petitgirard~\cite{ep1}.

\subsection{Two-loop integrals}
The only two-loop integral required is
\bqa\nonumber
I_{\rm Debye}&=& \int_{qr} {1\over(q^2+m^2)(r^2+m^2)(({\bf p+q+r})^2+m^2)} \bigg|_{p=i m}
\nonumber \\ &=&
{1\over4(4\pi)^2}\left[
{1\over\epsilon}+4\log{\mu\over2m}+6-8\log2
+
{\mathcal O}(\epsilon)
\right]\;.
\eqa
\comment{jm}

\subsection{Three-loop integrals}

The three-loop integrals required are
\bqa\nonumber
\int_{pqr}{1\over p^2+m^2}{1\over q^2+m^2}{1\over r^2+m^2}
{1\over({\bf p}+{\bf q}+{\bf r})^2+m^2}
&& 
\\ && \hspace{-6cm}
= -{m\over(4\pi)^3}\left[
{1\over\epsilon}+6\log{\mu\over2m}
+8-4\log2+{\mathcal O}(\epsilon)
\right] . \\ \nonumber
\int_{pqr}{1\over p^2+m^2}{1\over q^2+m^2}{1\over r^2+m^2}
{1\over[({\bf p}+{\bf q}+{\bf r})^2+m^2]^2}
&&
\\ && \hspace{-6cm}
= {1\over8m(4\pi)^3}\left[
{1\over\epsilon}+6\log{\mu\over2m}
+2-4\log2+{\mathcal O}(\epsilon)
\right] . 
\label{3l1}
\eqa
\comment{mj}

\subsection{Four-loop integrals}
The only four-loop integral needed is
\bqa\nonumber
&&\int_{pqrs}
{1\over p^2+m^2}{1\over q^2+m^2}{1\over r^2+m^2}{1\over s^2+m^2}
{1\over ({\bf p}+{\bf q}+{\bf r})^2+m^2} \\
&&\hspace{4cm}\times\;{1\over ({\bf p}+{\bf r}+{\bf s})^2+m^2} \nonumber \\
&&\hspace{1cm}=
{\pi^2\over32(4\pi)^4}\left[
{1\over\epsilon}+8\log{\Lambda\over2m}+2+4\log2-{84\over\pi^2}\zeta(3)
+{\mathcal O}(\epsilon)
\right]\;.
\eqa
\comment{mj}
This was calculated in Ref.~\cite{kajaloop}.
\section{$m/T$ expansions of sum-integrals}
We next consider the contribution to the free energy ${\mathcal F}_{\rm 2b}$ 
from the basketball diagram. There are two momentum scales in that the
sum-integral; the hard scale $T$ and the soft scale $gT$. The hard scale
includes all nonzero Matsubara frequencies for all momenta $p$
and also the region ${p}$ of order $T$ with $n=0$. The soft scale 
involves the region $p$ of order $gT$ with $n=0$.
Since there are three sum-integrals, all momenta can be hard, two momenta
can be hard and one soft, two momenta can be soft and one hard, or 
all three momenta can be soft. These regions are denoted by $(hhh)$, $(hhs)$,
$(hss)$, and $(sss)$, respectively. If a momentum is hard, one can expand
in powers of $m/T$. If a  momentum is soft, one must keep the mass in the
propagator.  We begin by writing
\bqa
{\mathcal F}_{\rm 2b} = - {g^4 \over 48} {\mathcal I}_{\rm ball} \; ,
\eqa
\comment{mj}
where
${\mathcal I}_{\rm ball}$ is the basketball sum-integral:
\begin{eqnarray}
{\mathcal I}_{\rm ball} &=& 
\sumint_{PQR} {1 \over P^2 + m^2} {1 \over Q^2 + m^2} 
	{1 \over R^2 + m^2} {1 \over S^2 + m^2} \,,
\label{IBall}
\end{eqnarray}
\comment{mj}
where $S = -(P+Q+R)$.  The various contributions to ${\mathcal I}_{\rm ball}$ are
\bqa
{\mathcal I}_{\rm ball}^{(hhh)}&=&
\sumint_{PQR}^{\prime}{1\over P^2Q^2R^2(P+Q+R)^2}\;,\\
{\mathcal I}_{\rm ball}^{(hhs)}&=&4T\int_p{1\over p^2+m^2}\;
\sumint_{QR}^{\prime}{1\over Q^2R^2(Q+R)^2}\;,\\
{\mathcal I}_{\rm ball}^{(hss)}&=&4T^2\int_{pq}{1\over p^2+m^2}{1\over q^2+m^2}
\sumint_R^{\prime}{1\over R^4}\;,\\
{\mathcal I}_{\rm ball}^{(sss)}&=&
T^3\int_{pqr}{1\over p^2+m^2}{1\over q^2+m^2}{1\over r^2+m^2}
{1\over ({\bf p}+{\bf q}+{\bf r})^2+m^2}\;.
\eqa
\comment{jm}
Using Eqs.~(\ref{tl}),~(\ref{1l1}), and the fact that $m\sim gT$, we obtain
Eq.~(\ref{32b}).

We need the $m/T$ expansions of various one-loop
sum-integrals appearing in the two and three-loop $\Phi$-derivable 
approximation to the free energy. 
We need the expansions for both soft and
hard momenta. Inserting the expansions~(\ref{hexp}) and~(\ref{sexp}),
we obtain for the one-loop sum-integrals
\bqa
\sumint_{P}^{(h)}\log\left(P^2+\Pi(P)\right)&=&
\sumint_P\left[\log P^2+{m^2\over P^2}-{1\over2}{m^4\over P^4}\right]
\nonumber \\
&& \hspace{1cm}
+g^4\sumint_P{\Pi_{4,0}(P)+\Pi_{4,1}(P)\over P^2}\;, \\
\sumint_{P}^{(s)}\log\left(P^2+\Pi(P)\right)&=& 
T\int_p\log\left(p^2+m^2\right)+g^4T\int_p{\sigma_{4,-2}(p)\over p^2+m^2}
\;,\\ 
\sumint_P^{(h)}{\Pi(P)\over P^2+\Pi(P)}&=&
\sumint_P\!\left[
{m^2\over P^2}-{m^4\over P^4}
\right]
+g^4\sumint_P{\Pi_{4,0}(P)+\Pi_{4,1}(P)\over P^2}
\,,\\ 
\sumint_P^{(s)}{\Pi(P)\over P^2+\Pi(P)}&=&
T\int_p{m^2\over p^2+m^2}
+g^4T\int_p{p^2\sigma_{4,-2}(p)\over(p^2+m^2)^2}
\;, \\
\sumint_P^{(h)}{1\over P^2+\Pi(P)}&=&\sumint_P
\left[{1\over P^2}-{m^2\over P^4}\right]
\;,\\
\sumint_P^{(s)}{1\over P^2+\Pi(P)}&=&
T\int_p{1\over p^2+m^2}
-g^4T\int_p{\sigma_{4,-2}(p)\over(p^2+m^2)^2}\;.
\eqa
\comment{jm}
Similarly, we need the $m/T$ expansions of some two-loop sum-integrals
\bqa
{\mathcal I}_{\rm Debye}^{(hh)}&=&
\sumint_{QR}{1\over Q^2R^2(Q+R)^2}+{\mathcal O}\left(g^2\right)\;,\\
{\mathcal I}_{\rm Debye}^{(hs)}&=&
3\sumint_P{1\over P^4}T\int_p{1\over p^2+m^2}+{\mathcal O}\left(g^2\right)\;,\\
{\mathcal I}_{\rm Debye}^{(ss)}&=&T^2I_{\rm Debye}\;.
\eqa
\comment{jm}
Finally, we need the $m/T$ expansion of a single three-loop sum-integral.
Through order $g^5$ we have 
\bqa\nonumber
&&\sumint_{PQR}^{(hhh)}{1\over P^2+\Pi(P)}
{1\over Q^2+\Pi(Q)}{1\over R^2+\Pi(R)}{1\over S^2+\Pi(S)}=
\\ &&\hspace{1cm}\sumint_{PQR}{1\over P^2Q^2R^2S^2}
+{\mathcal O}\left(g^2\right)\;,
\\\nonumber 
&&\sumint_{PQR}^{(sss)}{1\over P^2+\Pi(P)}
{1\over Q^2+\Pi(Q)}{1\over R^2+\Pi(R)}{1\over S^2+\Pi(S)}=\\ 
&&\hspace{5mm}\int_{pqr}{1\over p^2+m^2}{1\over q^2+m^2}{1\over r^2+m^2}
{1\over({\bf p}+{\bf q}+{\bf r})^2+m^2}+{\mathcal O}\left(g^2\right)\;.
\eqa
\comment{jm}
The $(hhs)$ and $(hss)$ terms first start to contribute at order $g^2$.